\documentclass[aps,prb,twocolumn,reprint,superscriptaddress,nofootinbib,preprintnumbers,longbibliography,floatfix,showpacs]{revtex4-2}
\usepackage{graphicx}
\usepackage{color}
\usepackage{amsmath}
\usepackage{mathtools}
\DeclarePairedDelimiter\ket{\lvert}{\rangle}

\DeclarePairedDelimiterX\braket[2]{\langle}{\rangle}{#1\,\delimsize\vert\,\mathopen{}#2}
\begin{document}
\title{Impact of Majorana fermions on the Kondo state in the carbon nanotube quantum dot}
\author{D. Krychowski}
\affiliation{%
Institute of Molecular Physics, Polish Academy of Sciences\\M. Smoluchowskiego 17,
60-179 Pozna\'{n}, Poland
}%
\date{\today}

\begin{abstract}
We have studied the quantum conductance of the Kondo state in the carbon nanotube quantum dot (CNTQD) with side-attached multi-Majorana fermion states in topological superconductors (TSCs). The zero-energy Majorana fermions interfere with the fourfold degenerate states of the CNTQD in the spin-orbital Kondo regime. Using the extended Kotliar-Ruckenstein slave-boson mean-field approach, we have analyzed the symmetry reduction of the SU(4) Kondo effect to the SU$^{\star}$(3) Kondo state with a fractional charge in the system by increasing the tunneling strength to a single Majorana fermion (TSC). We observed  the fractional quantum conductance, the residual impurity entropy, the enhancement of the thermoelectric power with two compensation points, the fractional linear and nonlinear Fano factor ($F_{K}$) and the spin polarization of the conductance. Two Majoranas (2TSC) in conjunction with the CNTQD have reduced the spin-orbital Kondo effect to the SU$^{\star}$(2) Kondo state with 2e in the system. $F_{K}$ contains information about the effective charge and the interaction between the quasiparticles, two- and three-body correlators and identifies the broken symmetry of the Kondo state. We discussed the local quadratic Casimir operator separately for the states associated with the Kondo effect and the Majorana fermion state to show the difference between the fluctuations of the pseudospin in both quantum channels. We have shown that the device coupled with three Majorana fermions (3TSC) achieves a quantized conductance $5/2(e^{2}/h)$, conserves the U$^{\star}$(1) charge symmetry at the electron-hole symmetry point and manifests the increase in nonlinear current and shot noise due to the entanglement in octuplets with opposite charge-leaking states. Furthermore, we have investigated the influence of the spin-orbit interaction in the CNTQD-TSC device on the quantum transport properties.
\end{abstract}

\keywords{Kondo effect; topological qubits; Majorana fermions; CNTQD; shot noise}
\pacs{72.10.Fk, 73.63.Kv, 71.27.+a, 85.35.Kt, 74.45.+c, 05.30.Pr}
\maketitle

\section{Introduction}

The quantum dot can be realized in the carbon nanotubes due to the quantization effect of the confinement potential \cite{Sapmaz2006,Laird2015}. One of the interesting features in the low-dimensional systems is the Kondo effect \cite{Nygard2000,Schmid2000,Cleuziou2013,Keller2014,Sasaki2000,Jespersen2011}, where the localized pseudospin or spin states on the quantum dot are screened by the conduction electrons in the leads. The system is in the new ground state, called the Kondo singlet, which opens the transport window in the Coulomb blockade region \cite{Schmid2000}.

The screening effect plays one of the most important roles in the formation of the Kondo singlet \cite{Kondo1964}. One of the popular methods to solve the Kondo cloud problem is the numerical renormalization group (NRG) method \cite{Wilson1975}, where the screening strongly depends on the number of sites in the Wilson chain and is based on deleting the states in the growing Hilbert space. In another method, the poor man's scaling method, we solve the Kondo Hamiltonian (the effective Hamiltonian of the Anderson model reduced by the Schrieffer-Wolff canonical transformation using the projection operator technique). In the solution we observe an asymptotic strengthening of the Kondo exchange coupling, logarithmically proportional to the Kondo temperature.  Other methods, which include the problem of spin-orbital fluctuations, the finite Coulomb interaction, the Fermi liquid behavior \cite{Nozieres1974} and the scaling problem \cite{Haldane1978}, are the symmetrized finite-U NCA (non-crossing approximation) \cite{Kroha2001}, the equation of motion (EOM) method \cite{Hamann1967,Roermund2010}, or the extended Kotliar-Ruckenstein slave-boson mean-field approach (KR-sbMFA)\cite{Kotliar1986} based on path integral methods \cite{Coleman2015}. On the other hand, the Bethe ansatz approach (BAA) is reserved to the strong coupling limit (infinite-U) and is also limited to the number of states in the chain \cite{Coleman1986}, which is a mapping between a set of quantum numbers and a set of momenta. This map is nonlinear and fully coupled. In this paper \cite{Wiegmann1983,Okiji1983}, the authors showed that the single impurity Anderson model is completely integrable with finite U using the Bethe ansatz approach and solved the transcendent equations with the energies lying not far from the Fermi level. Formally, for non-trivial values of the interaction, Bethe's equations form a transcendental system of equations that can't be solved in closed form. However, as is often the case in statistical mechanics, it is hoped that moving to a thermodynamically large system will simplify matters somewhat. In any case, all these methods have some limitations, but in most general aspects they describe the N-orbital Anderson model with modifications.

The Kondo singlet has been realized in the heterostructure \cite{Goldhaber1998,Sasaki2000} and molecular devices \cite{Park2002,Grove2007,Kurzmann2021,Liang2002}, but was first discovered in the dilute alloys \cite{Kondo1964}. In contrast to heterostructure quantum dots \cite{Schmid2000,Sasaki2000,Keller2014}, the intra- and inter-Coulomb interactions in carbon nanotube quantum dots (CNTQD) are comparable, which is an essential condition for the formation of the SU(4) Kondo effect \cite{Cleuziou2013,Laird2015,Herrero2005,Makarovski2007,Hata2019}.
In CNTQD, the Kondo temperature  \cite{Herrero2005,Makarovski2007,Nygard2000,Cleuziou2013} is three orders of magnitude higher than the milikelvin temperatures observed in GaAs quantum dots \cite{Schmid2000,Keller2014}. These two aspects are the features that determine the experimental attractiveness of CNTQDs in the strongly correlated electron regime \cite{Hewson1997,Haldane1978,Coleman2015}. The Seebeck coefficient for the Kondo-correlated single quantum dot transistor is suppressed
at the electron hole symmetry point and changes sign beyond this point \cite{Dutta2019,Molenkamp2005}. This is a result of the linear response theory \cite{Costi1994} and is a consequence of Fermi liquid (FL) behavior. For temperatures below the Kondo temperature $T_{K}$ only the linear coefficient contributes to the Seebeck effect and the thermoelectric power (TEP) is proportional to the ratio of the derivative of the quasiparticle densities to itself (the Mott's formula) \cite{Oguri2020}.
The Seebeck effect has also been observed in the Coulomb blockade regime \cite{Trocha2012}.
The authors observed the large enhancement of the thermoelectric power (TEP) in a double quantum dot system due to interference and Coulomb correlation effect.

The model of CNTQD is described by the linear approach of graphene bands \cite{Jespersen2011,Niklas2016}. The quantum numbers that characterize the carbon quantum dots are the spin, isospin (valley) and pseudospin (lattice) numbers. The source of the energy gap in semiconducting CNTQDs is the chirality of the nanotube (the gap is proportional to one over the diameter of the nanotube) and the perturbation effects (the perturbation gap is smaller than the geometry gap and proportional to one over the square of the diameter of the nanotube) \cite{Jespersen2011,Laird2015}. Despite the fact that the spin-orbit interaction (SOI) is negligible in graphene, the SOI has been uncovered in the semiconducting CNTQD \cite{Kuemmeth2008,Cleuziou2013,Laird2015}. In flat graphene, the symmetry forbids direct hopping between orbitals with opposite parity under the inversion.
In a nanotube, the symmetry is broken by the curvature, and the SOI arises from the direct hybridization between the non-orthogonal orbitals on the A and the B sites in graphene \cite{Ando2000,Hernando2006,Jeong2009,Laird2015}. The spin-orbit coupling breaks the fourfold degeneracy in the shell of the CNTQD and leads to the ground state with two Kramers doublets \cite{Jespersen2011} (opposite pairs in the spin and isospin sectors).
Increasing the SOI changes the symmetry of the Kondo effect from SU(4) to SU(2) \cite{Niklas2016,Krychowski2018,Schmid2015,Galpin2010,Mantelli2016}. The SOI in CNTQD has two contributions: Zeeman (diagonal part in the A(B) lattice basis) and orbital (off-diagonal part). The value and sign of these interactions determine the ground state in the energy spectrum of holes and electrons on the multishell quantum dot. In general, the SOI is of the order of tens of meV \cite{Steele2013,Laird2015}, but for the ultraclean semiconducting CNTQD the coupling is comparable to the Coulomb interaction. Special attention has been paid to the problem of the square dependence in the magnetic field of the CNTQD states \cite{Steele2013} as a consequence of the narrow band gap in semiconducting CNTQD, which contributed to the spinful SU(3) Kondo state \cite{Krychowski2020}. The observation of the SU(3) Kondo effect can be identified by localizing the quantum conductance to the characteristic value $(9/4)(e^{2}/h)$ \cite{Zitko2013}.

One of the most important differences between SU(2) and SU(4) Kondo states in CNTQD, besides the conductance measurements, is the nonlinear shot noise detection using the lock-in technique \cite{Ferrier2016,Hata2019,Teratani2016,Ferrier2020,Delattre2009}.
The electronic transport is described by the free non-interacting quasiparticles around the equilibrium state (low bias). In the non-equilibrium regime, two-particle scattering processes dominate due to the residual interaction. Using the Fermi liquid theory \cite{Nozieres1974,LeHur2009,Mora2009,Oguri2018,Teratani2020,Mora2008,Oguri2022,Moca2015}, the authors showed that the interaction parameter  \cite{Nishikawa2013}, called the Wilson ratio, takes the quantized values $W=2$ for SU(2) and $W=4/3$ for SU(4) Kondo states and changes the corresponding effective charges e$^{\star}$ \cite{Sela2006}. For single Kondo QDs, the experimental measurements showed the three-body correlation function in the nonlinear conductance at finite magnetic field, validating the recent Fermi liquid theory in the nonequilibrium Kondo regime \cite{Hata2021}.

Majorana fermions are it own antiparticles originally proposed by Ettore Majorana \cite{Majorana1937}. They are called real fermion quasiparticles because of the real nature of the creation and annihilation operators \cite{Leijnse2012}. These quasiparticles exhibit non-Abelian braiding properties \cite{Flensberg2011,Leijnse2012,Ezawa2020} and the Majorana fermion states are associated with zero energy modes that occur in the Bogoliubov-de Gennes description of a paired condensate with non-Abelian exchange statistics \cite{Beenakker2013}. The search for Majorana quasi-particle bound states in condensed matter systems is motivated in part by their potential use as topological qubits and possible applications in quantum computation. The Majorana qubits are predicted for quantum states in the fault-tolerant non-Abelian quantum processors \cite{Kitaev2003,Beenakker2020,Nayak2008,Flensberg2011}. A pair of Majorana fermions can be combined into a complex Dirac fermionic state. The Majorana fermion in this composite complex fermion is half of a normal fermion, and is obtained as a superposition of two Majorana
fermions. Each of the Majorana fermions is basically split into a real and an imaginary part of a fermion. The Majorana fermions exist at the edges of proximitized quantum wire by p-wave superconductor \cite{Kitaev2001}. The states are spatially separated and protected from most types of decoherence. However, in \cite{Lai2020} as a result, the authors showed that the Majorana qubit coherence and the fermion parity conservation cannot be immune to local perturbations during the braiding operations.

The statistical thermodynamic calculation based on the NRG method showed that the entropy of the quantum dot coupled to a single Majorana fermion leads to $S_{tot}/k_{B}=\ln[2]/2$ and corresponds to the NFL behavior,  confirming the anyonic non-Abelian nature of the hybrid device \cite{Zitko2011}. The author presented the contribution of the impurity to the electron entropy as a function of the temperature for different values of the ratio of the Majorana tunneling rate.

Rasetti and Castagnoli argued that anyons could be used to perform quantum computations \cite{Castagnoli1993}. The idea of statistical mechanics in anyons was originated with Arovas et al. \cite{Arovas1985} and had previously been studied by Frank Wilczek \cite{Wilczek1982}. The author mentioned that the interchange of two particles orbiting around the magnetic flux manifests itself as an arbitrary phase between bosons and fermions, and called the exotic state anyons. Other authors showed that the excitation spectrum of a half-quantum vortex in a p-wave superconductor contains a zero-energy Majorana fermion with non-Abelian statistics \cite{Ivanov2001,Greiter2019,Greiter2024}. Haldane proposed that the reduction of the apparent Hilbert space dimension by non-orthogonality of states describes localized topological defects at different points in space is also seen in the fractional quantum Hall effect (FQHE), and seems to be the fundamental feature of the fractional statistics \cite{Haldane1991}. In \cite{Greiter2019} the authors suggest, that p-wave superfluids, and the Moore-Read state are predicted to support the simplest non-Abelian anyons - the Ising anyons. Their behavior can be understood in terms of Majorana fermion modes at the vortex cores, and it is argued that in 1d, non-Abelian and in particular SU(2) level-k statistics manifest  themselves in fractional statistics. For $k = 2$, the authors have observed for the Ising anyons that the state counting of the internal Hilbert space associated with the non-Abelian statistics is equivalent to that of the Majorana fermion states coupled to the spinons \cite{Greiter2019}.

The simplest model of the Majorana fermion is predicted by the Kitaev toy model which assumes the spinless topological superconductor (TSC) \cite{Kitaev2001}. In real TSC, we should consider the polarization of the Majorana fermions in the Rashba and Dresselhaus 1d wire. The authors \cite{Sticlet2012} introduced the Majorana pseudospin and showed that the local Majorana polarization is correlated with the transverse spin polarization. Other authors \cite{He2014,Maska2017} studied the selective equal spin Andreev reflections (SESARs) spectroscopy to detect the polarized Majorana quasiparticles appearing at the edges of the proximitized Rashba chain. In this paper the authors show under which conditions a pseudo-spin degree of freedom can be attributed to Majorana bound states (MBS). MBS correspond to class D and are related to the Z-topological invariant. Class DIII with mirror symmetry supports multiple MBS and is described by the  Z$_{2}$-invariant with an additional time-reversal symmetry \cite{Sedlmayr2015}.

Discussions remain divided regarding over the preparation and actual implementation of Majorana fermions in low-dimensional systems \cite{Mourik2012,Lee2014,Yu2021}. Regardless of the scientific dispute, the SU(2) Kondo effect in quantum dots can be used as a very precise detector of topologically protected Majorana states \cite{Lutchyn2014,Kim2016}. There are currently several candidates for host boundary Majorana quasiparticles: vortices in two-dimensional (2d) $p_{x}+ip_{y}$ spinless superconductors \cite{Fu2008}, Moore-Reed type states in FQHE \cite{Moore1991,Read2000}, the surface of a 3d topological insulator in proximity to an s-wave superconductor \cite{Li2014}, 2d semiconductors with strong spin-orbit coupling coupled to an s-wave superconductor with broken time-reversal symmetry (using a local ferromagnet \cite{Sau2010,Cheng2014} or an external magnetic field \cite{Alicea2010}), domain walls in 1d p-wave topological superconductors \cite{Fu2009}, and helical Majorana modes appearing at the two ends of a 1d wire \cite{Oreg2010,Yazdani2019}. In particular, the authors of the references \cite{Deng2012} demonstrated in InSb nanowires with strong Rashba-type spin-orbit coupling, the artificial realization of a p-wave superconductor and the observation of a magnetic field-induced zero-bias conductance peak, as expected for a zero-energy Majorana fermion signature \cite{Zhang2019,Zhang2018}. In this setup, where the presence of a topological superconductor is controlled by the Zeeman gap, these systems require a delicate balance of the (spin-orbit coupling, magnetic field and chemical potential) to create the topological superconductor. The idea of 1d spinless p-wave superconductor based on the semiconducting nanowire, where spin-orbit coupling is used to shift the spin-up/down levels in the momentum space and Zeeman field, leading to spin splitting and spin texture at the Fermi level. A proximity induced SC gap within the spin-split levels would lead to an effective spinless p-wave SC \cite{DasSarma2023}.

Recently, a topological superconductor has been realized in 1d ferromagnetic atom chains \cite{Nadj2014}, where the 1d system with a strong spin-orbit interaction is placed in proximity to a conventional s-type superconductor. Using high-resolution spectroscopic imaging techniques, the authors have demonstrated the spatially resolved signature of edge-bound Majorana fermions in Fe atom chains and the appearance of zero-energy states in the electronic density of states of the chains. Majorana fermion states are expected to be realized in class D topological superconductors (TSCs with broken time-reversal symmetry)\cite{Sedlmayr2015}. However, Majorana zero modes can also appear in pairs in time-reversal invariant DIII class topological superconductors. These interesting types of Majorana fermions are called Majorana Kramers pairs \cite{Zhang2013,Gaidamauskas2014}. For example, chiral superconductors with $p_{x}+ip_{y}$ pairing state in 2d have a sharp topological distinction between the strong and weak pairing regime \cite{Zhang2013,Gaidamauskas2014,Kim2016}. In the weak pairing regime, the gapless chiral Majorana states at the edge are topologically protected. In two dimensions, a time-reversal invariant topological defect of a Z$_{2}$ non-trivial superconductor carries a Kramers pairs of Majorana fermions \cite{Yamazaki2020}.

One of the interesting papers focuses on the aspect of Majorana-Klein hybridization: in \cite{Beri2013,Cooper2012} the authors demonstrate a topological Kondo effect that implements the SO(M) Kondo problem for M Majorana lead couplings. These topological Kondo states give rise to robust non-Fermi liquid behavior, even for Fermi liquid leads, and to a quantum phase transition between the insulating and Kondo regimes when the leads form Luttinger liquids.

In another paper, the authors studied the interacting Majorana fermions \cite{Rahmani2019}. This is quite a challenge. The simplest interactions between the Majorana degrees of freedom show an unusual non-local structure involving four distinct Majorana sites \cite{Chiu2015}. The authors \cite{Rahmani2019} solved the Sachdev-Ye-Kitaev model and showed that correlated phases of matter with Majorana building blocks can lead to emergent spacetime supersymmetry (SUSY), topological order or Fibonacci topological phase, which are more exotic generalizations of Majorana fermions known as parafermions.

Returning to the issues discussed in this article, the previous theoretical work investigated the problem of the Majorana zero mode coupled to the spin Kondo state using NRG methods: in single \cite{Zitko2011,Lopez2013,Weymann2021} and double quantum dots \cite{Wojcik2020,Majek2021}. Several papers have discussed the thermoelectric effects of quantum dots coupled to side-attached TSCs in the Coulomb blockade regime \cite{Lopez2014,Leijnse2014,Chi2020} and in the T-shaped DQD system in the Kondo state side-attached to the Majorana fermion \cite{Majek2022}. In these thermoelectric quantum devices, the thermopower changes sign and is fully spin polarized. Measurements of the Seebeck coefficient beyond the e-h symmetry point show strong enhancement and a violation of the Wiedeman-Franz law.

The Kondo cloud plays the role of an interference detector for Majorana fermions. In this article, we discuss the influence of weakly and strongly coupled multi-Majorana fermions on the spin-orbital SU(4) Kondo effect in the carbon nanotube quantum dot. We have coupled three TSC devices proximitized by AB superconducting pairing coat. SOI and SC leads to the Majoranization of the wire states of the zigzag CNT in the absence of a magnetic field. The Majorana states are indexed by the spin-orbital numbers and coupled to the fourfold degenerate states of the CNTQD. The Majorana-Kondo effect is observed for the strong  coupling strength limit and manifests itself as the coexistence of the strongly correlated electrons and the topological Majorana states in the system. The spin-orbital type of the Majorana quasiparticles is chosen by the sign of the SOI. An alternative realization is proposed in the armchair CNT, where the electric field can induce the Majorana fermions \cite{Klinovaja2012}.

The shot noise for the QD-TSC device in the weak coupling limit and for the SU(2) Kondo quantum dot is discussed in \cite{Liu2015}, where the the shot noise power for the linear voltage is quantized to $(1/2)(e^{2}/h)$. Alternatively results, using the Keldysh field integral description are presented in \cite{Smirnov2017,Smirnov2022}, where the author suggests that in the Majorana state, for positive atomic level of the QD, we should observe two different fractional effective charges at low and high energies, e$^{\star}$/e$=1/2$ and e$^{\star}$/e$=3/2$, accessible at low and high bias voltages. In another paper \cite{Golub2011} the authors analyzed the shot noise in a 1d Majorana chain fermion coupled to a normal metal, and found that the Fano factor is quantized to $F=2$ for the single Majorana bound state (MBS) and to non-integer $F$ when both MBSs couple to the lead.

The paper is organized as follows. In Sec. II we discuss the Hamiltonian of the two-orbital Anderson model coupled with one, two and three topological superconducting wires (1TS, 2TS and 3TSC).  In the subsections of Sec. III we demonstrate the detection of the symmetry reduction of the spin-orbital SU(4) Kondo effect to exotic SU$^{\star}$(3) and SU$^{\star}$(2) Kondo states in the quantum transport measurements (i.e. quantum conductance, thermoelectric power, linear and nonlinear shot noise). In the last part of the results we study the influence of the SOI on the Majorana-Kondo states. Finally, we summarize  the results in the conclusions.

\section{Model of a CNTQD coupled to side-attached topological superconductors}
We address the calculation to the system with quantum dot tunneling coupled to multi-Majorna fermions (Fig. 1). We model the carbon nanotube quantum dot (CNTQD) by using the two-orbital Anderson Hamiltonian with side-attached topological superconductors (TSCs):
\begin{eqnarray}
&&{\mathcal{H}}=\sum_{ls}E_{ls}n_{ls}+\sum_{l}Un_{l\uparrow}n_{l\downarrow}+\sum_{ss'}Un_{+1s}n_{-1s'}+
\nonumber\\&&\sum_{k\alpha ls}E_{k\alpha ls}n_{k\alpha ls}+\sum_{k\alpha ls}t_{0}(c^{\dagger}_{k\alpha ls}d_{ls}+h.c)+\\
\nonumber&&\sum_{s}it_{+1s}\gamma_{+1s}(d^{\dagger}_{+1s}+d_{+1s})
+it_{-1\uparrow}\gamma_{-1\uparrow}(d^{\dagger}_{-1\uparrow}+d_{-1\uparrow}),
\end{eqnarray}
where the first term describes the energy of the spin-orbital quantum dot level ($E_{ls}=E_{d}(V_{g})+ls\Delta/2$). $\Delta$ is the spin-orbit interaction (SOI) observed in the CNTQD \cite{Jespersen2011, Laird2015}, which arises from the curvature of the carbon nanotube \cite{Jeong2009}. $s=\uparrow(\downarrow)$ and $l=\pm1$ are the spin and orbital numbers. The second and third terms are the intra- and inter-Coulomb interactions in the system. The next two parts of the Hamiltonian (1) are related to the energy of the left and right electrodes $E_{k\alpha ls}$ ($\alpha=L,R$) and the tunneling strength between the CNTQD and the normal electrodes ($t_{0}$). The last two terms are the tunneling terms of the Majorana fermions $\gamma_{ls}$ with the QD states. The tunneling strength is given by $t_{ls}$, and in the paper we considered three types of hybrid systems: CNTQD-TSC ($t_{+1\uparrow}=t$), CNTQD-2TSC ($t_{+1\uparrow}=t_{-1\uparrow}=t$) and CNTQD-3TSC device (where $t_{+1s}=t_{-1\uparrow}=t$). The Majorana fermions are indexed by their spin-orbital number. All energies are given in $\Gamma$ units, where $\Gamma=\frac{\pi t_{0}^{2}}{2W}$ is the tunnel coupling . $1/2W$ is the flat density of states in the electrode, inversely proportional to the half bandwidth ($W$).

Using the extended slave-boson Kotliar-Ruckenstein mean-field approach \cite{Kotliar1986,Krychowski2018,Krychowski2020}, the Hamiltonian (1) can be written in the effective form:
\begin{eqnarray}
\nonumber&& \widetilde{{\mathcal{H}}}=\sum_{ls}\widetilde{E}_{ls}n^{(f)}_{ls}+U\sum_{\nu}d^{\dagger}_{\nu}d_{\nu}
+3U\sum_{ls}t^{\dagger}_{ls}t_{ls}+
\\&& 6Uf^{\dagger}f+\lambda({\cal{I}}-1)-\lambda_{ls}\sum_{ls}Q_{ls}+
\\&&\nonumber \sum_{k\alpha ls}t_{0}(c^{\dagger}_{k\alpha ls}z_{ls}f_{ls}+h.c)+
it_{-1\uparrow}\gamma_{-1\uparrow}(z^{\dagger}_{-1\uparrow}f^{\dagger}_{-1\uparrow}+h.c)
\\&&\nonumber\sum_{s}it_{+1s}\gamma_{+1s}(z^{\dagger}_{+1s}f^{\dagger}_{+1s}+h.c)
\end{eqnarray}
where $\widetilde{E}_{ls}=E_{ls}+\lambda_{ls}$ is the renormalized energy level of the quasiparticle Kondo resonance. $\lambda$ and $\lambda_{ls}$ are the Lagrange multipliers associated with the completeness relation (${\cal{I}}=e^{\dagger}e+\sum_{ls}p^{\dagger}_{ls}p_{ls}+\sum_{\nu={20,02,s\overline{s}}}d^{\dagger}_{\nu}d_{\nu}
+\sum_{ls}\overline{t}^{\dagger}_{ls}\overline{t}_{ls}+f^{\dagger}f$) and charge conservation ($Q_{ls}=\widetilde{z}^{\dagger}_{ls}\cdot \widetilde{z}_{ls}=
p^{\dagger}_{ls}p_{ls}+d^{\dagger}_{l}d_{l}+\sum_{s'}d^{\dagger}_{ss'}d_{ss'}
+\overline{t}^{\dagger}_{ls}\overline{t}_{ls}+\sum_{s'}\overline{t}^{\dagger}_{\overline{l}s'}\overline{t}_{\overline{l}s'}+f^{\dagger}f$).  The $\cdot$ denotes the non-commutative multiplication in the charge operator.
In the effective Hamiltonian, the quantum dot operators $d_{ls}$ are replaced by the product of the bosonic operator $z_{ls}$ and the pseudofermionic operator $f_{ls}$ ($d_{ls}\equiv z_{ls}f_{ls}$). Thus, all physical states are obtained by creating electrons and auxiliary bosons on the vacuum state $\ket{\textrm{vac}}$. In this formalism, the empty and the fully occupied states are generated by the operators as follows: $\ket{e}=\ket{00}=e^{\dagger}\ket{\textrm{vac}}$ and $\ket{f}=\ket{22}=f^{\dagger}\prod_{ls}f^{\dagger}_{ls}\ket{\textrm{vac}}$. The single particle electron state is represented by $\ket{ p_{+s(-s)}}=\ket{ s0(0s)}=p^{\dagger}_{ls}f^{\dagger}_{ls}\ket{\textrm{vac}}$. The triple occupied states are given by $\ket{\overline{t}_{\pm s}}=\ket{ s2(2s)}=\overline{t}^{\dagger}_{\pm s}f^{\dagger}_{\pm s}f^{\dagger}_{\mp s}f^{\dagger}_{\mp \overline{s}}\ket{\textrm{vac}}$.
The auxiliary canonical particles for the double occupied state are represented by six states: $\ket{d_{20(02)}}=\ket{ 20(02)}=d^{\dagger}_{20(02)}f^{\dagger}_{\pm \uparrow}f^{\dagger}_{\pm \downarrow}\ket{\textrm{vac}}$ and $\ket{d_{ss(s\overline{s})}}=\ket{ ss'}=d^{\dagger}_{ss(s\overline{s})}f^{\dagger}_{+s(+s)}f^{\dagger}_{-s(-\overline{s})}\ket{\textrm{vac}}$.

The tunnel rates $\widetilde{t}_{0ls}=t_{0}z_{ls}$ and $\widetilde{t}_{ls}=t_{ls}z_{ls}$ are renormalized by
$z_{ls}=\widetilde{z}_{ls}/\sqrt{\delta n^{2}_{ls}}=(e^{\dagger}p_{ls}+p^{\dagger}_{l\overline{s}}d_{l}
+\sum_{ls'}p^{\dagger}_{ls'}d_{ss'}
+\sum_{l}d^{\dagger}_{l}\overline{t}_{ls}+\sum_{ls'}d^{\dagger}_{\overline{s}s'}\overline{t}_{ls'}+
\overline{t}^{\dagger}_{\overline{l}\overline{s}}f)/\sqrt{Q_{ls}(1-Q_{ls})}$. $z_{ls}$ is the renormalization of the width of the Kondo resonance (compare with the amplitude of the quasiparticle wave function \cite{Hewson1997,Coleman2015}), and determines the Kondo temperature $T_{K}=\min\{T_{K,ls}\}=\sqrt{\widetilde{E}^{2}_{ls}+\widetilde{\Gamma}^{2}_{ls}}$, where $\widetilde{\Gamma}_{ls}=\Gamma|z_{ls}|^2$ is renormalized tunnel coupling.
For the non-interacting system $U=0$, $z^{2}_{ls}\approx1$ and we can say that the spin-orbital fluctuations are comparable to the tunneling processes involved in the in Kondo effect $\delta n_{ls}\approx\widetilde{z}^{2}_{ls}$.
\begin{figure}[t!]
\includegraphics[width=\linewidth]{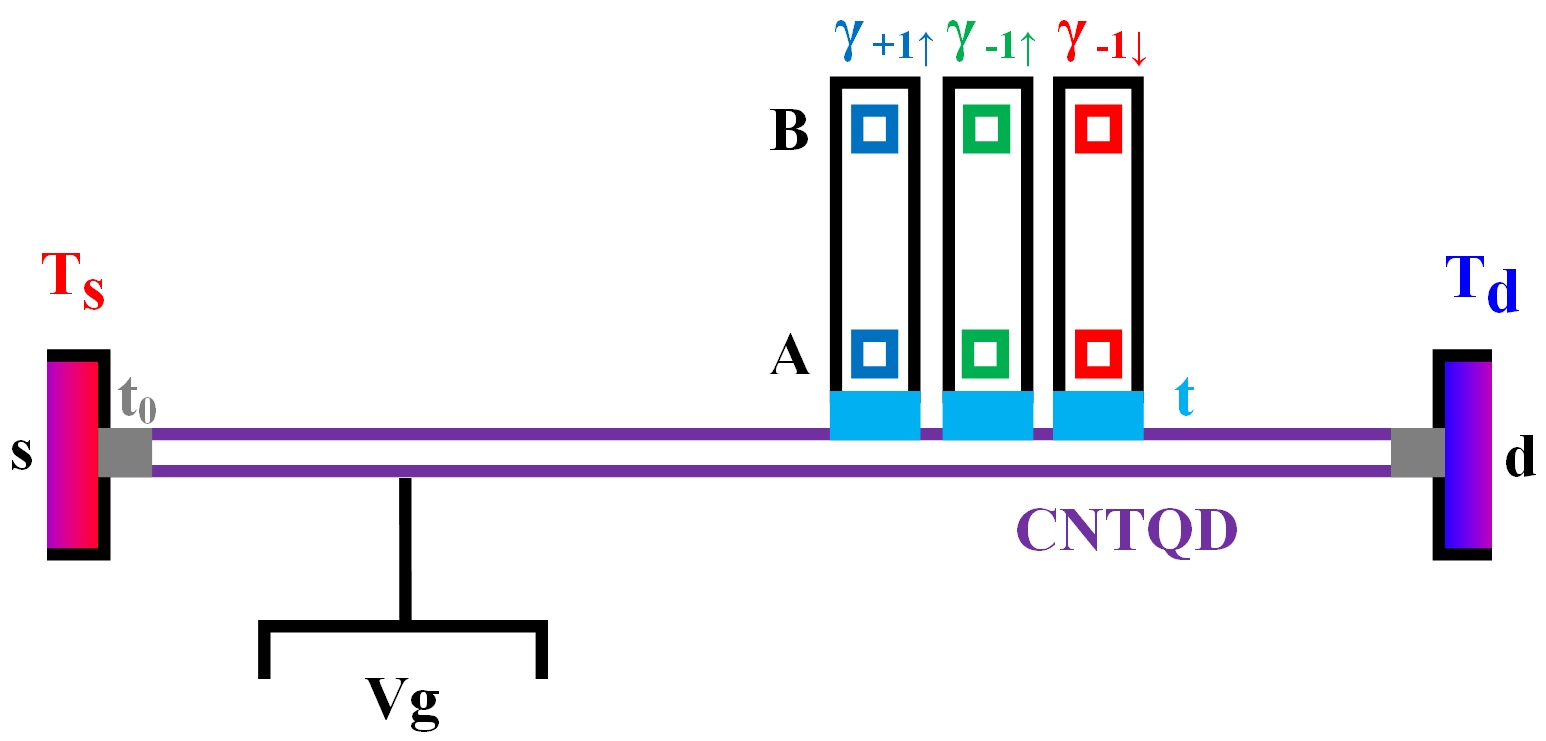}
\caption{\label{fig:epsart} (Color online) Carbon nanotube quantum dot (CTNQD) device tunnel-coupled through $V$ to the normal leads and through $t$ to the Majorana fermion quasiparticle states ($\gamma_{ls}$) in topological superconducting (TSC) wires. The backgate $V_{g}$ changes the number of the electrons on the CNTQD. $\delta T=T_{s}-T_{d}$ is the temperature gradient applied to the left and right electrodes s(d) in the thermoelectric power measurements.}
\end{figure}

In the calculations, we consider three models: a quantum dot coupled to a single Majorana fermion $\gamma_{+\uparrow}$ (CNTQD-TSC),
with two Majorana fermions $\gamma_{\pm\uparrow}$ (CNTQD-2TSC), and with side-attached three Majoranas $\gamma_{\pm \uparrow}$ and $\gamma_{+\downarrow}$ (CNTQD-3TSC). In all systems we take the same value for the coupling strength $t$ in the Hamiltonian (2). Using the saddle-point approximation, we solved the following self-consistent equations:
\begin{eqnarray}
&&\frac{\partial\langle\widetilde{{\mathcal{H}}}\rangle}{\partial b_{n}^{\dagger}}=\frac{\partial\langle\widetilde{{\mathcal{H}}}\rangle}{\partial b_{n}}=\Delta \widetilde{{\mathcal{H}}}_{{n}}+\Delta \widetilde{E}_{{n}}=0\\
&&\nonumber\frac{\partial\langle\widetilde{{\mathcal{H}}}\rangle}{\partial \lambda}={\cal{I}}-1=0, \frac{\partial\langle\widetilde{{\mathcal{H}}}\rangle}{\partial \lambda_{ls}}=\langle f^{\dagger}_{ls}f_{ls}\rangle^{<}-Q_{ls}=0
\end{eqnarray}
where $b^{\dagger}$ is represented by auxiliary boson operators: $b^{\dagger}_{n=1...16}=\{e^{\dagger},p^{\dagger}_{ls},d^{\dagger}_{\nu},\overline{t}^{\dagger}_{ls},f^{\dagger}\}$.
\begin{eqnarray}
&&\nonumber\Delta \widetilde{{\mathcal{H}}}_{n}=\sum_{k\alpha ls}t_{0}\left(\frac{\partial z_{ls}}{\partial b_{n}^{\dagger}}\langle c^{\dagger}_{k\alpha ls}f_{ls}\rangle^{<}+c.c.\right)
\\&&+it_{-1\uparrow}\left(\frac{\partial z^{\dagger}_{-1\uparrow}}{\partial b_{n}^{\dagger}}\langle \gamma_{-1\uparrow}f^{\dagger}_{-1\uparrow}\rangle^{<}+c.c.\right)
\\&&\nonumber+\sum_{s}it_{+1s}\left(\frac{\partial z^{\dagger}_{+1s}}{\partial b_{n}^{\dagger}}\langle \gamma_{+1s}f^{\dagger}_{+1s}\rangle^{<}+c.c.\right),
\end{eqnarray}
and $\Delta \widetilde{E}_{n}=\{\lambda e,(\lambda_{ls}+\lambda)p_{ls},(U+\sum_{s}\lambda_{ls}+\lambda)d_{l},(U+\lambda_{ls}+\lambda_{\overline{l}s'}+\lambda)d_{ss'}
,(3U+\lambda_{ls}+\sum_{s}\lambda_{\overline{l}s}+\lambda)\overline{t}_{ls},(6U+\sum_{ls}\lambda_{ls}+\lambda)f\}$.
The correlators in Eqs. (3-4) can be written in the form:
\begin{eqnarray}
&&\nonumber \langle f^{\dagger}_{ls}f_{ls}\rangle^{<}
=\int^{+W}_{-W}\frac{dE G^{<}_{ls,ls}}{2\pi i}\\
&& \sum_{k}\widetilde{t}_{0ls}\langle c^{\dagger}_{k\alpha ls}f_{ls}\rangle^{<}=\sum_{k}\widetilde{t}_{0ls}\int^{+W}_{-W}\frac{dE G^{<}_{k\alpha ls,ls}}{2\pi i}\\
&&\nonumber i\widetilde{t}_{ls}\langle \gamma_{ls}f^{\dagger}_{ls}\rangle^{<}=i\widetilde{t}_{ls}\int^{+W}_{-W}\frac{dE G^{<}_{\underline{ls},ls}}{2\pi i}
\end{eqnarray}
where $G^{<}_{ls,ls}$, $G^{<}_{k\alpha ls,ls}=\widetilde{t}_{0ls}(G^{R}_{ls,ls}g^{<}_{k\alpha ls}+G^{<}_{ls,ls}g^{A}_{k\alpha ls})$ and $G^{<}_{\underline{ls},ls}$ are the non-equilibrium Green's functions calculated using the
EOM and Keldysh formalism for the Hamiltonian (2) \cite{Elk1979,Balzer2012,Oguri2022}. The retarded and lesser Green's functions in the $\nu=ls$ channel (decoupled from the TSC) are given by $G^{R}_{\nu,\nu}(E)=(E-\widetilde{E}_{\nu}+i\widetilde{\Gamma}_{\nu})^{-1}$ and $G^{<}_{\nu,\nu}(E)=G^{R}_{\nu,\nu}\widetilde{\Sigma}^{<}_{\nu}G^{A}_{\nu,\nu}$. $\widetilde{\Sigma}^{<}_{\nu}=\sum_{\alpha}\widetilde{t}^{2}_{0\nu}g^{<}_{k\alpha\nu}=i\widetilde{\Gamma}_{\nu}\sum_{\alpha}f_{\alpha}$ is the lesser self-energy and
$f_{\alpha}=(e^{\frac{E\pm V_{\alpha}}{k_{B}T}}+1)^{-1}$ is the Fermi-Dirac function. The Green's functions in the $\nu'$ channel  (the channel coupled to the TSC) can be written in the matrix form as follows:
\begin{eqnarray}
&&\nonumber\hat{G}^{R}_{\nu'}=(E-\hat{E}_{\nu'}-\hat{\Sigma}^{R}_{\nu'})^{-1}\\&&=\left(
\begin{array}{ccc}
E-\widetilde{E}_{\nu'}+i\widetilde{\Gamma}_{\nu'} & 0 & -i\widetilde{t}_{\nu'} \\
0 & E+\widetilde{E}_{\nu'}+i\widetilde{\Gamma}_{\nu'} & -i\widetilde{t}_{\nu'} \\
i\widetilde{t}_{\nu'} & i\widetilde{t}_{\nu'} & E+i\delta \\
\end{array}
\right)^{-1}
\end{eqnarray}
where $\hat{E}_{\nu'}$ is the matrix of the diagonal energies $\{\pm\widetilde{E}_{\nu'},0\}$, the remaining elements are the matrix of retarded self-energy $\hat{\Sigma}^{R}$. $\delta$ is the lifetime of the Majorana fermion, and is the lowest energy in the system $\delta\ll T_{K}$ (in our calculations $\delta=10^{-8}$).
In practice this is the consequence of the disappearance of the overlap term between the Majorana fermions at the ends of the proximitized wire  ($i\Delta_{(0)}\gamma_{Als}\gamma_{Bls}$) in the self-energy of the TSC $\widetilde{\Sigma}^{R}_{t}=(\widetilde{t}^{2}z)/(z^2-\Delta^{2}_{(0)})$ \cite{Kitaev2001,Liu2015}. Therefore, in our model we used the self-energy with the finite lifetime of the Majorana fermion $\delta$, where in the asymptotic limit: $\lim_{\Delta_{(0)}\mapsto0}\widetilde{\Sigma}^{R}_{t}=\widetilde{t}^{2}/z=\widetilde{t}^{2}/(E+i\delta)$.
The lesser Green's function matrix can be written as $\hat{G}^{<}_{\nu'}=\hat{G}^{R}_{\nu'}\hat{\Sigma}^{<}_{\nu'}\hat{G}^{A}_{\nu'}$, where $\hat{\Sigma}^{<}_{\nu'}=\hat{\Sigma}^{R}_{\nu'}\sum_{\alpha}f_{\alpha}$. The mean-field slave-boson approach has self-consistent solutions for finite temperature $T$ and bias voltage $V_{\alpha=L(R)}=\pm V/2$ below and around the the Kondo temperature $T_{K}$.

\section{The Results}

\subsection{The Hilbert space and the states of the isolated CNTQD-TSC devices}
\begin{figure}[h!]
\includegraphics[width=0.75\linewidth]{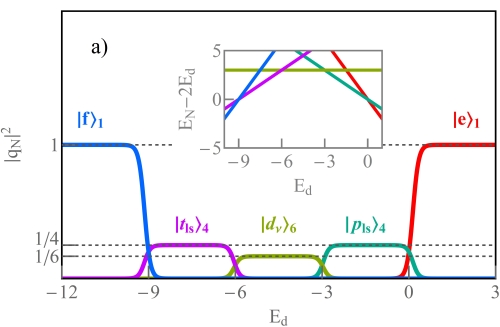}\\
\includegraphics[width=0.75\linewidth]{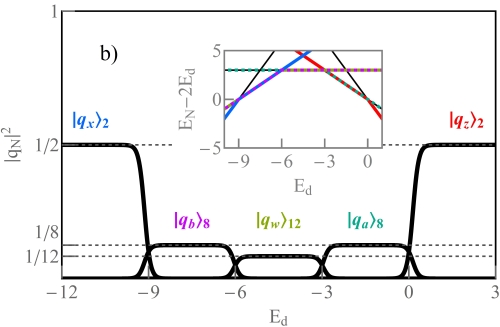}\\
\includegraphics[width=0.75\linewidth]{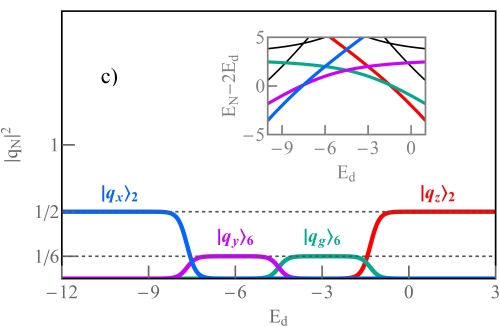}
\caption{\label{fig:epsart} (Color online) Isolated states of the CNTQD-TSC hybrid device: a-c) The quantum amplitudes $|q_{N}|^{2}$ of the ground states as a function of the quantum dot level $E_{d}$ for decoupled ($t=0$), weakly ($t=10^{-2}$) and strongly ($t=2$) coupled CNTQD with TSC.  The insets show the spectrum of the energies $E_{N}-2E_{d}$ with low energy ground states indicated by colors. $|q_{N}\rangle_{\underline{d}}$ represents the topological entangled quantum states except the pure quantum states in Fig. a ($T=10^{-1}$).}
\end{figure}

The tunneling term with the Majorana fermion (MF) can be written in the form : $it_{ls}\gamma_{ls}(d^{\dagger}_{ls}+d_{ls})
=it_{ls}(d^{\dagger}_{ls}\widetilde{c}^{\dagger}_{ls}-d_{ls}\widetilde{c}_{ls}-d^{\dagger}_{ls}\widetilde{c}_{ls}
-\widetilde{c}^{\dagger}_{ls}d_{ls})$, where $\gamma_{ls}=\gamma_{Als}=\widetilde{c}_{ls}+\widetilde{c}^{\dagger}_{ls}$ is the Majorana operator and $\widetilde{c}_{ls}$ is the complex Dirac fermion operator in one-dimensional topological superconductor (1d TSC)\cite{Kitaev2001}. As we can see, the term consists of the superconducting part $d^{\dagger}_{ls}\widetilde{c}^{\dagger}_{ls}$, proportional to the isospin and the normal tunneling part $\widetilde{c}^{\dagger}_{ls}d_{ls}$.
The Majorana fermion states are spatially separated at the ends A and B in the 1d TSC (Fig. 1). Taking two spatially separated MFs and the electron creation operator $\widetilde{c}^{\dagger}_{ls}=\widetilde{c}^{\dagger}_{Als}=(1/2)(\gamma_{Als}-i\gamma_{Bls})$, we can define the occupation number operator in the topological superconductor as: $\widetilde{n}_{ls}=\widetilde{c}^{\dagger}_{ls}\widetilde{c}_{ls}=(1+i\gamma_{Als}\gamma_{Bls})/2$, what is the consequence of the number of the states $\{\underline{0},\underline{\Uparrow}\}$. Therefore the occupation number $\widetilde{n}_{ls}$ is either $0$ or $1$. In the limit of the vanishing overlap between the Majorana fermions, there is still a nonlocal half-fermionic state in the single or zero quantum state, which is the main argument and attraction in the the topological quantum computation.
\begin{figure}[t!]
\includegraphics[width=0.75\linewidth]{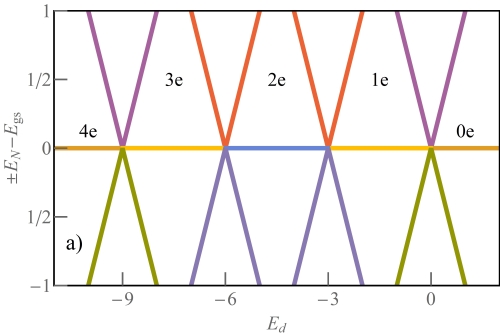}\\
\includegraphics[width=0.75\linewidth]{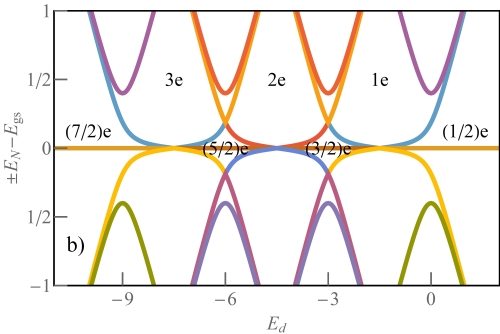}\\
\includegraphics[width=0.75\linewidth]{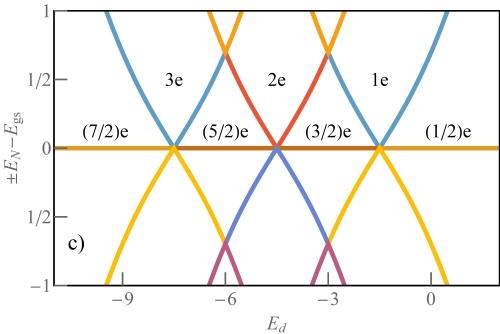}
\caption{\label{fig:epsart} (Color online) a-c) The difference between the excited states $\pm E_{N}$ and the ground state $E_{gs}$ for decoupled, weakly and strongly coupled CNTQD with a single Majorana fermion state $\gamma_{+\uparrow}$. The numbers represent the integer and fractional charge on the quantum dot near the zero energy.}
\end{figure}

The Majorana fermions obey the Clifford algebra $\{\gamma_{\nu},\gamma^{\dagger}_{\nu'}\}=\delta_{\nu,\nu'}$ ($[\gamma_{\nu},\gamma^{\dagger}_{\nu'}]=0$), where $\delta_{\nu,\nu'}$ is the Kronecker delta and $\nu,\nu'$ are Majorana indices. Moreover, unlike the complex fermions, Majoranas do not square to zero, but $\gamma^{2}_{\nu} = 1/2$ ($\gamma^{3}_{\nu}=\gamma_{\nu}/2$).
The quasiparticle parity $\widetilde{c}^{\dagger}_{ls}\widetilde{c}_{ls}$ is accessible by a joint measurement on both Majoranas. So we are talking about new half-fermions on the both sides of the topological wire, and  $\gamma_{ls}$ are the real operators and are own antiparticles. We neglect in our calculation the effect of overlapping between Majorana fermions in the form $i\Delta_{(0)}\gamma_{Als}\gamma_{Bls}$, which leads to a bowtie-like mismatch in the zero energy non-local Majorana state \cite{Prada2017,Krychowski2018}. $\Delta_{(0)}=e^{-w/\lambda_{K}}$, where $w$ is the separation length between Majoranas in the TSC, and $\lambda_{K}$ determines the quality of the MFs and is the superconducting coherence length, which strongly suppresses the overlap between two Majoranas. For the Hamiltonian ${\mathcal{H}}_{ls}=(ls\Delta-E_{g})\hat{\sigma}_{Z}\otimes\hat{\sigma}_{X}+\Delta_{(l)}\sum_{k=\pm}\mp\hat{\sigma}_{k}\otimes\hat{\sigma}_{Y}$ in the Nambu basis $\Psi=(c_{Als},c_{Bls},c^{\dagger}_{Als},c^{\dagger}_{Bls})$, where $\Delta$ is the spin-orbit coupling strength \cite{Jespersen2011}, $E_{g}=N\Delta_{l}$ is the perturbation gap  and $\Delta_{l}$ is the triplet AB-site superconducting order parameter (different for orbital $l=\pm$), we can find four independent Majorana bound state solutions at the zero energy level: $\gamma_{Als}=\Psi\cdot (1/\sqrt{2})\{1,0,1,0\}^{T}$ and $\gamma_{Bls}=\Psi\cdot (i/\sqrt{2})\{0,1,0,-1\}^{T}$ for $\Delta=\Delta_{l}/(N-ls)$. In this simple toy model, we consider the TSC wire with the Zeeman-like
SOI term \cite{Jespersen2011} and the orbital dependent AB triplet superconducting pairing strength. The proximitized term in the Hamiltonian ${\mathcal{H}}_{ls}$ is the crucial point of the toy model and will be a major challenge for experimental research \cite{Beenakker2013}.
\begin{figure}[t!]
\includegraphics[width=0.75\linewidth]{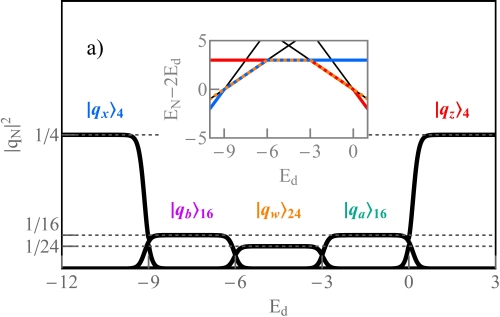}\\
\includegraphics[width=0.75\linewidth]{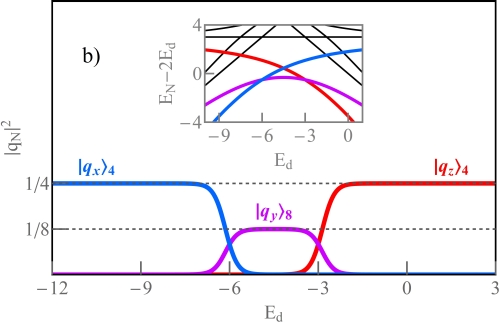}\\
\includegraphics[width=0.75\linewidth]{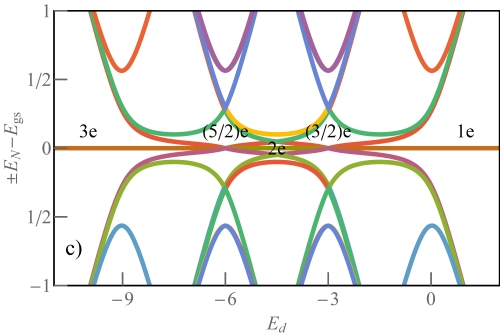}\\
\includegraphics[width=0.75\linewidth]{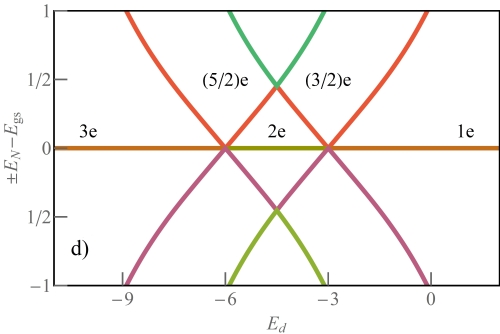}
\caption{\label{fig:epsart} Isolated states of the CNTQD-2TSC hybrid device: a, b) $|q_{N}|^{2}$ as a function of $E_{d}$ for weakly ($t=10^{-2}$) and strongly ($t=2$) coupled CNTQD with two Majorana fermion states ($\gamma_{\pm\uparrow}$). c, d) $E_{d}$ dependence of $\pm E_{N}-E_{gs}$ for the weak and strong tunnel coupling regime with TSCs.}
\end{figure}

Our proposed device, shown in Figure 1 is the hybrid carbon nanotube quantum dot (CNTQD) with side-attached topological superconductor (TSC) fabricated on the 1d nanotube in the spin-triplet p-wave superconducting coat. The real fermion particles $\gamma_{ls}=\gamma^{\dagger}_{ls}$ are located at the edges of the TSC. $\gamma_{ls}$ consist of equal parts of electrons and holes with the same spin orbital.
In contrast to the Bogoliubov quasiparticle operator $a^{\dagger}_{ks}=\Psi^{\dagger}_{k}\cdot  \{u_{k},v_{k}\}^{T}=u_{k}c^{\dagger}_{ks}+\textrm{sgn}(s)v_{k}c_{-k\overline{s}}$, where $a^{\dagger}_{ks}\neq a_{ks}$.

In our case, the Majorana fermions are well-prepared quantum states and are indexed by spin and orbital number, which determines the selective tunneling coupling term to $ls$ states on CNTQD. The main discussion in the experiments is about the preparation of the states \cite{Mourik2012,Lee2014,Yu2021}. We should always look on the both sides of the wire and selectively detect the non-local Majoranas, e.g. using the doubling effect in the supercurrent \cite{Kwon2004,Rokhinson2012}. For this reason, we focus on the problem of the side-attached TSC to the CNTQD in the Kondo state, which plays the role of a detector of non-Abelian MFs, especially observed in the quantum conductance, thermoelectric power and in the fractional noise measurements.
\begin{figure}[t!]
\includegraphics[width=0.75\linewidth]{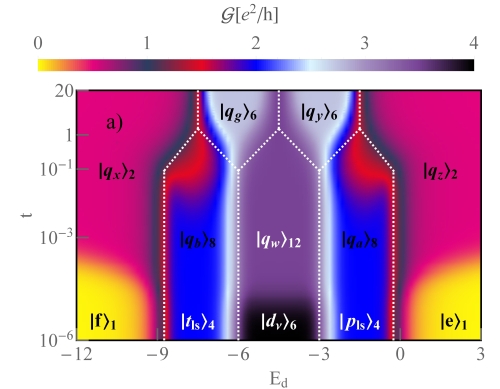}\\
\includegraphics[width=0.75\linewidth]{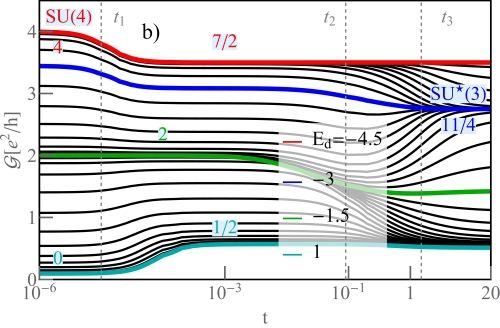}\\
\includegraphics[width=0.75\linewidth]{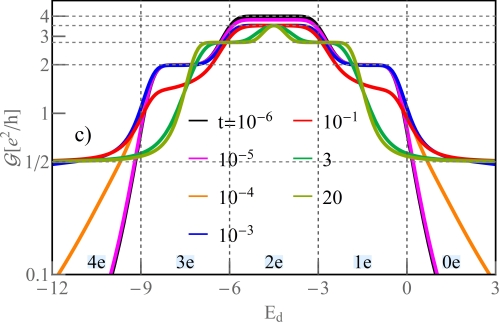}\\
\includegraphics[width=0.75\linewidth]{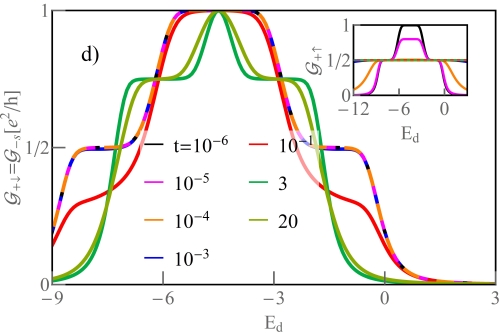}
\caption{\label{fig:epsart} (Color online) Quantum conductance for the fractional SU$^{\star}$(3) Kondo state:
a) The density plot of $\cal{G}$ as a function of $E_{d}$ and $t$.
$|q_{N}\rangle_{\underline{d}}$ represents the ground state configuration in the CNTQD-TSC device.
b) The landscape plot of $\cal{G}$  as a function of $t$. The black lines are plotted with an increment of $\delta E_{d}=0.15$ from $E_{d}=-4.5$ to $E_{d}=1$. c, d) $E_{d}$ dependence of the total and spin-orbital conductances with increasing $t$. The inset in Fig. d shows ${\cal{G}}_{+\uparrow}$ ($U=3$, $\Gamma=0.03$, $T=10^{-8}$).}
\end{figure}
\begin{figure}[b!]
\includegraphics[width=0.75\linewidth]{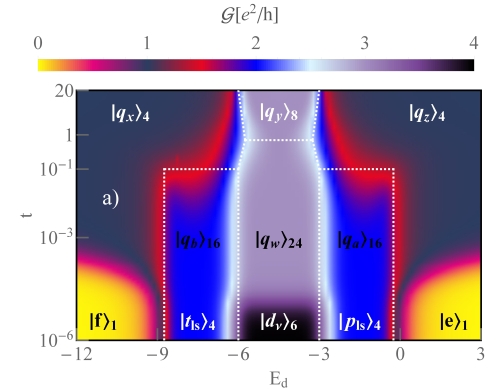}\\
\includegraphics[width=0.75\linewidth]{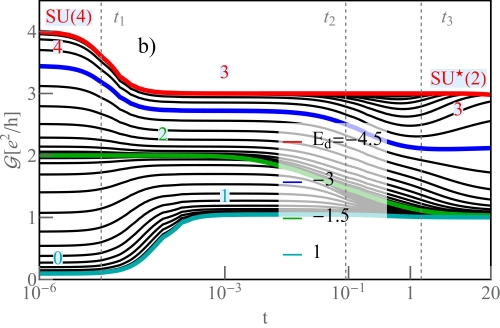}\\
\includegraphics[width=0.75\linewidth]{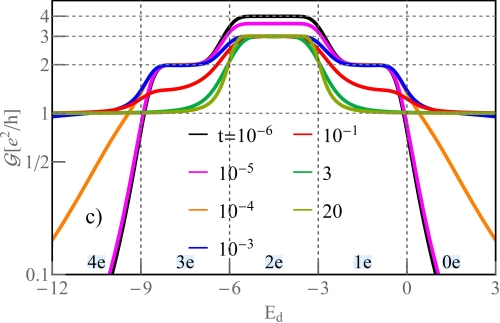}\\
\includegraphics[width=0.75\linewidth]{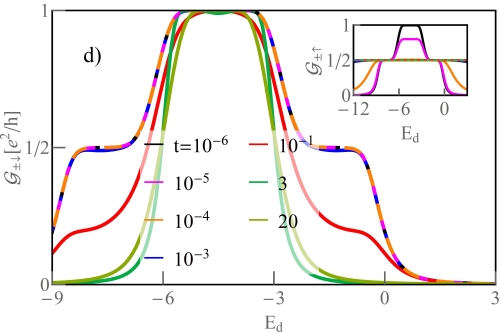}
\caption{\label{fig:epsart} (Color online) Quantum conductance for the integer SU$^{\star}$(2) Kondo state in the CNTQD-2TSC system:
a) The density plot of $\cal{G}$ as a function of $E_{d}$ and $t$. b) The landscape plot of $G$ as a function of $t$ ($\delta E_{d}=0.15$).
c, d) $E_{d}$ dependence of $\cal{G}$  and ${\cal{G}}_{ls}$.}
\end{figure}
\begin{figure}[t!]
\includegraphics[width=0.75\linewidth]{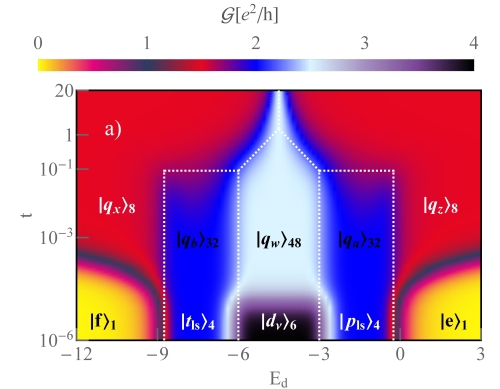}\\
\includegraphics[width=0.75\linewidth]{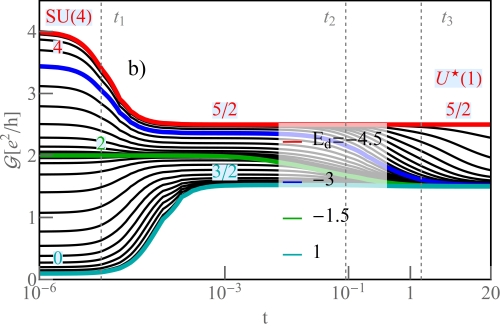}\\
\includegraphics[width=0.75\linewidth]{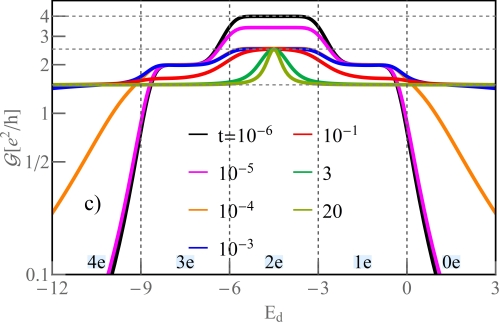}\\
\includegraphics[width=0.75\linewidth]{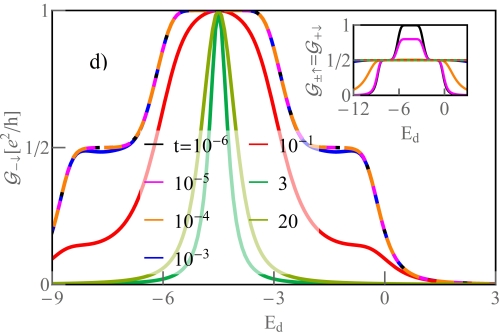}
\caption{\label{fig:epsart} (Color online) Quantum conductance in the CNTQD-3TSC device:
a, b) The density plot and the landscape plot of $\cal{G}$ as a function of $E_{d}$ and $t$ ($\delta E_{d}=0.15$).
c, d) $\cal{G}$  and ${\cal{G}}_{ls}$ as a function of $E_{d}$ with increasing $t$.}
\end{figure}

Let's first discuss the effect of the MF states on the quantum states in the isolated CNTQD ($t_{0}=0$). For finite Coulomb interaction $U=3$, the two-orbital Anderson model can be  written in the representation of occupation number on both orbitals $\ket{n_{+}n_{-}}$. The system describe the sixteenth quantum states: $\ket{e}=\ket{00},\ket{p_{ls}}=\ket{s0}=\ket{0s},\ket{d_{\nu}}=\ket{\nu}
=\ket{20}=\ket{02}=\ket{ss}=\ket{s\overline{s}},\ket{t_{ls}}=\ket{s2}=\ket{2s},\ket{f}=\ket{22}$.
Figure 2a shows the probability of the quantum amplitudes $|q_{N}|^{2}$ for decoupled CNTQD with TSC ($t=0$).
$|q_{N}|^{2}=e^{-E_{N}/(k_{B}T)}/Z_{N}$ are calculated for $U=3$ and $T=0.1$. $Z_{N}=\sum_{N}e^{-E_{N}/(k_{B}T)}$ is the partition function and $E_{N}$ are the energies of the individual quantum states. Above $E_{d}=0$ and below $E_{d}=-3U$, the empty $\ket{e}_{1}$ and the fully occupied states $\ket{f}_{1}$ without (0e) and with four electrons on CNTQD (4e) dominate. In the three-electron (3e) and one-electron (1e) charge region, the ground states $\ket{t_{ls}}_{4}$ and $\ket{p_{ls}}_{4}$  are the quadruplets with $|q_{N}|^{2}=1/4$. For two electrons (2e), the two-orbital Anderson model determines the six quantum states $\ket{d_{\nu}}_{6}$ with $|q_{N}|^{2}=1/6$. The lower index $\underline{d}$ in $\ket{q_{N}}_{\underline{d}}$ is the number of degenerate states. The inset shows the energies $E_{N}-2E_{d}$. The lowest energies represent the ground state energy $E_{gs}$ of the system and are represented by colored lines in the insets.

Taking the isolated CNTQD from the normal leads with tunneling term to the Majorana fermion state, we plotted $|q_{N}|^{2}$ and the energies $E_{N}$ with the ground states for weak ($t=10^{-3}$) and strong coupling limit to TSC ($t=2$) (Fig. 2). The quantum states are spanned by the basis vectors, which consist from the normal part (quantum dot) $\ket{n_{+}n_{-}}$ and the topological segment $\ket{n_{1}n_{2}n_{3}}$. In the normal part, the states are defined by $n_{\pm}=\{0,\uparrow,\downarrow,2\}$. The topological segment with two topological edge states is defined by a wave function describing one-qubit states for $\ket{\underline{0}}$ and $\ket{\underline{\Uparrow}}$, and for the last two topological states $n_{2(3)}$, the allowed configurations are ${\ket{\overline{0}},\ket{\overline{\Uparrow}}}$ and ${\ket{\underline{\underline{0}}},\ket{\underline{\underline{\Downarrow}}}}$.
The states are orthogonal and degenerate at zero energy, forming a two-dimensional Hilbert space for a single qubit state. Maximally, in the topological segment, the Hilbert space is the direct product of three Hilbert spaces and the many-body ground states are given by: $\ket{n_{1}n_{2}n_{3}}=\ket{n_{1}}\otimes\ket{n_{2}}\otimes\ket{n_{3}}$. The ground state degeneracy of 1d topological superconductors is $2^{N_{TS}}$. On CNTQD we have $2^{n}=2^{4}=16$ quantum states ($\ket{n_{+}n_{-}}$). If we couple TSC to our setup, the number of states grows to the number $2^{n+N_{TS}}$. The quantum amplitudes in Fig. 2b reach the half value for $\ket{ q_{z}}_{\underline{d}}$ and $\ket{q_{x}}_{\underline{d}}$. In the extended Hilbert space ($2^5=32$, $\ket{n_{+}n_{-}n_{1}}$) for $E_{d}>0$ and $E_{d}<-3U$ dominates $\ket{q_{z}}_{2}=\ket{ q_{-z_{1(\overline{1})}}}=\frac{1}{\sqrt{2}}(-\ket{00\underline{0}(\underline{\Uparrow})}+\ket{\uparrow0\underline{\Uparrow}(\underline{0})})$ and $\ket{q_{x}}_{2}=\ket{q_{-x_{1(\overline{1})}}}=\frac{1}{\sqrt{2}}(-\ket{\downarrow2\underline{0}(\underline{\Uparrow})}
+\ket{22\underline{\Uparrow}(\underline{0})})$. The minus sign in the lower index refers to the lower energy, the plus sign is reserved for the excited quantum states. In the following discussion, we have omitted the minus sign in the lower index, because we are focusing on the ground state. This is interesting because TSC entangles the pure quantum states from the even $\ket{00(22)}$ and the odd $\ket{\uparrow0(\downarrow2)}$ charge number sectors and opens the possibility of manipulating the states at the boundary of the integer charge numbers ($Q=0$e, 1e, 2e, 4e). The topological qubit forms two doublets for fractional charge numbers $Q=(1/2)$e and $Q=(7/2)$e. This scenario is observed for weak and strong coupling strengths $t$ (Fig. 2a, b). Fig.2b shows the highly degenerate states for the weak coupling limit: two octuplets $\ket{q_{a}}_{8}$($\ket{q_{b}}_{8}$) and one duodecuplet $\ket{q_{w}}_{12}$.
The low energy octuplets $\ket{q_{a}}_{8}$ in the 1e charge sector are given by:
\begin{eqnarray}
&&\nonumber\ket{ q_{-a_{1(\overline{1})}}}=-\textrm{a}'\ket{00\underline{0}(\underline{\Uparrow})}+\textrm{a}\ket{\uparrow0\underline{\Uparrow}(\underline{0})}
\\&&\ket{q_{-a_{2(\overline{2})}}}=-\textrm{a}\ket{0\uparrow\underline{0}(\underline{\Uparrow})}
+\textrm{a}'\ket{\uparrow\uparrow\underline{\Uparrow}(\underline{0})}
\\&&\nonumber\ket{q_{-a_{3(\overline{3})}}}=-\textrm{a}\ket{\downarrow0\underline{0}(\underline{\Uparrow})}
+\textrm{a}'\ket{20\underline{\Uparrow}(\underline{0})}
\\&&\nonumber\ket{q_{-a_{4(\overline{4})}}}=-\textrm{a}\ket{0\downarrow\underline{0}(\underline{\Uparrow})}
+\textrm{a}'\ket{\uparrow\downarrow\underline{\Uparrow}(\underline{0})}
\end{eqnarray}
and the high energy octuplets $\ket{q_{b}}_{8}$ can be written as follows:
\begin{eqnarray}
&&\nonumber\ket{q_{-b_{1(\overline{1})}}}=-\textrm{a}\ket{\downarrow2\underline{0}(\underline{\Uparrow})}
+\textrm{a}'\ket{22\underline{\Uparrow}(\underline{0})}
\\&&\ket{q_{-b_{2(\overline{2})}}}=-\textrm{a}'\ket{\downarrow\uparrow\underline{0}(\underline{\Uparrow})}
+\textrm{a}\ket{2\uparrow\underline{\Uparrow}(\underline{0})}
\\&&\nonumber\ket{q_{-b_{3(\overline{3})}}}=-\textrm{a}'\ket{02\underline{0}(\underline{\Uparrow})}
+\textrm{a}\ket{\uparrow2\underline{\Uparrow}(\underline{0})}
\\&&\nonumber\ket{q_{-b_{4(\overline{4})}}}=-\textrm{a}'\ket{\downarrow\downarrow\underline{0}(\underline{\Uparrow})}
+\textrm{a}\ket{2\downarrow\underline{\Uparrow}(\underline{0})}
\end{eqnarray}
In the 2e charge region, the ground state of the system represents the duodecuplet $\ket{q_{w}}_{12}$ in the form:
\begin{eqnarray}
&&\nonumber\ket{q_{-w_{1(\overline{1})}}}=-\textrm{a}'\ket{0\uparrow\underline{0}(\underline{\Uparrow})}
+\textrm{a}\ket{\uparrow\uparrow\underline{\Uparrow}(\underline{0})}
\\&&\nonumber\ket{q_{-w_{2(\overline{2})}}}=-\textrm{a}'\ket{\downarrow0\underline{0}(\underline{\Uparrow})}
+\textrm{a}\ket{20\underline{\Uparrow}(\underline{0})}
\\&&\ket{q_{-w_{3(\overline{3})}}}=-\textrm{a}'\ket{0\downarrow\underline{0}(\underline{\Uparrow})}
+\textrm{a}\ket{\uparrow\downarrow\underline{\Uparrow}(\underline{0})}
\\&&\nonumber\ket{q_{-w_{4(\overline{4})}}}=\textrm{a}'\ket{2\uparrow\underline{0}(\underline{\Uparrow})}
-\textrm{a}\ket{\downarrow\uparrow\underline{\Uparrow}(\underline{0})}
\\&&\nonumber\ket{q_{-w_{5(\overline{5})}}}=\textrm{a}'\ket{\uparrow2\underline{0}(\underline{\Uparrow})}
-\textrm{a}\ket{02\underline{\Uparrow}(\underline{0})}
\\&&\nonumber\ket{q_{-w_{6(\overline{6})}}}=\textrm{a}'\ket{2\downarrow\underline{0}(\underline{\Uparrow})}
-\textrm{a}\ket{\downarrow\downarrow\underline{\Uparrow}(\underline{0})}
\end{eqnarray}
$\textrm{a}(\textrm{a}')$ are the amplitudes as the function of $E_{d}$, $U$ and the coupling strength $t$.
Despite the fact that $\textrm{a}'\ll\textrm{a}$ in the weak coupling limit to TSC, the states are entangled and the Hilbert space is extended (this is particularly important and visible in the $Q=2$e charge region). Fig. 2b shows $|q_{N}|^{2}=1/8$ for the two octuplets and $1/12$ for the single duodecuplet.

In the strong coupling limit $\textrm{a}\approx\textrm{a}'$ for $Q=(3/2)$e and $Q=(5/2)$e, two sextuplets are the ground states of the system. The low energy quantum states $\ket{q_{g}}_{6}$ can be expressed as:
\begin{eqnarray}
&&\nonumber\ket{q_{-g_{1(\overline{1})}}}=-\textrm{a}\ket{\downarrow\uparrow\underline{0}(\underline{\Uparrow})}
+\textrm{a}'\ket{2\uparrow\underline{\Uparrow}(\underline{0})}
\\&&\ket{q_{-g_{2(\overline{2})}}}=-\textrm{a}'\ket{02\underline{0}(\underline{\Uparrow})}
+\textrm{a}\ket{\uparrow2\underline{\Uparrow}(\underline{0})}
\\&&\nonumber\ket{q_{-g_{3(\overline{3})}}}=-\textrm{a}'\ket{\downarrow\downarrow\underline{0}(\underline{\Uparrow})}
+\textrm{a}\ket{2\downarrow\underline{\Uparrow}(\underline{0})}
\end{eqnarray}
and the high energy ground states $\ket{q_{y}}_{6}$ are described by: $\ket{ q_{-y_{1(\overline{1})}}}=-\textrm{a}\ket{0\uparrow\underline{0}(\underline{\Uparrow})}
+\textrm{a}'\ket{\uparrow\uparrow\underline{\Uparrow}(\underline{0})}$, $\ket{ q_{-y_{2(\overline{2})}}}=-\textrm{a}\ket{\downarrow0\underline{0}(\underline{\Uparrow})}
+\textrm{a}'\ket{20\underline{\Uparrow}(\underline{0})}$ and $\ket{ q_{-y_{3(\overline{3})}}}=-\textrm{a}\ket{0\downarrow\underline{0}(\underline{\Uparrow})}
+\textrm{a}'\ket{\uparrow\downarrow\underline{\Uparrow}(\underline{0})}$.
The insets in Fig. 2 show the energies $E_{N}$. The lower colored lines are the ground state energies. As the tunneling strength increases, the lines on the insets, especially for intermediate couplings $t$, are the nonlinear function of $E_{d}$.

Figure 3 shows the difference between the energies $E_{N}$ and the ground state energy for low bias $\pm E_{N}+E_{gs}$.
At the zero energy line the system is determined by the ground state $E_{gs}$ and all lines above and below this point show the excited states $\pm E_{N}+E_{gs}$. The excited states can be observed in the range of finite bias voltages $|V_{s(d)}|>0$, higher than the Kondo temperature of the strongly correlated system. For decoupled CNTQD with TSC we observed the integer charge regions $Q=1e,2e,3e$, where for $\pm E_{N}-E_{gs}=0$ the SU(4) Kondo state is realized. In the range of weak coupling strength regime the fractional charge regions $Q=(1/2)$e, $(3/2)$e, $(5/2)$e and $(7/2)e$ are formed. For $Q=(1/2)$e and $(7/2)$e the ground state of the system is determined by two doublets $\ket{q_{z}}_{2}$ and $\ket{q_{x}}_{2}$, opening the Majorana channel in transport measurements. For the strong coupling $t$, the Majorana channel is independent and separate from the channels involved in the fractional SU$^{\star}$(3) Kondo effect. The Kondo state is denoted by $\star$ because, in the contrast to the standard SU(3) Kondo effect \cite{Zitko2013,Krychowski2020}, the quasiparticle state is formed for the fractional charges $Q=(3/2)e$ and $Q=(5/2)e$, which is non-trivial and the main result of the paper. The SU$^{\star}$(3) Kondo state is a signature of sixfold degenerate states: low $\ket{q_{g}}_{6}$ and $\ket{q_{y}}_{6}$ high energy sextuplets.

For the CNTQD-2TSC device, two spin-orbital channels $+\uparrow$ and $-\uparrow$ are correspondingly coupled to two selected Majorana quasiparticles $\gamma_{+\uparrow}$ and $\gamma_{-\uparrow}$. The Hilbert space for the isolated system CNTQD-2TSC is spanned by $2^{4+2}=64$ quantum states. In the weak coupling regime (Fig. 4a), the probability amplitudes $|q_{a(b)}|^{2}$ lead to $1/16$ for $1$e and $3$e on the quantum dot. In the $2$e charge region, the amplitudes reach the value $|q_{N}|^{2}=1/24$ and the lowest energy state is represented by the twenty-fourfold degenerate state $\ket{q_{w}}_{24}$. The sum of all degeneracies in the weak coupling limit leads to the number $\underline{d}=4+16+24+16+4=64$, which is the number of all quantum states in the system. A similar relation can be written for CNTQD-TSC, where $\underline{d}=2+8+12+8+2=32$. With increasing $t$ (Fig. 4c, d) the four charge regions are reduced, and we observe the three quantum integer charge numbers $Q=1$e, $2$e and $3$e. Empty $\ket{e}_{1}$ and full $\ket{f}_{1}$ occupied quantum states are switched to two quartets
$\ket{q_{x}}_{4}$ and $\ket{q_{z}}_{4}$, which are visible in the transport measurements with the channels coupled to Majorana fermions. In the strong coupling regime the quantum states $\ket{q_{z}}_{4}$ are represented by:
\begin{eqnarray}
&&\nonumber\ket{q_{-z_{1(\overline{1})}}}=\frac{1}{2}(\mp\ket{00\underline{0}\overline{0}(\underline{\Uparrow}\overline{\Uparrow})}
\pm\ket{0\uparrow\underline{0}\overline{\Uparrow}(\underline{\Uparrow}\overline{0})}\\
&&+\ket{\uparrow0\underline{\Uparrow}\overline{0}(\underline{0}\overline{\Uparrow})}
+\ket{\uparrow\uparrow\underline{\Uparrow}\overline{\Uparrow}(\underline{0}\overline{0})})\\
&&\nonumber\ket{q_{-z_{2(\overline{2})}}}=\frac{1}{2}(\pm\ket{00\underline{0}\overline{\Uparrow}(\underline{\Uparrow}\overline{0})}
\pm\ket{0\uparrow\underline{0}\overline{0}(\underline{\Uparrow}\overline{\Uparrow})}
\\&&\nonumber-\ket{\uparrow0\underline{\Uparrow}\overline{\Uparrow}(\underline{0}\overline{0})}
+\ket{\uparrow\uparrow\underline{\Uparrow}\overline{0}(\underline{0}\overline{\Uparrow})})
\end{eqnarray}
and the high energy quartets $\ket{q_{x}}_{4}$ have the following forms: $\ket{q_{-x_{1(\overline{1})}}}=\frac{1}{2}(\ket{22\underline{\Uparrow}\overline{\Uparrow}(\underline{0}\overline{0})}
+\ket{2\downarrow\underline{\Uparrow}\overline{0}(\underline{0}\overline{\Uparrow})}
\mp\ket{\downarrow2\underline{0}\overline{\Uparrow}(\underline{\Uparrow}\overline{0})}
\pm\ket{\downarrow\downarrow\underline{0}\overline{0}(\underline{\Uparrow}\overline{\Uparrow})})$ and $\ket{q_{-x_{2(\overline{2})}}}=\frac{1}{2}(\ket{22\underline{\Uparrow}\overline{0}(\underline{0}\overline{\Uparrow})}
-\ket{2\downarrow\underline{\Uparrow}\overline{\Uparrow}(\underline{0}\overline{0})}
\mp\ket{\downarrow2\underline{0}\overline{0}(\underline{\Uparrow}\overline{\Uparrow})}
\mp\ket{\downarrow\downarrow\underline{0}\overline{\Uparrow}(\underline{\Uparrow}\overline{0})})$.
The probability amplitudes for these states have the following value $|q_{N}|^{2}=1/4$ (Fig. 4b). As we can see in the inset of Fig. 4b, the energy ground states $E_{N}-2E_{d}$ are the quadratic function of the atomic level $E_{d}$.
For $Q=2$e the ground state is the octuplet $\ket{q_{y}}_{8}$ with eightfold degenerate states.
All states contributing to the SU$^{\star}$(2) Kondo effect. The strongly correlated state is realized for even number of electron in the system,
which is typical e.g for the charge Kondo state with polarons. The octuplet quantum states can written in the form:
\begin{eqnarray}
&&\nonumber\ket{q_{-y_{1(\overline{1})}}}=\textrm{a}\ket{20\underline{\Uparrow}\overline{0}(\underline{0}\overline{\Uparrow})}
+\textrm{a}'\ket{2\uparrow\underline{\Uparrow}\overline{\Uparrow}(\underline{0}\overline{0})}\\
&&\nonumber\mp\textrm{a}'\ket{\downarrow0\underline{0}\overline{0}(\underline{\Uparrow}\overline{\Uparrow})}
\pm\textrm{a}\ket{\downarrow\uparrow\underline{0}\overline{\Uparrow}(\underline{\Uparrow}\overline{0})}\\
&&\nonumber\ket{q_{-y_{2(\overline{2})}}}=-\textrm{a}\ket{20\underline{\Uparrow}\overline{\Uparrow}(\underline{0}\overline{0})}
+\textrm{a}'\ket{2\uparrow\underline{\Uparrow}\overline{0}(\underline{0}\overline{\Uparrow})}\\&&
\pm\textrm{a}'\ket{\downarrow0\underline{0}\overline{\Uparrow}(\underline{\Uparrow}\overline{0})}
\pm\textrm{a}\ket{\downarrow\uparrow\underline{0}\overline{0}(\underline{\Uparrow}\overline{\Uparrow})}\\
&&\nonumber\ket{q_{-y_{3(\overline{3})}}}=\mp\textrm{a}\ket{02\underline{0}\overline{\Uparrow}(\underline{\Uparrow}\overline{0})}
\pm\textrm{a}'\ket{0\downarrow\underline{0}\overline{0}(\underline{\Uparrow}\overline{\Uparrow})}\\&&\nonumber
+\textrm{a}'\ket{\uparrow2\underline{\Uparrow}\overline{\Uparrow}(\underline{0}\overline{0})}
+\textrm{a}\ket{\uparrow\downarrow\underline{\Uparrow}\overline{0}(\underline{0}\overline{\Uparrow})}\\
&&\nonumber\ket{q_{-y_{4(\overline{4})}}}=\mp\textrm{a}\ket{02\underline{0}\overline{0}(\underline{\Uparrow}\overline{\Uparrow})}
\mp\textrm{a}'\ket{0\downarrow\underline{0}\overline{\Uparrow}(\underline{\Uparrow}\overline{0})}\\&&\nonumber
+\textrm{a}'\ket{\uparrow2\underline{\Uparrow}\overline{0}(\underline{0}\overline{\Uparrow})}
-\textrm{a}\ket{\uparrow\downarrow\underline{\Uparrow}\overline{\Uparrow}(\underline{0}\overline{0})}
\end{eqnarray}
The states are the combination of one single, one triple and two double quantum states spanned by Majorana fermion quantum states in TSCs. For the coupling strength $\widetilde{t}_{\nu}\gg\widetilde{\Gamma}_{\nu}$, the amplitudes in Eq. (12) are comparable $\textrm{a}'\approx\textrm{a}$ ($|q_{N}|^{2}=1/8$). All these eight states are the linear combination of the four extended states $\ket{n_{+}n_{-}(n_{1}n_{2})}$.
Figures  4c, d show that except for the zero energy state, which is represented by the integer charge $Q=1$e, 2e, 3e, the excited states
are determined by fractional charges $Q=(3/2)$e and $(5/2)$e. All excited states
can be observed in tunneling spectroscopy measurements. The fractional charges are manifested in the spin dependent conductances of a quantum dot side-attached to the topological superconductor \cite{Cuniberti2018}.
For example, in \cite{Flensberg2011} the authors have shown that non-Abelian rotations within the degenerate ground state manifold of a set of
Majorana fermions and the quantum dot in the Coulomb blockade regime can be realized by adding or removing a single electron, and by exchanging electrons we can generate rotations similar to braiding operations. In the paper \cite{Flensberg2011} the authors proposed the scheme to manipulate the state of a set of two Majorana fermions by changing the even/odd parity and degeneracy of the dot qubit states
with the quantum flux $\varphi_{1}=2n\pi$.

The carbon nanotube quantum dot with side-attached three Majorana fermions (CNTQD-3TSC) remains in the strongly correlated Kondo phase only in
the weak coupling strength regime. The CNTQD-3TSC device is determined in the weak coupling limit for $Q=1(3)$e by the states $\ket{q_{a(b)}}_{32}$.
$\ket{q_{w}}_{48}$ is the ground state for $Q=2$e and is squeezed to the e-h symmetry point in the strong coupling limit.
The squeezing mechanism created the new type of the U$^{\star}$(1) charge symmetry for even number of electrons in the system. In the normal
state, this symmetry exists only for the points with the fractional charge in the quantum dot system.
Beyond this line, the system defines the low and high energy octuplets $\ket{q_{x(z)}}_{8}$.
One of the interesting points is the opposite charge-leaking mechanism (observed in the general susceptibilities, nonlinear current and shot noise) - directly visible in the structure of the topological qubit states (the leaking states are marked in red in Eqs. (13-14)).
The high energy octuplet $\ket{q_{z}}_{8}$ can be expressed, as follows:
\begin{eqnarray}
&&\nonumber\ket{q_{-z_{1}}}=\frac{1}{\sqrt{8}}(-\ket{00\underline{0}\overline{0}\underline{\underline{0}}}
+\ket{0\uparrow\underline{0}\overline{\Uparrow}\underline{\underline{0}}}
-\ket{20\underline{\Uparrow}\overline{0}\underline{\underline{\Downarrow}}}\\&&\nonumber\color{red}
+\ket{2\uparrow\underline{\Uparrow}\overline{\Uparrow}\underline{\underline{\Downarrow}}}\color{black}
-\ket{\downarrow\uparrow\underline{0}\overline{\Uparrow}\underline{\underline{\Downarrow}}}
+\ket{\uparrow0\underline{0}\overline{0}\underline{\underline{\Downarrow}}}\\&&\nonumber
+\ket{\uparrow0\underline{\Uparrow}\overline{0}\underline{\underline{0}}}
-\ket{\uparrow\uparrow\underline{\Uparrow}\overline{\Uparrow}\underline{\underline{0}}})\\
&&\nonumber\ket{q_{-z_{2}}}=\frac{1}{\sqrt{8}}(-\ket{00\underline{0}\overline{0}\underline{\underline{\Downarrow}}}
+\ket{0\uparrow\underline{0}\overline{\Uparrow}\underline{\underline{\Downarrow}}}
-\ket{20\underline{\Uparrow}\overline{0}\underline{\underline{0}}}\\&&\nonumber\color{red}
+\ket{2\uparrow\underline{\Uparrow}\overline{\Uparrow}\underline{\underline{0}}}\color{black}
+\ket{\downarrow0\underline{0}\overline{0}\underline{\underline{0}}}
-\ket{\downarrow\uparrow\underline{0}\overline{\Uparrow}\underline{\underline{0}}}\\&&
+\ket{\uparrow0\underline{\Uparrow}\overline{0}\underline{\underline{\Downarrow}}}
-\ket{\uparrow\uparrow\underline{\Uparrow}\overline{\Uparrow}\underline{\underline{\Downarrow}}})\\
&&\nonumber\ket{q_{-z_{3}}}=\frac{1}{\sqrt{8}}(-\ket{00\underline{0}\overline{\Uparrow}\underline{\underline{0}}}
+\ket{0\uparrow\underline{0}\overline{0}\underline{\underline{0}}}
-\ket{20\underline{\Uparrow}\overline{\Uparrow}\underline{\underline{\Downarrow}}}\\&&\nonumber\color{red}
+\ket{2\uparrow\underline{\Uparrow}\overline{0}\underline{\underline{\Downarrow}}}\color{black}
+\ket{\downarrow0\underline{0}\overline{\Uparrow}\underline{\underline{\Downarrow}}}
-\ket{\downarrow\uparrow\underline{0}\overline{0}\underline{\underline{\Downarrow}}}\\&&\nonumber
+\ket{\uparrow0\underline{\Uparrow}\overline{\Uparrow}\underline{\underline{0}}}
-\ket{\uparrow\uparrow\underline{\Uparrow}\overline{0}\underline{\underline{0}}})\\
&&\nonumber\ket{q_{-z_{4}}}=\frac{1}{\sqrt{8}}(-\ket{00\underline{\Uparrow}\overline{0}\underline{\underline{0}}}
+\ket{0\uparrow\underline{\Uparrow}\overline{\Uparrow}\underline{\underline{0}}}
-\ket{20\underline{0}\overline{0}\underline{\underline{\Downarrow}}}\\&&\nonumber\color{red}
+\ket{2\uparrow\underline{0}\overline{\Uparrow}\underline{\underline{\Downarrow}}}\color{black}
+\ket{\downarrow0\underline{\Uparrow}\overline{0}\underline{\underline{\Downarrow}}}
-\ket{\downarrow\uparrow\underline{\Uparrow}\overline{\Uparrow}\underline{\underline{\Downarrow}}}\\&&\nonumber
+\ket{\uparrow0\underline{0}\overline{0}\underline{\underline{0}}}
-\ket{\uparrow\uparrow\underline{0}\overline{\Uparrow}\underline{\underline{0}}})
\end{eqnarray}
$\ket{q_{-z_{\overline{1}}}}$, $\ket{q_{-z_{\overline{2}}}}$, $\ket{q_{-z_{\overline{3}}}}$ and $\ket{q_{-z_{\overline{4}}}}$ are the states with opposite configuration in the topological part. The \color{red}$\ket{2\uparrow n_{1}n_{2}n_{3}}$\color{black} states penetrate into the forbidden charge sectors in the CNTQD-3TSC quantum device. The states leak from the triple occupied states on the quantum dot to the charge states above the e-h symmetry point. The mechanism is related to the entanglement of eight spanned states by the tunneling strength between CNTQD and the three Majorana fermions. The charge-leaking states appear around the e-h symmetry point, indicating the strong dependence on the Coulomb interaction. The low energy states, which show the same charge-leaking mechanism can be written as follows:
\begin{eqnarray}
&&\nonumber\ket{q_{-x_{1}}}=\frac{1}{\sqrt{8}}(\ket{02\underline{0}\overline{\Uparrow}\underline{\underline{0}}}\color{red}
-\ket{0\downarrow\underline{0}\overline{0}\underline{\underline{0}}}\color{black}
+\ket{22\underline{\Uparrow}\overline{\Uparrow}\underline{\underline{\Downarrow}}}\\&&\nonumber
-\ket{2\downarrow\underline{\Uparrow}\overline{0}\underline{\underline{\Downarrow}}}
-\ket{\downarrow2\underline{0}\overline{\Uparrow}\underline{\underline{\Downarrow}}}
+\ket{\downarrow\downarrow\underline{0}\overline{0}\underline{\underline{\Downarrow}}}\\&&\nonumber
-\ket{\uparrow2\underline{\Uparrow}\overline{\Uparrow}\underline{\underline{0}}}
+\ket{\uparrow\downarrow\underline{\Uparrow}\overline{0}\underline{\underline{0}}})\\
&&\nonumber\ket{q_{-x_{2}}}=\frac{1}{\sqrt{8}}(\ket{02\underline{0}\overline{\Uparrow}\underline{\underline{\Downarrow}}}\color{red}
-\ket{0\downarrow\underline{0}\overline{0}\underline{\underline{\Downarrow}}}\color{black}
+\ket{22\underline{\Uparrow}\overline{\Uparrow}\underline{\underline{0}}}\\&&\nonumber
-\ket{2\downarrow\underline{\Uparrow}\overline{0}\underline{\underline{0}}}
-\ket{\downarrow2\underline{0}\overline{\Uparrow}\underline{\underline{0}}}
+\ket{\downarrow\downarrow\underline{0}\overline{0}\underline{\underline{0}}}\\&&\nonumber
-\ket{\uparrow2\underline{\Uparrow}\overline{\Uparrow}\underline{\underline{\Downarrow}}}
+\ket{\uparrow\downarrow\underline{\Uparrow}\overline{0}\underline{\underline{\Downarrow}}})\\
&&\nonumber\ket{q_{-x_{3}}}=\frac{1}{\sqrt{8}}(\ket{02\underline{0}\overline{0}\underline{\underline{0}}}\color{red}
-\ket{0\downarrow\underline{0}\overline{\Uparrow}\underline{\underline{\Downarrow}}}\color{black}
+\ket{22\underline{\Uparrow}\overline{0}\underline{\underline{\Downarrow}}}\\&&
-\ket{2\downarrow\underline{\Uparrow}\overline{\Uparrow}\underline{\underline{\Downarrow}}}
-\ket{\downarrow2\underline{0}\overline{0}\underline{\underline{\Downarrow}}}
+\ket{\downarrow\downarrow\underline{0}\overline{\Uparrow}\underline{\underline{\Downarrow}}}\\&&\nonumber
-\ket{\uparrow2\underline{\Uparrow}\overline{0}\underline{\underline{0}}}
+\ket{\uparrow\downarrow\underline{\Uparrow}\overline{\Uparrow}\underline{\underline{0}}})\\
&&\nonumber\ket{q_{-x_{4}}}=\frac{1}{\sqrt{8}}(\ket{02\underline{\Uparrow}\overline{\Uparrow}\underline{\underline{0}}}\color{red}
-\ket{0\downarrow\underline{\Uparrow}\overline{0}\underline{\underline{0}}}\color{black}
+\ket{22\underline{0}\overline{\Uparrow}\underline{\underline{\Downarrow}}}\\&&\nonumber
-\ket{2\downarrow\underline{0}\overline{0}\underline{\underline{\Downarrow}}}
-\ket{\downarrow2\underline{\Uparrow}\overline{\Uparrow}\underline{\underline{\Downarrow}}}
+\ket{\downarrow\downarrow\underline{\Uparrow}\overline{0}\underline{\underline{\Downarrow}}}\\&&\nonumber
-\ket{\uparrow2\underline{0}\overline{\Uparrow}\underline{\underline{0}}}
+\ket{\uparrow\downarrow\underline{0}\overline{0}\underline{\underline{0}}})
\end{eqnarray}
$\ket{q_{-x_{\overline{1}}}}$, $\ket{q_{-x_{\overline{2}}}}$, $\ket{q_{-x_{\overline{3}}}}$ and $\ket{q_{-x_{\overline{4}}}}$ are the states with opposite configuration in the topological sector of ket states. Three Majorana fermions do not allow the formation of the Kondo state. These three channels  $\{\pm\uparrow,+\downarrow\}$ are involved in interference with the Majorana fermion quantum states. The effect is qualitatively similar to result for the SU(2)-Kondo dot with a side-attached MF, where in the strong coupling limit the quantum conductance at the e-h symmetry point reaches ${\mathcal{G}}=(3/2)(e^{2}/h)$\cite{Lopez2013,Liu2015}.
\begin{figure}[b!]
\includegraphics[width=0.75\linewidth]{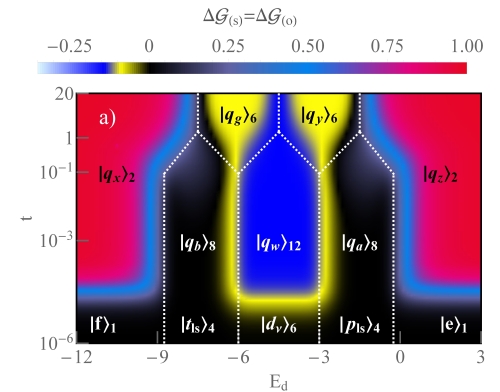}\\
\includegraphics[width=0.75\linewidth]{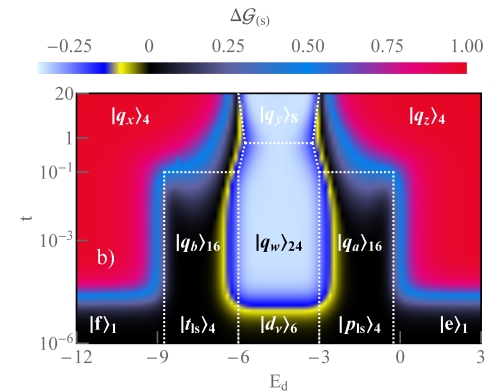}\\
\includegraphics[width=0.75\linewidth]{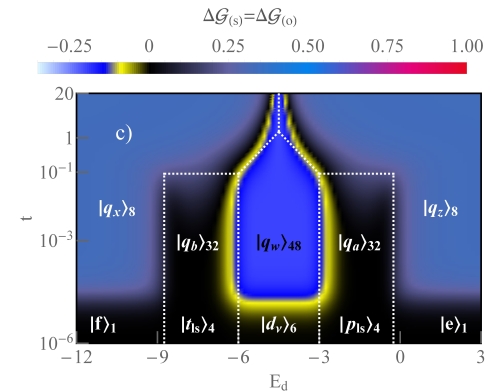}\\
\includegraphics[width=0.75\linewidth]{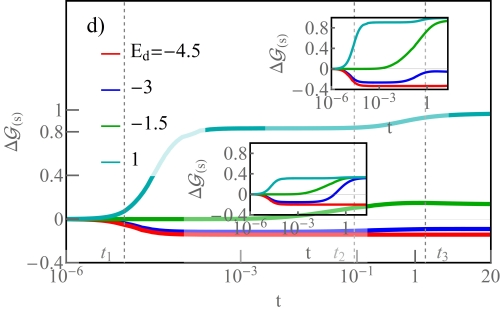}
\caption{\label{fig:epsart} (Color online) a-c) The density plot of the spin $\Delta{\cal{G}}_{(s)}$ and orbital $\Delta{\cal{G}}_{(o)}$ polarization of the conductance as a function of $E_{d}$ and $t$ for CNTQD coupled to single, double and triple MF states. d) $t$ dependence of $\Delta{\cal{G}}_{s}$ for CNTQD-TSC. The upper and lower insets  in Fig. d show $\Delta{\cal{G}}_{(s)}$ for QD coupled to 2TSC and 3TSC.}
\end{figure}
\begin{figure}[b!]
\includegraphics[width=0.75\linewidth]{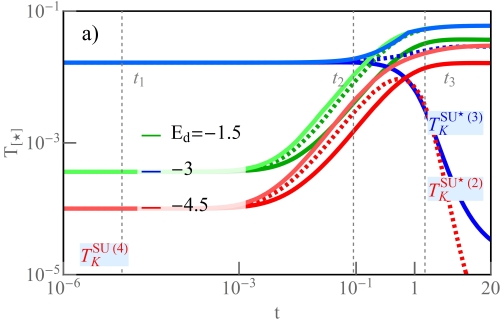}\\
\includegraphics[width=0.75\linewidth]{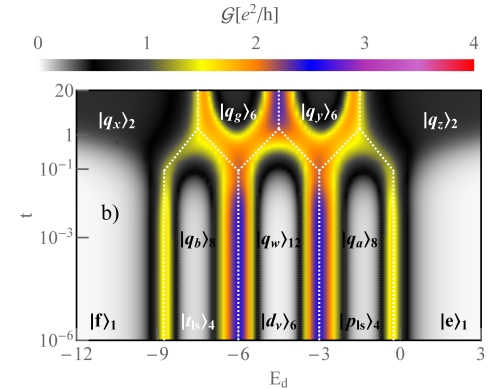}\\
\includegraphics[width=0.75\linewidth]{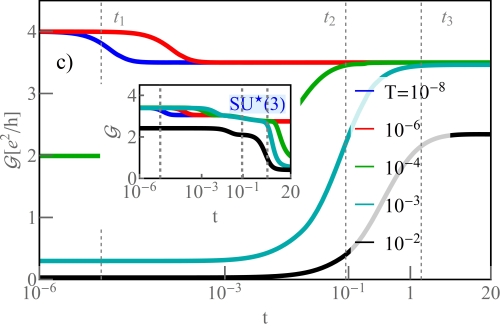}
\caption{\label{fig:epsart} (Color online) a) The characteristic temperature $T_{[\star]}$ as a function of $t$. $T^{SU(4)}_{K}$, $T^{SU^{\star}(3)}_{K}$ and $T^{SU^{\star}(2)}_{K}$ are the Kondo temperatures. The dark, dark dashed, and light lines present $T_{[\star]}$ for CNTQD-TSC, CNTQD-2TSC and CNTQD coupled to 3MFs. b) $\cal{G}$ of the CNTQD-TSC device as a function of $E_{d}$ and $t$ for finite temperature $T=10^{-3}>T_{K}$. c) $t$ dependence of $\cal{G}$ with increasing $T$ for $E_{d}=-4.5$ and $E_{d}=-3$ (inset). The quantum conductances are obtained for CNTQD associated with a single MF.}
\end{figure}

\subsection{Thermodynamics of the Kondo system}
The thermodynamic potential $\widetilde{F}$ in the KR-sbMFA approach is given by the partition function $\widetilde{Z}$ at the saddle point of the action function $\widetilde{S}$ (see \cite{Igoshev2015}):
\begin{eqnarray}
&&\nonumber \widetilde{F}=-T ln(\widetilde{Z})=\widetilde{F}_{b}+\widetilde{F}_{f}+\Delta\widetilde{F}=U\sum_{\nu}d^{\dagger}_{\nu}d_{\nu}
\\
&&\nonumber+3U\sum_{ls}t^{\dagger}_{ls}t_{ls}+6Uf^{\dagger}f+\lambda({\cal{I}}-1)-\lambda_{ls}\sum_{ls}Q_{ls}\\
&&+\widetilde{F}_{f}+\Delta\widetilde{F}
\end{eqnarray}
where $\widetilde{F}_{b(f)}$ are the bosonic and fermionic parts of the free energy.
$\Delta\widetilde{F}$ is the correction to the thermodynamic potential (it includes the two- and three-body fluctuations introduced by the FL theory \cite{Mora2009,Moca2015,Oguri2018}). The $\widetilde{F}_{f}$ can be written in terms of the Matsubara Green's functions
using the contour integral method with cut along
the real frequency axis \cite{Coleman1987,Kim2016}:
\begin{eqnarray}
&& \widetilde{F}_{f}=-T\sum_{\nu,iw_{n}}ln[\Lambda_{\nu}-iw_{n}]
\\&&\nonumber-T\sum_{m\nu',iw_{n}}a_{m}ln[\Lambda_{m\nu'}-iw_{n}]=\sum_{\nu}\int^{\Lambda_{\nu}}_{-\infty}Im\{X[z]\}dz
\\&&\nonumber+\sum_{m\nu'}\int^{\Lambda_{m\nu'}}_{-\infty}Im\{a_{m}X[z]\}dz
\end{eqnarray}
$iw_{n}$ is the Matsubara frequency, and $\Lambda_{\nu(m\nu')}$ are the complex poles of the quasiparticle Kondo resonance. The poles of the channel $\nu'$ coupled to the TSC are represented by $\Lambda_{m\nu'}$ where $m=0,\pm$ ($m=0$ is associated with the zero Majorana bound state and $m=\pm$ represents the states excited by $t$). The complex poles can be written in the form $\Lambda_{0\nu'}=(-2i\widetilde{\Gamma}_{\nu'}-i\delta+c)/3+b/(3c)$ and $\Lambda_{\pm\nu'}=(-2i\widetilde{\Gamma}_{\nu'}-i\delta\pm ie^{m i(\pi/6)})/3\pm (ie^{\overline{m} i(\pi/6)}b)/(3c)$, where $c=\sqrt[3]{d+\sqrt{b^{3}+d^{2}}}$, $b=-3\widetilde{E}^{2}_{\nu'}-6\widetilde{t}^{2}+\widetilde{\Gamma}^{2}_{\nu'}$ and $d=-2i\widetilde{\Gamma}_{\nu'}(9\widetilde{E}^{2}_{\nu'}-9\widetilde{t}^{2}+\widetilde{\Gamma}^{2}_{\nu'})$.
The coefficients are defined by $a_{m}=(-\widetilde{t}^{2}+(\Lambda_{m\nu'}+i\delta)(\Lambda_{m\nu'}+\widetilde{E}_{\nu'}
+i\delta+i\widetilde{\Gamma}_{\nu'}))/\prod_{m'\neq m}(\Lambda_{m\nu'}-\Lambda_{m'\nu'})$.
Here $\nu'=+\uparrow,-\uparrow,+\downarrow$ are the quantum numbers addressed to one (two) and three TSCs coupled with CNTQD. $X[z]=(1/(2\pi))\sum_{\alpha={L,R}}\{\Psi_{0}[1/2+(z\pm V_{\alpha})/(2\pi iT)]-ln[W/(2\pi iT)]\}$ where $\Psi_{0}$ is the hypergeometric digamma function, and $X[z]$ is written for the non-equilibrium case.

The Fermi liquid theory describes the low-energy regime and is based on the following assumptions: the Kondo singlet elastically scatters conduction electrons, the dressed polarization of the singlet leads to the weak interactions between electrons with different spin orbitals, and the energy of the system is a function of the bare energies $E_{k}$ and the relative quasiparticle occupancy numbers $\delta n_{\nu}$. Using the FL theory \cite{Nozieres1974} and adopting the results of \cite{Mora2009,Oguri2022}, $\Delta\widetilde{F}$ can be expressed in the following general form:
\begin{eqnarray}
&&\nonumber \Delta\widetilde{F}=-\frac{1}{\pi T_{K}}\sum_{\nu,E}\left(\alpha_{1,\nu}E+\frac{\alpha_{2,\nu}E^{2}}{T_{K}}\right)\delta n_{\nu}
+\\&&\nonumber \frac{1}{\pi T_{K}}\sum_{\nu<\nu',EE'}\left(\varphi_{1,\nu\nu'}+\frac{\varphi_{2,\nu\nu'\nu'}(E+E')}{2T_{K}}\right)\delta n_{\nu}\delta n_{\nu'}+\\&&-\frac{1}{\pi T_{K}}\sum_{\nu<\nu'<\nu'',EE'E''}\frac{\varphi_{2,\nu\nu'\nu''}}{(N-2)T_{K}}\delta n_{\nu}\delta n_{\nu'}\delta n_{\nu''}
\end{eqnarray}
where $\alpha_{1,\nu}/\pi=\widetilde{\chi}_{\nu\nu}$ ($\alpha_{2,\nu}/\pi=-(1/2)\widetilde{\chi}^{[3]}_{\nu\nu\nu}$), $\varphi_{1,\nu\nu'}/\pi=-\widetilde{\chi}_{\nu\nu'}$ ($\varphi_{2,\nu\nu'\nu''}/\pi=2\widetilde{\chi}^{[3]}_{\nu\nu'\nu''}$) are the FL coefficients [see \cite{Oguri2018}\cite{Moca2015}], which are the functions of the renormalized spin-orbital two- ($\widetilde{\chi}_{\nu_{1}\nu_{2}}$) and three-body static correlators ($\widetilde{\chi}^{[3]}_{\nu_{1}\nu_{2}\nu_{3}}$).

In this approach, we can integrate the energies $E$ by $\delta n_{\nu}$ in Eq. (17) and find $\Delta\widetilde{F}$ in the form intended for sbMFA calculations, determined by the general susceptibilities:
\begin{eqnarray}
&&\nonumber \Delta\widetilde{F}=-\sum_{\nu}\left(\frac{\widetilde{\chi}_{\nu\nu}\delta \widetilde{E}_{\nu}}{2}
+\frac{\widetilde{\chi}^{[3]}_{\nu\nu\nu}\delta \widetilde{E}^{2}_{\nu}}{6}\right)\delta \widetilde{E}_{\nu}+
\\&&\nonumber\sum_{\nu<\nu'}\left(-\widetilde{\chi}_{\nu\nu'}-\frac{\widetilde{\chi}^{[3]}_{\nu\nu'\nu'}(\delta \widetilde{E}_{\nu'}+\delta \widetilde{E}_{\nu'})}{4}\right)\delta \widetilde{E}_{\nu}\delta \widetilde{E}_{\nu'}+
\\&&\sum_{\nu<\nu'<\nu''}\frac{\widetilde{\chi}^{[3]}_{\nu\nu'\nu''}}{2}\delta \widetilde{E}_{\nu}\delta \widetilde{E}_{\nu'}\delta \widetilde{E}_{\nu''}
\end{eqnarray}
$\widetilde{\chi}_{\nu\nu'}=-\frac{\partial^{2} \Delta\widetilde{F}}{\partial \delta\widetilde{E}^{2}_{\nu}}$ and $\widetilde{\chi}^{[3]}_{\nu\nu'\nu'}=-\frac{\partial^{3} \Delta\widetilde{F}}{\partial \delta\widetilde{E}_{\nu}\partial \delta\widetilde{E}^{2}_{\nu'}}$. $\delta\widetilde{E}_{\nu}=\sqrt{\langle\widetilde{E}_{\nu}^{2}\rangle-\widetilde{E}_{\nu}^{2}}$ is the fluctuation of the quasiparticle level and $\delta\widetilde{E}_{\nu}\ll\widetilde{E}_{\nu}$. In the weak coupling ansatz ($\widetilde{t}\ll t$), $\Delta \widetilde{F}$ does not fundamentally change the solution of the equations in the self-consistent sbMFA procedure. Formally, we can prove this by considering $\delta\widetilde{E}_{\nu}=(\Delta\widetilde{\Gamma}_{\nu})\cot[\pi Q_{\nu}]$, where $\Delta\widetilde{\Gamma}_{\nu}=\Gamma_{\nu}\delta z^{2}_{\nu}\ll\Gamma_{\nu}z^{2}_{\nu}$ and $\delta z_{\nu}=\sqrt{\sum_{\nu}z^{\dagger}_{\nu}\cdot\sum_{\nu}z_{\nu}-z^{2}_{\nu}}$.
The expected values of the boson fields $b_{n}$ and the constraints $\lambda, \lambda_{ls}$
we found by solving the non-equilibrium self-consistent equations from Eq.(3), modified by the $\Delta \widetilde{F}$ and completed with an additional equation:
\begin{eqnarray}
&&\nonumber \frac{\partial\widetilde{F}}{\partial b_{n}}=\Delta \widetilde{{\mathcal{H}}}_{{n}}+\Delta \widetilde{E}_{{n}}+\frac{\partial\Delta\widetilde{F}}{\partial b_{n}}=0\\
&&\frac{\partial\widetilde{F}}{\partial \lambda}={\cal{I}}-1=0\\
&&\nonumber\frac{\partial\widetilde{F}}{\partial \lambda_{ls}}=\langle f^{\dagger}_{ls}f_{ls}\rangle^{<}-Q_{ls}
+\frac{\partial\Delta\widetilde{F}}{\partial\lambda_{ls}}=0\\
&&\nonumber\frac{\partial\widetilde{F}}{\partial \delta \widetilde{E}_{\nu}}=0
\end{eqnarray}

Another approach that can be applied to the sbMFA method is presented in \cite{Lavagna1990}, where quantum fluctuations are taken into account at the level of individual boson fields and Lagrange multipliers. By integrating over the Grassmann variables
and expanding to second order in the boson variables, one obtains Gaussian corrections to the saddle-point action. The alternative methods are based on the spin-rotation invariant (SRI) representation of the auxiliary bosons, where by finding the Z-component and the transverse components of the spin operators, the charge and spin density fluctuations can be obtained in terms of the auxiliary boson fields \cite{Fresard1997}.

Differentiating $\widetilde{F}_{f}$ with respect to $\widetilde{E}_{\nu}$ we get the spin-orbital occupation number $n_{\nu=ls}$ in the following way:
\begin{eqnarray}
&&  n_{\nu}=Q_{\nu}=\langle f^{\dagger}_{ls}f_{ls}\rangle^{<}=\frac{\partial \widetilde{F}_{f}}{\partial \widetilde{E}_{\nu}}=\frac{\delta_{\nu}}{\pi}
=\\&&\nonumber\sum_{\alpha}\textrm{Im} \left\{\frac{-ln\left[\frac{W}{2\pi iT}\right]+\Psi_{0}\left[1/2+\frac{\widetilde{E}_{\nu}+i\widetilde{\Gamma}_{\nu}\pm V_{\alpha}}{2\pi iT}\right]}{2\pi}\right\}.
\end{eqnarray}
In further calculations (especially for the shot noise and the current in the nonlinear voltage range), the general two- and three-body susceptibilities will be relevant. Thermodynamically, we can define two- and three-body correlation functions as follows  $\widetilde{\chi}_{\nu_{1}\nu_{2}}=\int^{1/T}_{0}d\tau\langle\delta n_{\nu_{2}}(\tau)\delta n_{\nu_{1}}(0)\rangle^{<}$ and $\widetilde{\chi}^{[3]}_{\nu_{1}\nu_{2}\nu_{3}}=-\int^{1/T}_{0}d\tau_{3}\int^{1/T}_{0}d\tau_{2}\langle T_{[\tau]}\delta n_{\nu_{3}}(\tau_{3})\delta n_{\nu_{2}}(\tau_{2})\delta n_{\nu_{1}}(0)\rangle^{<}$ (for diagonal parts $\nu_{1}=\nu_{2}=\nu_{3}=\nu$). According to the definitions, the static nonequilibrium partial susceptibilities of the renormalized quasiparticle cloud can be expressed by the following (as used in the NRG calculations \cite{Oguri2022}):
\begin{eqnarray}
&&\nonumber \widetilde{\chi}_{\nu\nu}=-\frac{\partial^{2} \widetilde{F}_{f}}{\partial \widetilde{E}^{2}_{\nu}}=\\
&& \sum_{\alpha}\textrm{Im} \left\{\frac{-\Psi_{1}\left[1/2+\frac{\widetilde{E}_{\nu}+i\widetilde{\Gamma}_{\nu}\pm V_{\alpha}}{2\pi iT}\right]}{4\pi^{2} iT}\right\}
\stackrel{V=T=0}=
\\&&\nonumber\frac{\widetilde{\Gamma}_{\nu}}{\pi(\widetilde{E}_{\nu}^{2}+\widetilde{\Gamma}_{\nu}^{2})}
=\frac{\widetilde{\Gamma}_{\nu}}{\pi T^{2}_{K,\nu}}=\frac{\sin^{2}[\delta_{\nu}]}{\pi \widetilde{\Gamma}_{\nu}}
=\frac{\sin^{2}[\delta_{\nu}]\delta n_{\nu}}{\pi \Gamma \widetilde{z}^{2}_{\nu}}\\
&&\nonumber \widetilde{\chi}^{[3]}_{\nu\nu\nu}=-\frac{\partial^{3} \widetilde{F}_{f}}{\partial \widetilde{E}^{3}_{\nu}}=\\
&& \sum_{\alpha}\textrm{Im} \left\{\frac{\Psi_{2}\left[1/2+\frac{\widetilde{E}_{\nu}+i\widetilde{\Gamma}_{\nu}\pm V_{\alpha}}{2\pi iT}\right]}{8\pi^{3} T^{2}}\right\}
\stackrel{V=T=0}=\\&&\nonumber\frac{-2\widetilde{\Gamma}_{\nu}\widetilde{E}_{\nu}}{\pi(\widetilde{E}_{\nu}^{2}+\widetilde{\Gamma}_{\nu}^{2})^{2}}
=\frac{-2\widetilde{\Gamma}_{\nu}\widetilde{E}_{\nu}}{\pi T^{4}_{K,\nu}}=\frac{-2\cos[\delta_{\nu}]\sin^{3}[\delta_{\nu}]}{\pi \widetilde{\Gamma}^{2}_{\nu}}
\end{eqnarray}
The diagonal $\nu'=\nu$ static susceptibilities, according to  Yamada and Yosida \cite{Yamada1975} are determined by the renormalization factor $1/\widetilde{z}^{2}_{\nu}$. In the channel directly coupled to the  Majorana fermion $\nu'$, two and three-body correlation functions can be expressed as follows:
\begin{eqnarray}
&& \widetilde{\chi}_{\nu'\nu'}\stackrel{V=T=0}=\frac{\widetilde{\Gamma}_{\nu'}+\frac{\widetilde{t}^{2}_{\nu'}}{\delta}}
{\pi(\widetilde{E}_{\nu'}^{2}+\widetilde{\Gamma}_{\nu'}^{2}+\frac{2\widetilde{t}^{2}_{\nu'}\widetilde{\Gamma}_{\nu'}}{\delta})}\\
&&\widetilde{\chi}^{[3]}_{\nu'\nu'\nu'}=\frac{\partial\widetilde{\chi}_{\nu'\nu'}}{\partial\widetilde{E}_{\nu'}}
=\frac{-2(\widetilde{\Gamma}_{\nu'}+\frac{\widetilde{t}^{2}_{\nu'}}{\delta})\widetilde{E}_{\nu'}}
{\pi(\widetilde{E}_{\nu'}^{2}+\widetilde{\Gamma}_{\nu'}^{2}+\frac{2\widetilde{t}^{2}_{\nu'}\widetilde{\Gamma}_{\nu'}}{\delta})^{2}}
\end{eqnarray}
In both cases the two-body static correlation functions are equal to the quasiparticle density of states at the Fermi level:
$\widetilde{\chi}_{\nu\nu}=\langle\langle f_{\nu};f^{\dagger}_{\nu}\rangle\rangle|_{E=0}=\widetilde{\varrho}_{\nu}(0)$ ($\widetilde{\chi}_{\nu'\nu'}=\widetilde{\varrho}_{\nu'}$). The diagonal and off-diagonal two- and three-body correlation functions result from the FL theory and the spin-orbital fluctuations can be obtained by using the derivatives in the following form:
\begin{eqnarray}
&& \nonumber\widetilde{\chi}_{\nu\nu}=-\frac{\partial^{2} \Delta\widetilde{F}}{\partial \delta\widetilde{E}^{2}_{\nu}}
= \nonumber\widetilde{\chi}_{\nu\nu}+\sum_{\nu'}\widetilde{\chi}^{[3]}_{\nu'\nu\nu}\delta \widetilde{E}_{\nu'}\stackrel{\delta \widetilde{E}_{\nu'}\approx0}\approx\widetilde{\chi}_{\nu\nu}\\
&& \nonumber\widetilde{\chi}^{[3]}_{\nu\nu\nu}=-\frac{\partial^{3} \Delta\widetilde{F}}{\partial \delta\widetilde{E}^{3}_{\nu}}=\widetilde{\chi}^{[3]}_{\nu\nu\nu}=\frac{\partial\widetilde{\chi}_{\nu\nu}}{\partial\widetilde{E}_{\nu}}\\
&& \nonumber\widetilde{\chi}_{\nu\nu'}=-\frac{\partial^{2} \Delta\widetilde{F}}{\partial \delta\widetilde{E}_{\nu}\partial\delta\widetilde{E}_{\nu'}}
= \nonumber\widetilde{\chi}_{\nu\nu'}+\widetilde{\chi}^{[3]}_{\nu\nu'\nu'}\delta \widetilde{E}_{\nu'}+\widetilde{\chi}^{[3]}_{\nu'\nu\nu}\delta \widetilde{E}_{\nu}
\\&&\nonumber-(1/2)\sum_{\nu''\neq(\nu,\nu')}\widetilde{\chi}^{[3]}_{\nu\nu'\nu''}\delta \widetilde{E}_{\nu''}\stackrel{\delta \widetilde{E}_{\nu(\nu',\nu'')}\approx0}\approx\widetilde{\chi}_{\nu\nu'}=\\&&-(W_{\nu\nu'}-1)
\sqrt{\widetilde{\chi}_{\nu\nu}\widetilde{\chi}_{\nu'\nu'}}\\
&& \nonumber\widetilde{\chi}^{[3]}_{\nu\nu'\nu'}=-\frac{\partial^{3} \Delta\widetilde{F}}{\partial \delta\widetilde{E}_{\nu}\partial \delta\widetilde{E}^{2}_{\nu'}}=\widetilde{\chi}^{[3]}_{\nu\nu'\nu'}=\frac{\partial\widetilde{\chi}_{\nu\nu'}}{\partial\widetilde{E}_{\nu'}}
=\\&&\nonumber-(W_{\nu\nu'}-1)\frac{\partial \sqrt{\widetilde{\chi}_{\nu\nu}\widetilde{\chi}_{\nu'\nu'}}}{\partial \widetilde{E}_{\nu'}}+ \dot{W}_{\nu\nu'}\sqrt{\widetilde{\chi}_{\nu\nu}\widetilde{\chi}_{\nu'\nu'}}=\\&&\nonumber
-K_{\nu\nu'}\widetilde{\chi}^{[3]}_{\nu'\nu'\nu'}
\end{eqnarray}
where $W_{\nu\nu'}\equiv1-\widetilde{\chi}_{\nu\nu'}/\sqrt{\widetilde{\chi}_{\nu\nu}\widetilde{\chi}_{\nu'\nu'}}$ is the Wilson ratio \cite{Wilson1975,Coleman1987,Hewson1997,Nishikawa2013,Coleman2015}. By definition, $W_{\nu\nu'}$ is expressed by the susceptibilities and, in its original form, is experimentally determinable by the ratio of the spin susceptibility $\chi_{(s)}$ and the linear coefficient of the specific heat $\gamma_{N}$, as will be discussed later in this subsection. Two-body correlation functions written on the basis of $\Delta \widetilde{F}$ are more general and include the correction for the fluctuation $\delta \widetilde{E}_{\nu}$. For the systems where $\delta \widetilde{E}_{\nu}\sim \widetilde{E}_{\nu}$ the additive part can play a crucial role. In practice, to compute 2-body even correlation functions $\widetilde{\chi}_{\nu\nu'}$ and 3-body odd correlation functions $\widetilde{\chi}_{\nu\nu'\nu'}$, we can formally adopt the Random Phase Approximation (RPA) method and its correction for non-zero frequency susceptibility \cite{Coleman1987,Lee2004}. This alternative approach introduces the imaginary and real parts of the higher-order correlations and will be useful for discussing of the frequency dependent shot noise and the current. In this paper we have proposed to use the weak coupling approach to calculate the Wilson ratios and consequently the higher-order correlation functions. The weak coupling approach is based on the low renormalization coupling strength of the Kondo resonance to the normal electrodes ($z^{2}_{\nu}$) ($\widetilde{t}_{0}\ll t_{0}$). Finally, in the general case for SU(N) Anderson model, we found that the Wilson ratio $W_{\nu\nu'}=1-\widetilde{\chi}_{\nu\nu'}/\sqrt{\widetilde{\chi}_{\nu\nu}\widetilde{\chi}_{\nu'\nu'}}\approx1+1/(N-1)$ and as we can see $-\widetilde{\chi}_{\nu\nu'}/\sqrt{\widetilde{\chi}_{\nu\nu}\widetilde{\chi}_{\nu'\nu'}}=1/(N-1)$ is the correction from two-particle correlators.  $W_{\nu\nu'}>1$ is the consequence of the finite Coulomb (residual) interaction  between the quasipaticles and depends on the degree of degeneracy $N$\cite{Nishikawa2013}.  In general terms, using the weak coupling ansatz and exact expression for the partition function $Z_{N}$, the Wilson ratio can be written as follows:
\begin{eqnarray}
&&W_{\nu\nu'}-1=\frac{n_{\nu}n_{\nu}-n_{\nu\nu'}}{\sqrt{\delta n^{2}_{\nu}\delta n^{2}_{\nu'}}}=
\\&&\nonumber\frac{Q_{\nu}Q_{\nu'}-Q_{\nu\nu'}}{\sqrt{Q_{\nu}(I-Q_{\nu})Q_{\nu'}(I-Q_{\nu'})}}=\frac{\delta Q_{\nu\nu'}}{\Delta Q_{\nu\nu'}}\\&&K_{\nu\nu'}=\frac{\delta Q_{\nu\nu'}}{\delta n^{2}_{\nu'}}
\end{eqnarray}
where $Q_{\nu}$ is the charge expressed by the boson fields operators (averaged over the time, in the static susceptibilities) and $Q_{\nu\nu'}=\sum_{\nu\nu'}b^{2}_{\nu\nu'}I$ is the sum
over all boson fields amplitudes at which the two-particle state $\nu\nu'$ exists. Surprisingly, the ansatz quantitatively reproduces the NRG result, which use the self-energy and the Ward identities to calculate 2(3)-body quantities \cite{Zawadowski1978,Oguri2022}.
$\dot{W}_{\nu\nu'}$ in Eq. (25) is the derivative of the Wilson ratio and plays the important role in the 3-body correlation function. Formally, $\dot{W}_{\nu\nu'}$ can be expressed in terms of $K_{\nu\nu'}$, but using the weak coupling approach, it can be defined as:
\begin{eqnarray}
&&\dot{W}_{\nu\nu'}=-\frac{\partial W_{\nu\nu'}}{\partial \widetilde{E}_{\nu'}}=\\&&\nonumber-\frac{\delta n^{2}_{\nu'}\delta Q_{\nu\nu'}[Q_{\nu}(Q_{\nu}+Q_{\nu'}-I-2Q_{\nu\nu'})+Q_{\nu\nu'}]}
{2\Delta Q^{3}_{\nu\nu'}}
\end{eqnarray}
\begin{figure}[b!]
\includegraphics[width=0.75\linewidth]{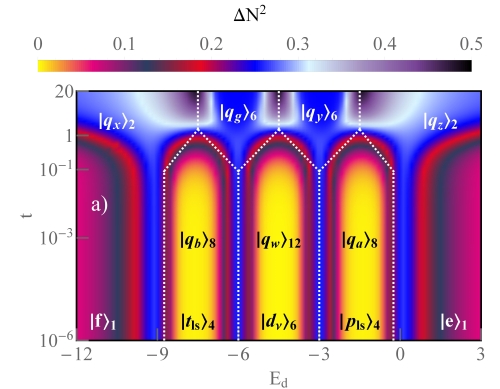}\\
\includegraphics[width=0.75\linewidth]{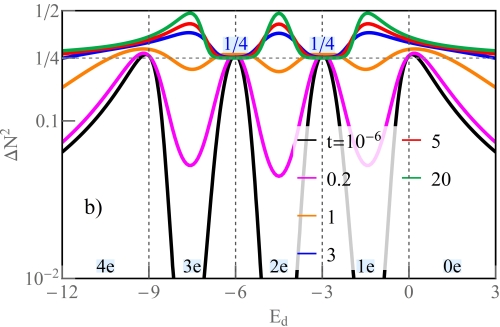}\\
\caption{\label{fig:epsart} (Color online) a) The density plot of the charge fluctuation $\Delta N^{2}$ as function of $E_{d}$ and $t$ for the CNTQD-TSC hybrid device. b) The $E_{d}$ dependence of $\Delta N^{2}$ with increasing of tunnel coupling to a single TSC. $\Delta N^{2}=1/4$ shows the saturation value in the strong coupling with MF.}
\end{figure}

Using the two-body correlators, we can write down the charge, spin and pseudospin susceptibilities in the forms (following A.C. Hewson et al. in \cite{Nishikawa2013}):
$\chi_{(c)}=\int^{1/k_{B}T}_{0}d\tau\langle\delta Q(\tau)\delta Q(0)\rangle^{<}=\sum_{lsl's'}\int^{1/k_{B}T}_{0}d\tau\langle\delta n_{ls}(\tau)\delta n_{l's'}(0)\rangle^{<}=\sum_{\nu=ls}(1-\sum_{\nu'\neq\nu}\widetilde{U}^{\nu\nu'}\widetilde{\varrho}_{\nu'})\widetilde{\varrho}_{\nu}
=\sum_{\nu\nu'}\widetilde{\chi}_{\nu\nu'}$, $\chi_{(s)}=\sum_{l}\int^{1/k_{B}T}_{0}d\tau\langle\delta S^{Z}_{l}(\tau)\delta S^{Z}_{l}(0)\rangle^{<}=\frac{1}{4}\sum_{ls}s\overline{s}\int^{1/k_{B}T}_{0}d\tau\langle\delta n_{ls}(\tau)\delta n_{ls}(0)\rangle^{<}=\frac{1}{4}\sum_{\nu=ls}(1+\widetilde{U}^{\nu\nu'}\widetilde{\varrho}_{\nu'=l\overline{s}})\widetilde{\varrho}_{\nu}
=\frac{1}{4}\sum_{lss'}ss'\widetilde{\chi}_{lsls'}$ and $\chi_{(ps)}=\frac{1}{4}\int^{1/k_{B}T}_{0}d\tau\langle\delta T^{Z}(\tau)\delta T^{Z}(0)\rangle^{<}=\frac{1}{4}\sum_{lsl's'}ll'\int^{1/k_{B}T}_{0}d\tau\langle\delta n_{ls}(\tau)\delta n_{l's'}(0)\rangle^{<}=\frac{1}{4}\sum_{\nu=ls}(1-\widetilde{U}^{\nu\nu'=l\overline{s}}\widetilde{\varrho}_{\nu'=l\overline{s}}
+\sum_{\nu'=\overline{l}s}\widetilde{U}^{\nu\nu'}\widetilde{\varrho}_{\nu'})\widetilde{\varrho}_{\nu}
=\frac{1}{4}\sum_{\nu=ls\nu'=l's'}ll'\widetilde{\chi}_{\nu\nu'}$, where $\delta n_{\nu}=n_{\nu}-\langle n_{\nu}\rangle$, $\delta Q=\sum_{\nu}n_{\nu}-\langle \sum_{\nu}n_{\nu}\rangle$, $\delta S^{Z}_{l}=(1/2)(n_{l\uparrow}-n_{l\downarrow}-\langle n_{l\uparrow}-n_{l\downarrow}\rangle)$ and $\delta T^{Z}=(1/2)(n_{+}-n_{-}-\langle n_{+}-n_{-}\rangle)$ are the total charge, spin and pseudospin fluctuations. Consequently, the residual quasiparticle interaction is given by $\widetilde{U}^{\nu\nu'}\equiv-\widetilde{\chi}_{\nu\nu'}/(\widetilde{\chi}_{\nu\nu}\widetilde{\chi}_{\nu'\nu'})$ \cite{Nishikawa2013,Oguri2020,Oguri2022}.
The linear coefficient of the quasiparticle specific heat for SU(4) Kondo symmetry in a Fermi liquid theory is given by $\gamma_{N}=\frac{\pi^{2}}{3}\sum_{\nu=ls}\widetilde{\varrho}_{\nu}$. For a fully symmetric SU(4) Kondo effect, the spin, charge and pseudospin susceptibilities are determined by the quasiparticle two-body correlation function as follows: $\chi_{(c)}=\widetilde{\chi}_{\nu\nu}[1-(\widetilde{U}^{\nu\nu}+2\widetilde{U}^{\nu\overline{\nu}})\widetilde{\chi}_{\nu\nu}]
=\widetilde{\chi}_{\nu\nu}[1-3/(N-1)]\approx0$, $\chi_{(s)}=\widetilde{\chi}_{\nu\nu}[1+\widetilde{U}^{\nu\nu}\widetilde{\chi}_{\nu\nu}]
=\widetilde{\chi}_{\nu\nu}[1+1/(N-1)]=\widetilde{\chi}_{\nu\nu}W_{(s)}$ and $\chi_{(ps)}=\widetilde{\chi}_{\nu\nu}[1-(\widetilde{U}^{\nu\nu}-2\widetilde{U}^{\nu\overline{\nu}})\widetilde{\chi}_{\nu\nu}]
=\widetilde{\chi}_{\nu\nu}[1+1/(N-1)]=\widetilde{\chi}_{\nu\nu}W_{(ps)}$.
For the fully symmetric SU(4) Kondo state, both Wilson ratios are the equal $W_{ps}=W_{s}=W_{\nu\nu'}
=(\pi^{2}N\chi_{(s)})/(3\gamma_{N})=(4/3)$ \cite{Nishikawa2013}.

To discuss the expected value of the local pseudospin for SU(4) symmetry, we used a quadratic Casimir operator, which is the bilinear sum of $N^{2}-1$ generators belonging to the Lie group. The quadratic Casimir operator is proportional to the fluctuations of the local pseudospin momentum. For the SU(4) Kondo effect, the pseudospin is screened by the conduction electrons. Based on the Lie algebra generators $\hat{\mathcal{O}}$ we can define the total local Casimir operator: $C=\sum_{i\nu\nu'}d^{\dagger}_{\nu}{\hat{\mathcal{O}}}^{i}_{\nu\nu'}d_{\nu'}$, where $i=1...N^{2}-1$ \cite{Zitko2013,Mantelli2016}. The Z-component of the Casimir operator can be constructed from $N-1$ diagonal Lie generators of SU($N=4$) symmetry as follows: $C_{Z}=\sum_{i=3,8,15\nu\nu'}d^{\dagger}_{\nu}{\hat{\mathcal{O}}}^{i}_{\nu\nu'}d_{\nu'}$ \cite{Lee1965}. Finally, we can express the total quadratic Casimir operator and its Z-component in the following way: $C^{2}=(15/8)(\sum_{\nu}Q_{\nu}-(2/3)\sum_{\nu}Q_{\nu}\cdot Q_{\overline{\nu}})$, and $C^{2}_{Z}=(3/8)(\sum_{\nu}Q_{\nu}-(2/3)\sum_{\nu}Q_{\nu}\cdot Q_{\overline{\nu}})=(1/4)(\sum_{\nu\nu}\chi_{\nu\nu}-\sum_{\nu}\chi_{\nu\overline{\nu}})$.
The local Z-component, depending on the coupling strength with 1TSC, 2TSC or 3TSC. For further discussion, we can express $C^{2}_{Z}$ in the boson fields operators and separate in the following sum $C^{2}_{Z}=C^{2}_{Z(K)}+C^{2}_{Z(M)}$, where  $C^{2}_{Z(K)}$ describes the fluctuations in the normal (Kondo-like) channel and $C^{2}_{Z(M)}$ is related to the Majorana fermion part. For the CNTQD-1TSC device, the relations of the quadratic Casimir operators can be expressed as:
\begin{eqnarray}
&& \nonumber C^{2}_{Z(K)}=(1/3)(p^{2}_{+\downarrow}+\sum_{s}p^{2}_{-s}+\sum_{\nu}d^{2}_{\nu}\\&&+\overline{t}^{2}_{+\uparrow}+\sum_{s}\overline{t}^{2}_{-s})\\
&& \nonumber C^{2}_{Z(M)}=(1/24)(9p^{2}_{+\uparrow}+p^{2}_{+\downarrow}+\sum_{s}p^{2}_{-s}+4\sum_{\nu}d^{2}_{\nu}+\\&&
+\overline{t}^{2}_{+\uparrow}+9\overline{t}^{2}_{+\downarrow}+\sum_{s}\overline{t}^{2}_{-s}).
\end{eqnarray}
$C^{2}_{Z}$ for two Majoranas that are coupled to CNTQD can be obtained in the following way:
\begin{eqnarray}
&& \nonumber C^{2}_{Z(K)}=(1/4)(\sum_{l}p^{2}_{l\downarrow}+\sum_{l}d^{2}_{l}+d^{2}_{\uparrow\downarrow}+d^{2}_{\downarrow\uparrow}
\\&&+\overline{t}^{2}_{+\uparrow}+\overline{t}^{2}_{-\downarrow})\\
&& C^{2}_{Z(M)}=(1/8)(3\sum_{l}p^{2}_{l\uparrow}+\sum_{l}p^{2}_{l\downarrow}+2\sum_{l}d^{2}_{l}
\\&&\nonumber+2d^{2}_{\uparrow\downarrow}+2d^{2}_{\downarrow\uparrow}+4\sum_{s}d^{2}_{ss}+\overline{t}^{2}_{+\uparrow}+3\overline{t}^{2}_{+\downarrow}
+3\overline{t}^{2}_{-\uparrow}+\overline{t}^{2}_{-\downarrow}).
\end{eqnarray}
The quadratic Casimir operator for the CNTQD-3TSC system can be written using the following formulas:
\begin{eqnarray}
&& \nonumber C^{2}_{Z(K)}=(1/24)(\sum_{s}p^{2}_{+s}+p^{2}_{-\uparrow}+9p^{2}_{-\downarrow}+4\sum_{\nu}d^{2}_{\nu}
\\&&+\sum_{s}\overline{t}^{2}_{+s}+\overline{t}^{2}_{-\uparrow}+9\overline{t}^{2}_{-\downarrow})\\
&& \nonumber C^{2}_{Z(M)}=(1/3)(\sum_{s}p^{2}_{+s}+p^{2}_{-\uparrow}+\sum_{\nu}d^{2}_{\nu}
\\&&+\sum_{s}\overline{t}^{2}_{+s}+\overline{t}^{2}_{-\uparrow}).
\end{eqnarray}
In the calculations we used the invariant of the two-body susceptibility $\chi_{(z)}$,
which is equal to $C^{2}_{Z}$ in the isolated (local) case (or when $T\gg T_{K}$).
$\chi_{(z)}$ is the quantum metric in the two-body correlation space and can be expressed as follows
$\chi_{(z)}=(1/4)(\sum_{\nu}\widetilde{\chi}_{\nu\nu}-\sum_{\nu'\neq\nu}\widetilde{\chi}_{\nu\nu'})$.
The charge susceptibility is given by: $\chi_{(c)}=\sum_{\nu\nu'}\widetilde{\chi}_{\nu\nu'}$.
The three-body susceptibility can be written as $\chi_{(z)}^{[3]}=(1/4)(\sum_{\nu}\widetilde{\chi}^{[3]}_{\nu\nu\nu}-\sum_{\nu'\neq\nu}\widetilde{\chi}^{[3]}_{\nu\nu'\nu'})$
- in fact, the cubic Casimir operator $C^{3}$ would be more appropriate. However, after expanding this in terms of the conformal Lie group generators, it turns out that it is simply proportional to the quadratic Casimir (by the explicit relation $C^{3}=(d/2)C^{2}$). Formally, the conformal group in $d$ dimensions (in our case $d=4$) consists of a single dilation operator $\hat{D}$, $d$ translations $\hat{P}$, $d$ special
conformal transformations $\hat{K}$ and $d(d-1)/2$ rotations $\hat{J}$ \cite{Iachello2007}.

Using the quadratic Casimir operator we have computed $T_{[\star]}\chi_{(z)}$, $T_{[\star]}\chi_{(c)}$ and $T_{[\star]}^{2}\chi_{(z)}^{[3]}$ which are quantized at zero temperature. $T_{[\star]}=\min\{T_{K,\nu}\}$ is the scaling characteristic energy. In the Fermi liquid phase it corresponds to the Kondo temperature. $T_{[\star]}\chi_{(z,c)}(T=0K)$ and $-T_{[\star]}^{2}\chi_{(z)}^{[3]}(0)$ are the frozen effective spin/charge and the three-body correlator in the system \cite{Hata2021}.
Last year the physicists crossed the Rubicon, and measured the spin susceptibility in the SU(2) Kondo quantum dot, in fact it was the fundamental measurement of the spin of the Kondo impurity, using the charge-sensing method \cite{Piquard2023}. One of the most important results in this matter is the measurement of the three-body correlations, indirectly using the lock-in technique to detect linear and nonlinear shot noise \cite{Hata2021}.

\subsection{Transport measurements}
At the Fermi level, the system in Fig. 1 satisfies the Friedel sum rule, and the linear conductance for the normal (Kondo) channel can be written as follows:
\begin{eqnarray}
&&\nonumber{\cal{G}}_{\nu}(V)=\frac{e^{2}}{h}\sum_{\alpha}\textrm{Re}\left\{\frac{\widetilde{\Gamma}_{\nu}\Psi_{1}\left[\frac{1}{2}+\frac{\widetilde{E}_{\nu}+i\widetilde{\Gamma}_{\nu}\pm V_{\alpha}}{2\pi iT}\right]}{4\pi T}\right\}
\\&&\stackrel{V=T=0}=\frac{e^{2}}{h}\frac{\widetilde{\Gamma}^{2}_{\nu}}{T^{2}_{K,\nu}}=\frac{e^{2}}{h}\sin^{2}[\delta_{\nu}]
\end{eqnarray}
where the quantum conductance in the $\nu'$ channels that are coupled to the TSC is given by :
\begin{eqnarray}
&&\nonumber{\cal{G}}_{\nu'}(V)=\frac{e^{2}}{h}\sum_{\alpha m}\textrm{Re}\left\{\frac{\widetilde{\Gamma}_{\nu'}a_{m}\Psi_{1}\left[\frac{1}{2}+\frac{\Lambda^{*}_{m\nu'}\pm V_{\alpha}}{2\pi iT}\right]}{4\pi T}\right\}
\\&&\stackrel{V=T=0}=\frac{e^{2}}{h}\frac{\widetilde{\Gamma}_{\nu'}(\widetilde{\Gamma}_{\nu'}+\frac{\widetilde{t}^{2}_{\nu'}}{\delta})}
{\widetilde{E}_{\nu'}^{2}+\widetilde{\Gamma}_{\nu'}^{2}+\frac{2\widetilde{t}^{2}_{\nu}\widetilde{\Gamma}_{\nu'}}{\delta}}\\
&&\nonumber=\frac{e^{2}}{h}\frac{\delta\widetilde{\Gamma}_{\nu'}+\widetilde{t}^{2}_{\nu'}}
{\delta\widetilde{\Gamma}_{\nu'}\csc^{2}[\delta_{\nu'}]+2\widetilde{t}^{2}_{\nu'}}.
\end{eqnarray}
The total conductance can be expressed as ${\cal{G}}=\sum_{\nu}{\cal{G}}_{\nu}+\sum_{\nu'}{\cal{G}}_{\nu'}$ where $\nu'=\pm1\uparrow,+1\downarrow$ and depends on the number of the spin-orbital channels coupled to the TSC.  The quantum conductance ${\mathcal{G}}$ develops from the current formula in the following way ${\mathcal{G}}(0)=dI/dV|_{V\mapsto0}$. The thermal fluctuations (thermal noise) can be related to the linear conductance via the fluctuation-dissipation theorem and the nonlinear temperature part, as shown in \cite{Oguri2022}: $S_{T}=4k_{B}T{\sum_{\nu}(\mathcal{G}_{\nu}}(0)-(e^{2}/h)\sum_{\nu}c_{T,\nu}(\pi k_{B}T)^{2}+..)$, where $c_{T,\nu}$ is the temperature coefficient explicitly expressed by the higher-order correlations. In our analysis we have focused on the zero temperature shot noise and, therefore the equilibrium fluctuations are negligible in the further discussion of the shot noise $S$. Using the Keldysh formalism for nonequilibrium Green's functions, and zero-frequency limit for the shot noise, we can write the zero temperature current $I$ and the shot noise $S$ as a function of the unitary transmission $T_{\nu}$ for the $\nu$ channel, in the Landauer-B\"{u}tikker form:
\begin{eqnarray}
&&I=\langle\hat{I}\rangle=(e/h)\int^{|V|/2}_{-|V|/2}\sum_{\nu}T_{\nu}(E)dE
\\&&\nonumber S(t,t')=(1/4)\sum_{\alpha}S^{>}_{\alpha\alpha}(t,t')+S^{<}_{\alpha\alpha}(t,t')
\\&&-S^{>}_{\alpha\overline{\alpha}}(t,t')-S^{<}_{\alpha\overline{\alpha}}(t,t')\nonumber=S(\tau,0)
\\&&\nonumber=(1/2)\langle\{\hat{I}(\tau),\hat{I}(0)\}\rangle-I^{2}
\\&& S=\lim_{w\mapsto0}2\int^{+\infty}_{-\infty}e^{iw\tau}S(\tau,0)d\tau
\\&&\nonumber=2(e^{2}/h)\int^{|V|/2}_{-|V|/2}\sum_{\nu}T_{\nu}(E)[1-T_{\nu}(E)]dE
\end{eqnarray}
where $\hat{I}=\frac{ie}{h}[\widetilde{{\cal{H}}},\sum_{k\alpha=L(R)\nu}\pm(1/2)c^{\dagger}_{k\alpha\nu}c_{k\alpha\nu}]$ is the current operator. Applying the Wick theorem, to compute contour-ordered auto and cross-correlation functions, the shot noise can expressed in following way $S^{>(<)}_{\alpha\alpha'}(t,t')=2\left(\frac{ie}{\hbar}\right)^{2}\sum_{kq\nu\nu'}\widetilde{t}_{0\nu}\widetilde{t}_{0\nu'}[G^{>(<)}_{\nu,q\alpha'\nu'}(t,t')G^{<(>)}_{\nu',k\alpha\nu}(t',t)
+G^{>(<)}_{k\alpha\nu,\nu'}(t,t')G^{<(>)}_{q\alpha'\nu',\nu}(t',t)-G^{>(<)}_{\nu,\nu'}(t,t')G^{<(>)}_{q\alpha'\nu',k\alpha\nu}(t',t)
-G^{>(<)}_{k\alpha\nu,q\alpha'\nu'}(t,t')G^{<(>)}_{\nu',\nu}(t',t)]$. Two-correlation functions in $S$ are decoupled in Hartree-Fock approximation (HFA) for two-particle Green's functions \cite{Kadanoff1962}. $T_{\nu}$ in Eqs. (37-38) is the transmission, expressed by the formulas for decoupled and coupled channel to TSC in the following form:  $T_{\nu}(E)=\widetilde{\Gamma}^{2}_{\nu}/[(E-\widetilde{E}_{\nu})^{2}+\widetilde{\Gamma}^{2}_{\nu}]$ and  $T_{\nu'}(E)=-\widetilde{\Gamma}_{\nu'}\textrm{Im}[G^{R}_{\nu'\nu'}]$, where $G^{R}_{\nu'\nu'}=\hat{G}^{R}_{11}$ is the retarded Green's function in the matrix Eq. (6).
In general, we have developed the current and the shot noise in the series:
$I=I_{0}V-\sum_{\nu}c_{V,\nu}V^{3}+0[V^{5}]$ and $S=S_{0}V+\sum_{\nu}c_{S,\nu}V^{3}+0[V^{5}]$
where $F_{0}=\lim_{V\mapsto0}S_{0}/2eI_{0}=\frac{\sum_{\nu}T_{\nu}(0)(1-T_{\nu}(0))}{\sum_{\nu}T_{\nu}(0)}$ is the linear Fano factor expressed by the linear shot noise $S_{0}$ and the current $I_{0}$. For identical transmissions in both spin-orbital channel ($\nu=\nu'$), the linear Fano factor can be written as follows: $F_{0}=1-T_{\nu}(0)$. The nonlinear contribution is described by $F_{K}=|S-S_{0}|/(2e|I-I_{0}|)=|S_{K}|/(2e|I_{K}|)=\frac{\delta S_{K}}{\delta I_{K}}=\frac{e^{\star}}{e}$.
$S_{K}$ and $I_{K}$ measurements contain the information about the effective
charge e$^{\star}$ of the current-carrying particles.
The charge differs from the electron charge e.
The nonlinear shot noise $S_{K}$ and the nonlinear current $I_{K}$ are defined as the absolute values and scaled by the characteristic temperature expression: $\delta S_{K}=T^{2}_{[\star]}|S_{K}|$ and $\delta I_{K}=2eT^{2}_{[\star]}|I_{K}|$. These definitions simplify the following discussion, emphasize the Fermi liquid behavior, and are formulated by the expressions of the two- and three-body correlation functions. The coefficients can be written as separate parts of the sum: $c_{V,\nu}=c_{V,\nu}|_{W_{\nu\nu'}-1=0}+\delta c_{V,\nu}|_{W_{\nu\nu'}-1>0}$ and
$c_{S,\nu}=c_{S,\nu}|_{W_{\nu\nu'}-1=0}+\delta c_{S,\nu}|_{W_{\nu\nu'}-1>0}$, where $\delta c_{V(S),\nu}|_{W_{\nu\nu'}-1>0}$ is the correction
developed from the residual interaction $\widetilde{U}^{\nu\nu'}$ between the Kondo  quasiparticles. $\delta c_{S,\nu}|_{W_{\nu\nu'}-1=0}$ is related to the elastic scattering processes and $\delta c_{S,\nu}|_{W_{\nu\nu'}-1>0}$ includes the elastic and inelastic scattering contribution \cite{Sela2006,Mora2009,Mora2008}. If we write the equations in the series: $T_{\nu}\approx i_{1}+i_{2}E+i_{2}E^{2}+0[E^{3}]$, $T_{\nu}(1-T_{\nu})\approx s_{1}+s_{2}E+s_{2}E^{2}+0[E^{3}]$, we can find that $I_{0}=i_{1}$, $c_{V,\nu}|_{W_{\nu\nu'}-1=0}=i_{2}/12$ and $S_{0}=s_{1}$, $c_{S,\nu}|_{W_{\nu\nu'}-1=0}=s_{2}/12$.
Based on the main results of \cite{Oguri2022}, where the authors found the nonlinear transport coefficients to the shot noise and the current for the SU(N) Anderson model using vertex corrections, we adopted the general expressions to calculate the nonlinear Fano factor $F_{K}=|c_{S}|/|c_{V}|=\sum_{\nu}|c_{S,\nu}|/\sum_{\nu}|c_{V,\nu}|$. The transport coefficients are determined by the static linear and nonlinear susceptibilities at low energies. The authors showed that the Ward identities, between the casual self-energies and the Feynman diagrams for the Keldysh vertex function of the zero temperature formalism, can be expressed in terms of the collision integrals. The formulas derived in \cite{Oguri2022}, as suggested by the authors are applicable to a wide class of quantum dots without particle-hole or time-reversal symmetry.
According to \cite{Oguri2022}, the coefficients $c_{V,\nu}$ and $c_{S,\nu}$ can be expressed in the following way:
\begin{eqnarray}
&&\nonumber c_{S,\nu}=\frac{\pi^{2}}{12}(\cos[4\delta_{\nu}]\chi^{2}_{\nu\nu}+(2+3\cos[4\delta_{\nu}])\sum_{\nu'\neq\nu}\chi^{2}_{\nu\nu'}
+\\ && 4\sum_{\nu'\neq\nu}\cos[2\delta_{\nu}]\cos[2\delta_{\nu'}]\chi^{2}_{\nu\nu'}
\\&&\nonumber+3\sum_{\nu'\neq\nu}\sum_{\nu''\neq\nu,\nu'}\sin[2\delta_{\nu}]\sin[2\delta_{\nu'}]\chi_{\nu\nu''}\chi_{\nu'\nu''}
\\&&\nonumber-(\chi_{\nu\nu\nu}+3\sum_{\nu'\neq\nu}\chi_{\nu\nu'\nu'})\frac{\sin[4\delta_{\nu}]}{4\pi})
\\&& c_{V,\nu}=\frac{\pi^{2}}{12}(-\cos[2\delta_{\nu}](\chi^{2}_{\nu\nu}+5\sum_{\nu'\neq\nu}\chi^{2}_{\nu\nu'})+
\\&&\nonumber(\chi_{\nu\nu\nu}+3\sum_{\nu'\neq\nu}\chi_{\nu\nu'\nu'})\frac{\sin[2\delta_{\nu}]}{2\pi}).
\end{eqnarray}

The factors are calculated for symmetric coupling to the normal electrodes  $\widetilde{\Gamma}_{L}=\widetilde{\Gamma}_{R}$ and are derived from the Keldysh vertex corrections to the current and the shot noise using the principles of Fermi liquid theory. The main contribution to the current and shot noise coefficients is determined by the charge (in the phase shift $\delta_{\nu}$), 2-body (susceptibilities) and 3-body correlation functions. In connection with the previous results, the authors introduced the higher-order fluctuations into the shot-noise formula and expressed the transport coefficients in the elegant form of the general static susceptibilities \cite{Oguri2022}. In this article we propose to calculate the dressed susceptibilities ($\widetilde{\chi}_{\nu_{1}\nu_{2}}$) and the 3-body correlations ($\widetilde{\chi}^{[3]}_{\nu_{1}\nu_{2}\nu_{3}}$) using the extended K-R slave boson mean-field approach \cite{Coleman1987,Choi2004,Krychowski2018} and the weak coupling ansatz to calculate the Wilson ratio ($\widetilde{t}_{0}\ll t_{0}$).

In this paper, we also theoretically investigate the thermoelectric power using the Onsager equations \cite{Costi1994}.
In the  linear response theory, the electric and thermal currents can be expressed as: $I=e^{2}\sum_{\nu}L^{(0)}_{\nu}\delta V-(e/T)\sum_{\nu}L^{(1)}_{\nu}\delta T$ and $I_{Q}=-e\sum_{\nu}L^{(1)}_{\nu}\delta V+(1/T)\sum_{\nu}L^{(2)}_{\nu}\delta T$, where $\delta V=V_{L}-V_{R}$ and $\delta T=T_{d}-T_{s}$ are the difference of the bias voltage and the temperature gradient. Finally, we can write the conductance $\mathcal{G}$ and the thermoelectric power ${\mathcal{S}}=(\delta V/\delta T)|_{I=0}$ using the integral $L^{(n)}_{\nu}=T\sum_{\alpha}\int^{+\infty}_{-\infty}(E-V_{\alpha})^{n}T_{\nu}(E)\left(-\frac{\partial f_{\alpha}}{\partial E}\right)dE$ - in the following forms:
\begin{eqnarray}
&&\nonumber \mathcal{G}=dI/dV=\sum_{\nu}L^{(0)}_{\nu}/T\\
&&\mathcal{S}=-\frac{k_{B}}{|e|T}\frac{\sum_{\nu}L^{(1)}_{\nu}}{\sum_{\nu}L^{(0)}_{\nu}}=\\
&&\nonumber-\frac{k_{B}}{|e|T}\frac{\sum_{\alpha m\nu}Im\left\{\frac{\widetilde{\Gamma}_{\nu}a_{m}(\Lambda_{m\nu}+V_{\alpha})}{4\pi i}\Psi_{1}\left[\frac{1}{2}+\frac{\Lambda_{\nu}-V_{\alpha}}{2\pi iT}\right]\right\}}{\sum_{\alpha m\nu}Re\left\{\frac{\widetilde{\Gamma}_{\nu}a_{m}}{4\pi}\Psi_{1}\left[\frac{1}{2}+\frac{\Lambda_{m\nu}-V_{\alpha}}{2\pi iT}\right]\right\}}
\end{eqnarray}
where for the normal channels: $a_{m}=1$, $\Lambda_{m\nu}=\Lambda_{\nu}$ and the sums run only over the $\alpha\nu$. For $t=0$ the system is in the symmetric SU(4) Kondo state, and the thermoelectric power is given by:
\begin{eqnarray}
&&\mathcal{S}=-\frac{\pi^{2}}{3|e|}
\frac{\sum_{\nu}\widetilde{\chi}_{\nu\nu}(\sin[2\delta_{\nu}]/\widetilde{\Gamma}_{\nu})}
{\sum_{\nu}(\sin^{2}[\delta_{\nu}]/(\pi\widetilde{\Gamma}_{\nu}))}T+0[T^{3}]
\end{eqnarray}
If we measure ${\mathcal{S}}$ below the Kondo temperature $T_{K}$, the Seebeck effect of the quasiparticles is determined by the linear part of the thermoelectric power and FL corrections give the same results as the sbMFA ($\gamma_{S}\approx-\cos[\delta_{\nu}]$). Finally we can introduce the linear thermoelectric power coefficient in the form $\gamma_{(S)}=({\mathcal{S}}T_{K})/(2\pi T)$. For the SU(4) Kondo state, the coefficient leads to $\gamma_{(S)}=-(k_{B}/|e|)(\pi/3)(\widetilde{E}_{\nu}/T_{K})=-(k_{B}/|e|)(\pi/3)\cos[\delta_{\nu}]$. The quantity $\gamma_{(S)}$ is given by the phase shift $\delta_{\nu}$ and changes its sign at the electron-hole symmetry point \cite{Lopez2014}.
The coefficient contains the information about the position and the width of the quasiparticle resonance and the SU(N) symmetry of the Kondo state. For the SU(4) Kondo effect,  the linear TEP coefficient is related to the numbers $\mp\pi/(3\sqrt{2})$ in the $1e(3e)$ charge sector, and for the fully symmetric SU(3) Kondo state, $\gamma_{(S)}$ reaches $\mp\pi/6$ \cite{Krychowski2020}. Generally,
TEP developed to the lowest order in terms of temperature is given by the Mott's formula $\mathcal{S}=-\frac{\pi^2}{3|e|}\frac{\sum_{\nu}d\widetilde{\varrho}_{\nu}/dE|_{E=0}}{\sum_{\nu}\widetilde{\varrho}_{\nu}}T$. In particular, for the fractional SU$^{\star}$(3) Kondo state in the whole range of the coupling strength $t$, we obtain $\gamma_{(S)}$ as follows:
\begin{eqnarray}
&&\gamma_{(S)}=\\&&\nonumber\frac{-\pi T_{K}\left\{3\dot{\widetilde{\varrho}}_{\nu}(0)
+\frac{\cot[\delta_{\nu'}][\widetilde{\Gamma}_{\nu'}\delta^{2}+\widetilde{t}^{2}_{\nu'}(\delta-\widetilde{\Gamma}_{\nu'})]}
{\pi\widetilde{\Gamma}_{\nu'}(2\widetilde{t}^{2}_{\nu'}+\widetilde{\Gamma}_{\nu'}\delta\csc^{2}[\delta_{\nu'}])^{2}}\right\}}
{3\widetilde{\varrho}_{\nu}(0)+\widetilde{\varrho}_{\nu'}(0)}
\end{eqnarray}
where $\dot{\widetilde{\varrho}}_{\nu}(0)=-\widetilde{\chi}^{[3]}_{\nu\nu\nu}/2
=\frac{\cos[\delta_{\nu}]\sin^{3}[\delta_{\nu}]}{\pi\widetilde{\Gamma}^{2}_{\nu}}$ (see Eq.(22)). In the limit of the strong coupling strength to the Majorana fermion ($\widetilde{t}_{\nu'}\mapsto\infty$), the linear TEP factor reaches $\gamma_{(S)}=\mp\frac{\pi\sin[\delta_{\nu}]\sin[2\delta_{\nu}]}{-4+3\cos[2\delta_{\nu}]}=\mp3\pi/22\approx\mp0.42..<\mp\pi/6$ \cite{Krychowski2020}. $\gamma_{(S)}$ in the strong coupling limit leads to the number $\mp3\pi/22$ and, in contrast to the SU(3) Kondo state, the value is modified by the coupling term to TSC. The result is different from the broken Kondo state e.g. by the magnetic field, because the topological channel with an increase of $\widetilde{t}_{\nu}$ is active in ${\mathcal{S}}$. The results suggest that the symmetry of the Kondo effect is violated and is associated with the fractional SU$^{\star}$(3) Kondo state.

In our device we use the finite bias $V_{s(d)}$ and the temperature gradient $\delta T$ to study the quantum conductance, shot noise and thermoelectric power. In all the results we can distinguish three sectors of the coupling strengths: $t_{1}=\sqrt{\delta\Gamma/6}$, which shares the normal and crossover region in the transport, $t_{2}=\Gamma U$ is the upper limit of the transport in the intermediate coupling strength, and $t_{3}=U/2$ is the starting point of the strong coupling regime, where the new Kondo phase is realized.

\subsection{SU$^{\star}$(3) and SU$^{\star}$(2) Kondo states and the U$^{\star}$(1) charge symmetry phase in CNTQD-TSC devices}
Figure 5a shows the density plot of the quantum conductance for the weak, intermediate and strong coupling regions in the CNTQD-TSC system. For $t\approx0$ the unitary conductance can be expressed directly from the phase shift $\delta_{\nu}$ of the Kondo quasiparticle resonance ${\mathcal{G}}=(e^{2}/h)\sum_{\nu}\sin^{2}[\delta_{\nu}]$.

In this limit, the full SU(4) Kondo effect is realized in the device. For $Q=1$e and $3$e, the total conductance reaches ${\mathcal{G}}=2(e^2/h)$ (black lines in Figure 5c-d and blue area in Figure 5a). SU(4) Kondo effect emerges from the fourfold degeneracy of the states $\ket{p_{ls}}_{4}$ and $\ket{t_{ls}}_{4}$. In the case of $Q=2$e (two electrons on the quantum dot), the conductance is quantized to $4e^{2}/h$ for $t=0$ (dark black area in Figure 5a). The half-filling region is determined by the six states $\ket{d_{\nu}}_{6}$. The SU(4) Kondo effect in the CNTQD was first observed by \cite{Herrero2005} and confirmed by the measurements of the other groups \cite{Makarovski2007,Cleuziou2013}. In comparison with the results of the sbMFA method \cite{Krychowski2018}, the NRG calculation also showed the two-stage quantized conductance for the SU(4) Kondo state in CNTQD \cite{Galpin2010,Mantelli2016}.

As the coupling strength increases, the multiplet states are formed in the system. Below $t_{1}$, as we can see in Figure 5b, in the intermediate coupling range there is a decrease in the conductance for the $Q=2e$ region, the conductance reaches $(7/2)(e^{2}/h)$ (red line on the landscape plot). The Kondo effect is determined by the duodecuplet states $\ket{q_{w}}_{12}$. The conductance in the channel coupled to the Majorana fermion is quantized to ${\mathcal{G}}_{+\uparrow}=(1/2)(e^{2}/h)$ (inset in Figure 5d), the remaining conductance value comes from the Kondo channels and reaches $3(e^{2}/h)$. With increasing the coupling to the TSC, the conductance at the e-h symmetry point ($E_{d}=E_{e-h}=-3U/2=-4.5$) does not change, this is due to the number of twelve states $\ket{q_{w}}_{12}$ and $\ket{q_{y}}_{6}$, $\ket{q_{g}}_{6}$ involved in the quantum transport (red curve in Figure 5b and dark purple area in Figure 5a).

The quantum conductance for the low and high energy octuplets $\ket{q_{a}}_{8}$ and $\ket{q_{b}}_{8}$ in the weak coupling regime reaches $2(e^{2}/h)$, of which in the Majorana-coupled channel ${\mathcal{G}}_{+\uparrow}=(1/2)(e^{2}/h)$, and the other three normal channels contribute to ${\mathcal{G}}={\mathcal{G}}_{+\downarrow}+\sum_{s}{\mathcal{G}}_{-s}=(3/2)(e^{2}/h)$. In the strong coupling regime above $t_{2}$, especially near $t_{3}$, the conductance for $E_{d}=-1.5$ (respectively $E_{d}=-7.5$) reaches $(3/2)(e^{2}/h)$ (green line in Figure 5b).  As can be seen from the partial contributions of the total conductance (Figure 5d), the conductance in the Majorana channel remains unchanged ${\mathcal{G}}_{+\uparrow}=(1/2)(e^{2}/h)$ but in the other channels it reaches ${\mathcal{G}}_{+\downarrow(-s)}=(1/3)(e^{2}/h)$, due to the charge degeneracy between the $d$ and $p$ states. For $E_{d}=1$, above the $t_{1}$ (cyan line in Figure 5b), the conductance reaches a constant value of ${\mathcal{G}}_{+\uparrow}=1/2(e^{2}/h)$ and originates only from the Majorana fermion-coupled channel (a value previously confirmed by calculations within NRG \cite{Lopez2013,Wojcik2020,Majek2021,Weymann2021}, EOM and sbMFA method \cite{Cuniberti2018}). The half-quantum conductance originates from the ground state of the doublet $\ket{q_{x}}_{2}$ and $\ket{q_{z}}_{2}$.
The transition to the strong coupling strength regime depends on the ratio of $U/\Gamma$. $t_{1}=\sqrt{\delta\Gamma/6}$ separates the normal and the entangled qubit states by TSC. The quantum measurements above $t_{3}=U/2$ show only well-defined quantum states in the strong coupling regime.  $t_{1}$ and $t_{2}$, shown as vertical dashed gray lines in Figure 5b, can be shifted by reducing the coupling to the normal electrodes $\widetilde{\Gamma}$.

The most significant result predicted by the theory  is the transition with increasing the coupling strength $t$ from the charge degeneracy line between $Q=1$e, 2e and  $Q=2$e, 3e to the SU$^{\star}$(3) Kondo state in the fractional charge region. The symmetry type of the Kondo effect is upper indexed by $\star$ due to the fact that the Kondo state appears in three channels with quantized conductance ${\mathcal{G}}_{+\downarrow(-s)}=(3/4)(e^{2}/h)$ for the half-occupancy region $Q=3/2$e and $Q=5/2$e (green lines in Figs. 5c, d). This is a surprising result in contrast to the fully SU(3) Kondo effect \cite{Zitko2013}, where it occurs for integer charges $Q=1$e and $2$e. This is mainly due to the degeneracy of the six quantum states, the low and high energy sextuplets $\ket{q_{g}}_{6}$ and $\ket{q_{y}}_{6}$ (Eq.(10)). The Kondo state does not follow from the degeneracy of the three pure quantum states $\ket{p_{ls}}_{3}$ or $\ket{d_{ls}}_{3}$ \cite{Krychowski2020}, but from the degeneracy of the six entangled states $\ket{q_{y}}_{6}$ for $Q=(5/2)$e and six high energy quantum states $\ket{q_{g}}_{6}$ for $Q=(3/2)$e. In the quantum conductance map, we observe the light violet sector of the SU$^{\star}$(3) Kondo effect, where the total conductance  reaches ${\mathcal{G}}=({\mathcal{G}}_{+\downarrow}+\sum_{s}{\mathcal{G}}_{-s})+{\mathcal{G}}_{+\uparrow}=(9/4)(e^{2}/h)+(1/2)(e^{2}/h)=(11/4)(e^{2}/h)$ (blue line in Fig. 5b). The conductance in the channel coupled to the Majorana fermion is independent of the coupling strengths $t$ and $E_{d}$. The contribution of ${\mathcal{G}}_{\nu'}$ to the total conductance is $(1/2)(e^{2}/h)$.

Figure 6 shows the quantum conductance for the CNTQD-2TSC system. The quantum dot is coupled to two Majorana fermions $\gamma_{+\uparrow}$ and $\gamma_{-\uparrow}$ (Fig. 1). Two half-fermions can also be prepared, e.g. in DIII class superconductors, where the quantum state with two Majoranas can be realized as a Majorana Kramers pairs at the edge of a single topological superconducting wire \cite{Zhang2013,Gaidamauskas2014,Kim2016}. The conductance of the CNTQD-2TSC device reaches ${\mathcal{G}}=\sum_{l}{\mathcal{G}}_{l\uparrow}+\sum_{l}{\mathcal{G}}_{l\downarrow}=(e^{2}/h)+2(e^{2}/h)=3(e^{2}/h)$ for $E_{d}=-4.5$ and $t>t_{1}$. For the subsequent growth of the coupling strength, ${\mathcal{G}}$ is unchanged  (red curve in Fig. 6b, light violet region in Fig. 6a). Around $t_{1}$ we observe a transition in the state configuration from $\ket{d_{\nu}}_{6}$ to $\ket{q_{w}}_{24}$. In the weak coupling regime, the quantum state of the system is determined by twenty-four states $\ket{q_{w}}$ in $2^{4+2}=64$ dimensional Hilbert space. The SU$^{\star}$(2) Kondo state is mainly realized for $t>t_{2}$  and includes the eight entanglement quantum states. The ground state is the octuplet $\ket{q_{y}}_{8}$ (Eq. (12)). The channels coupled to the Majorana states contribute $(e^{2}/h)$ to the quantum conductance, the other two channels are related to the Kondo effect. It is difficult to determine the moment of the transition between strongly and weakly coupled systems with TSCs - based only on the quantum conductance. However, the quantum transition will be detectable in the nonlinear shot noise and the current measurements, in the temperature dependent effective pseudospin or in the entropy detection \cite{Piquard2023}, which we will discuss later.
In the $Q=1$e and $3$e charge region, the total conductance is quantized to ${\mathcal{G}}=2(e^{2}/h)$ for the intermediate coupling strength. The conductance in the channels associated with the Kondo states contributes ${\mathcal{G}}_{\pm\downarrow}=(1/2)(e^{2}/h)$ and is determined by the degeneracy of the sixteen quantum states $\ket{q_{b}}_{16}$ and $\ket{q_{a}}_{16}$. Beyond the Kondo solutions for weak and strong coupling with TSCs, the total conductance reaches $e^{2}/h$, for two quartets $\ket{q_{x}}_{4}$ and $\ket{q_{z}}_{4}$ as the ground states. For the strong coupling to the Majorana fermion, the conductance for $Q=1,3$e in the channels associated with the Kondo state is suppressed to $e^{2}/h$, because the next two quantum channels are operated by Majorana fermions (green lines in Figures 6c,d).

Figure 7 shows the quantum conductance as a function of $E_{d}$ and the effect of breaking the SU(4) Kondo state by increasing the coupling strength to three Majorana fermion in the CNTQD-3TSC device. For $Q=0$e and $Q=4$e with increasing the coupling strength we observe a transition from empty and fully occupied states to high and low energy octuplets:  $\ket{q_{z}}_{8}$ and $\ket{q_{x}}_{8}$. The conductance reaches ${\mathcal{G}}=\sum_{s}{\mathcal{G}}_{+s}+{\mathcal{G}}_{-\uparrow}=(3/2)(e^{2}/h)$, when the transport goes through the channels coupled to three Majorana fermions (red region in Figure 7a). The number of available states in the system is $2^{4+3}=128$, and all quantum states are  spanned by the basis vectors $|n_{+}n_{-}n_{1}n_{2}n_{3}>$.
The total charge on the quantum dot is $Q=5/2e$ and $Q=3/2e$ for two octuplets $\ket{q_{x}}_{8}$ and $\ket{q_{z}}_{8}$, defined in Eqs. (13-14). Each of these states is a linear combination of eight pure quantum states mutually mixed with a topological segment.
The transitions from $\ket{p_{ls}}_{4}$ via $\ket{q_{a}}_{32}$ to $\ket{q_{z}}_{8}$ and from $\ket{t_{ls}}_{4}$ via $\ket{q_{b}}_{32}$ to $\ket{q_{x}}_{8}$ are observed  as an increase mechanism from integer charges $Q=1e$ and $Q=3e$ to the fractional charges $Q=(3/2)e$ and $Q=(5/2)e$.

The conductance for $E_{d}=-1.5$ (green line in Fig. 7b) changes the quantized value from $2(e^{2}/h)$ to $(3/2)(e^{2}/h)$ (contributed by the channels in the conjunction with the MFs). For $E_{d}=-4.5$ we observe a transition from $\ket{d_{\nu}}_{6}$ to the entangled quantum states with the highest degeneracy in the hybrid system $\ket{q_{w}}_{48}$. In this case the conductance decreases from $4(e^{2}/h)$ to $(5/2)(e^{2}/h)$ (red curve in Fig. 7b). In the normal channel ${\mathcal{G}}_{-\downarrow}=1(e^{2}/h)$, the remaining contribution from the channels coupled to the TSC is quantized to $(3/2)(e^{2}/h)$ and dominates in the quantum transport measurements. In the strong coupling regime for $t>t_{3}$ at the e-h symmetry point, the conductance is fixed and quantized to $5/2(e^{2}/h)$. Two quantum octuplets $\ket{q_{x}}_{8}$ and $\ket{q_{z}}_{8}$ degenerate on this line. The system at this point is determined by the charge degrees of freedom and we observe a U$^{\star}$(1) charge symmetry solution. U(1) symmetry is observed in the normal state between the charge regions for the fractional charge on the quantum dot. U$^{\star}$(1) symmetry appears for $Q=2$e. The analogous result was found for $Q=1$e in the QD coupled to 1TSC \cite{Wojcik2020}, since in this system we have only two channels $s=\uparrow,\downarrow$, where $s=\uparrow$ is coupled to the single Majorana fermion, consequently the total conductance at this point reaches a value of $(3/2)(e^{2}/h)$ \cite{Weymann2017}. It is worth noting that saturation occurs at the e-h symmetry point, so applying an external magnetic field or polarization would allow this point to be moved along the $E_{d}$ axis (see the effect of the exchange field on the Kondo state \cite{Hauptmann2008}). The analogy can be found in the charge Kondo effect with the polarons, where the Kondo temperature for the charge Kondo state decreases with increasing coupling strength to the phonon bath \cite{Krychowski2022}. For the CNTQD-3TSC device, the characteristic energy scale $T_{[\star]}$ is saturated (light red curve in Fig. 9a). Comparing the results from Figures 5-7b, we observe the effect of squeezing the conductance around the value $2\sin^{2}[\pi/2]=2(e^{2}/h)$. From this we can conclude that for the Kondo effect with even SU(N) symmetry, the squeezed conductance will be seen around the value $(N/2)(e^{2}/h)$ when all quantum states in the quantum dot are coupled to TSC segments.

Let us introduce the magnitude of the spin and orbital polarization in terms of $\Delta {\mathcal{G}}_{s}=\sum_{l}({\mathcal{G}}_{l\uparrow}-{\mathcal{G}}_{l\downarrow})/{\mathcal{G}}$ and $\Delta {\mathcal{G}}_{o}=\sum_{s}({\mathcal{G}}_{+s}-{\mathcal{G}}_{-s})/{\mathcal{G}}$. TSC strongly polarizes the conduction channels. In two cases, namely coupling with 1TSC and 3TSC, the orbital and spin polarization are equal $\Delta{\mathcal{G}}_{s}=\Delta{\mathcal{G}}_{o}$. In Figure 8a we see a negatively polarized conduction for $Q=2$e and and for twelvefold degenerate quantum states $\ket{q_{w}}_{12}$. The value of the spin (orbital) polarization for $\ket{q_{w}}_{12}$ reaches $-1/7$ and saturates for $t>t_{1}$ (red lines in Figure 8d and lower inset in Figure 8d). In the case where the sextuplets $\ket{q_{g(y)}}_{6}$ are the ground states of the fractional SU$^{\star}$(3) Kondo effect, the spin polarization of the conductance is negative and corresponds to the rational number $-1/11$.
For $Q=1(3)$e and $t<t_{2}$ the polarizations are equal to zero. For $t>t_{2}$, $\Delta {\mathcal{G}}_{s}$ is positive and reached $+1/9$ for the CNTQD-1TSC device and $+1/3$ for the CNTQD-3TSC (green lines in Fig. 8d). The highest spin(orbital) polarization in the hybrid systems is dominated by the conductance contribution of the Majorana channels and occurs for doublet $\ket{q_{x(z)}}_{2}$ and quartet $\ket{q_{x(z)}}_{4}$ states. The spin and orbital polarizations reach a value of $+1$ for these ground states. For the Majorana-Kondo state, the polarizations are always negative, in contrast to cases where the channels coupled to the Majorana fermions dominate, where they are positive. The reversal of the polarization sign is observed for $E_{d}=-3$, where the value of $\Delta {\mathcal{G}}_{s}$ changes from negative to positive enhancement for the CNTQD-3TSC device and is suppressed for the CNTQD-2TSC system (blue lines in the insets in Fig. 8d). For the CNTQD-2TSC device, due to the type of the coupling strength ($\widetilde{t}_{\pm\uparrow}$), the orbital polarization is equal to zero, and we only observed the spin polarization of the conductances in the system (Fig. 8b). The spin and orbital polarization of the conductance for the CNTQD-3TSC system reaches a positive rational number $+1/3$ when the transport is determined by the two octuplets $\ket{q_{x}}_{8}$ and $\ket{q_{x}}_{8}$ (light blue region in Fig. 8c). For $\ket{q_{w}}_{48}$ the spin(orbital) polarization of the conductance of CNTQD with side-attached three Majorana fermions is saturated at the negative quantized value $-1/5$ (red curve in the lower inset in Fig. 8d and dark blue region in Fig. 8c).

The transport in the channel coupled to the Majorana fermion is determined by the anomalous Green's function, which significantly modifies the Kondo temperature $T_{K}$ in the system (Eq. (6)). Looking at the quantum conductance in terms of linear transport, we can see that the channels associated with the Kondo state are described by Eq. (35), while in the case of a channel coupled to the TSC we have found the relation in Eq. (36). Both formulas are determined by the characteristic temperature $T^{2}_{K,\nu}=\widetilde{E}^{2}_{\nu}+\widetilde{\Gamma}^{2}_{\nu}$ and for the channel coupled to MFs, $T^{2}_{K,\nu'}=\widetilde{E}^{2}_{\nu}+\widetilde{\Gamma}^{2}_{\nu}+(2\widetilde{t}^{2}_{\nu'}\widetilde{\Gamma}_{\nu'})/\delta$.
Within the sb-MFA formalism, $\widetilde{E}^{2}_{\nu}$ and $\widetilde{\Gamma}^{2}_{\nu}$ determine the position and width of the quasiparticle Kondo resonance. In an elegant way, following Coleman \cite{Coleman1987}, we can relate the complex pole of the quasiparticle Green function with $T_{K}$ and the charge $Q_{\nu}$ in the following form $\ln[\widetilde{E}_{\nu}-i\widetilde{\Gamma}_{\nu}]=\ln[T_{K}]-i\pi Q_{\nu}$, where in the context of FL theory, $Q_{\nu}$ is given by the phase shift $\delta_{\nu}$ (Eq. (20)). If we write the expression as $e^{i\pi Q_{\nu}}=T_{K}/\Lambda^{*}_{\nu}$, we get the Euler's formula, which materializes in the physics. This is why we talk about the logarithmic scaling of the Kondo effect at low temperatures. In my opinion, the wide range of mathematical functions (in particular the hypergeometric functions \cite{Rao2005}) opens up to experimental physics a multiversum of a new type of correlated states, not yet discovered. In summary, if we find a functional relation (correlation) between the charge (spin etc.) and the pole of the Green's function of a new quantum state - then we have a simple recipe to open the door to a new world of correlated systems of spin(electron) and other particles. Perhaps AI, with its uncompromising approach to finding solutions, will be a great tool in this research, the future will tell us.

Figure 9a shows the characteristic temperatures, defined as $T_{[\star]}=\min\{T_{K,\nu},T_{K,\nu'}\}$. In the decoupled CNTQD with TSC ($t=0$), the system is determined by the SU(4) Kondo temperature $T^{SU(4)}_{K}$. For $E_{d}=-4.5$ the Kondo resonance is centered on the Fermi level due to the phase shift $\delta_{\nu}=\pi/2$, and the sixfold degenerate quantum states $\ket{d_{\nu}}_{6}$ determines the Kondo temperature (red lines in Fig. 9a). However, the number of six states for 2e is higher than the fourfold degeneracy for $Q=1(3)$e, $T^{SU(4)}_{K,Q=1e(3)}>T^{SU(4)}_{K,Q=2e}$, as is well documented in the literature \cite{Galpin2010,Mantelli2016,Krychowski2017}. From the extended K-R sbMFA we obtained the relation $T^{SU(4)}_{K,2e}/T^{SU(4)}_{K,1e(3)}\equiv e^{\frac{-\pi U}{4\Gamma\Lambda}}$, where $\Lambda=60$ and follows from the derivatives of the renormalization of the tunneling rates with respect to the boson fields operators for two charge regions $Q=1(2)$e. The characteristic temperature $T_{[\star]}$ in the weak coupling limit at the charge degeneracy line for $Q=(3/2)$e and $E_{d}=-3$ (similarly for $Q=(5/2)$e) is proportional to the hybridization parameter $\Gamma$, which determines the width of the charge resonances (all blue lines in Fig. 9a). For $E_{d}=-1.5$, increasing the coupling strength to the TSC  leads to an increase in the characteristic temperature (the temperature in the Majorana channel determines the saturation of $T_{[\star]}$, green curves in Fig. 9a). We have shown that rise of the number of $N_{TS}$ topological segments in the strong coupling regime leads to the enhancement and saturation of  $T_{[\star]}$ for $E_{d}=-1.5$ (dark, dotted and light green lines represent the results for CNTQD-1TSC, CNTQD-2TSC and CNTQD-3TSC). Similar effects are observed for $E_{d}=-3$ and $E_{d}=-4.5$, in the case where the transport determines the channel directly coupled to the Majorana fermions.

A fractional Kondo effect with SU$^{\star}$(3) symmetry is observed in the CNTQD-1TSC system for $Q=(3/2)e$ and $Q=(5/2)e$. The transport is mainly determined by the channels associated with the Kondo effect (${\mathcal{G}_{\nu}}=(9/4)(e^{2}/h)$), which is manifested by a decrease in the characteristic temperature $T_{[\star]}$ (dark blue line for $E_{d}=-3$ in Fig. 9a). For $E_{d}=-4.5$, in the CNTQD coupled with two Majorana fermions, $T_{[\star]}$ decreases with increasing the coupling strength to the topological wire, and we observe a strong decrease of $T^{SU^{\star}(2)}_{K}$. By increasing $t$ for the CNTQD-2TSC device, we start from the SU(4) Kondo effect and end up in the SU$^{\star}$(2) Kondo state with the reduced characteristic temperature scale $T^{SU^{\star}(2)}_{K}$. Between these two types of strongly correlated states we observed a strong enhancement of $T_{[\star]}$, which is characteristic of crossover \cite{Krychowski2018}. The  similar effects of the Kondo temperature boost, have been observed with an increase of the spin-orbital interaction (SOI) in the CNTQD system, as indicated by sbMFA \cite{Krychowski2018} and the NRG method \cite{Mantelli2016}. Based on the NRG framework it is difficult to explain the increase of $T_{K}$, in the sb-MFA method we can relate it to the trends in the bosonic fields, more precisely to the products of operators in the renormalization parameter of the quasiparticle resonance $z_{ls}$. An increase of the Coulomb interaction U in CNTQD increases the values of these products and finally we observe the enhancement of $T_{K}$ in the crossover region \cite{Krychowski2018}. We expect the same effects for the CNTQD-2TSC device with an increase of the coupling strength $t$.

Furthermore, the fact that the characteristic temperature increases with $t$ determines the behavior of the quantum conductance at finite temperature T. Fig.9b shows the density plot of the quantum conductance for CNTQD coupled to a single Majorana fermion at finite temperature $T=10^{-2}$. The SU(4) and SU$^{\star}$(3) Kondo states are destroyed by the temperature effects, leading to a suppression of the quantum conductance in the unitary Kondo regions (${\mathcal{G}}\mapsto0$ for $\ket{q_{a(b)}}_{8}$, $\ket{q_{w}}_{12}$ and ${\mathcal{G}}\mapsto{\mathcal{G}}_{\nu'}=(1/2)(e^{2}/h)$ for $\ket{q_{g(y)}}_{6}$). The transport takes place along the line of charge degeneracy, e.g. between the quantum states $\ket{q_{a(b)}}_{8}$ and $\ket{q_{w}}_{12}$ and between $\ket{q_{g}}_{8}$ and $\ket{q_{g}}_{8}$ we observe a value of finite quantum conductance at a level of $(5/2)(e^{2}/h)$. Between $\ket{q_{a(b)}}_{8}$ and $\ket{q_{x(z)}}_{2}$ the conductance reaches a value of $(3/2)(e^{2}/h)$.
Since $T_{[\star]}$ is higher than $T$ for the $\ket{q_{x(z)}}_{2}$ states, the Majorana fermion channel remains active in the quantum transport.
The doublet states $\ket{q_{x(z)}}_{2}$ in the weak coupling regime are collapsed, and the transport is determined by two singlets $\ket{e}_{1}$ and by the fully occupied quantum state $\ket{f}_{1}$. Figure 9c shows the evolution of the conductance as a function of $t$ for different temperatures T. It can be seen that the transition from $\ket{d_{\nu}}_{6}$ to $\ket{q_{w}}_{12}$  is shifted at low temperatures (red line in Fig. 9c). For $T=10^{-4}$ the quantum conductance decreases to $2(e^{2}/h)$ and gradually approaches zero. In the strong coupling regime the quantum conductance reaches ${\mathcal{G}}=(7/2)(e^{2}/h)$, and after passing $T=10^{-3}$, ${\mathcal{G}}$ is quantized to $(5/2)(e^{2}/h)$ (black line in Fig. 9c). The inset shows the evolution of the quantum conductance with increasing  temperature for $E_{d}=-3$. For $T>T^{SU^{\star}(3)}_{K}$, the fractional SU$^{\star}$(3) Kondo state is destroyed and the quantum conductance reduces to $(1/2)(e^{2}/h)$. The inset of Fig. 9c shows the gradual decrease of the quantum conductance for $Q=(3/2)$e and shift to the characteristic coupling strength $t_{3}$. In the weak coupling regime, $\mathcal{G}$ leads to $(5/2)(e^{2}/h)$ for the charge degeneracy line (black curve in the inset of Fig.9c).
\begin{figure}[t!]
\includegraphics[width=0.75\linewidth]{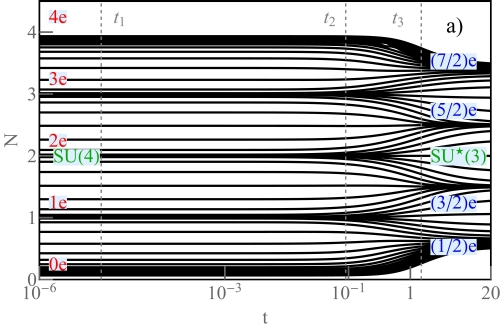}\\
\includegraphics[width=0.75\linewidth]{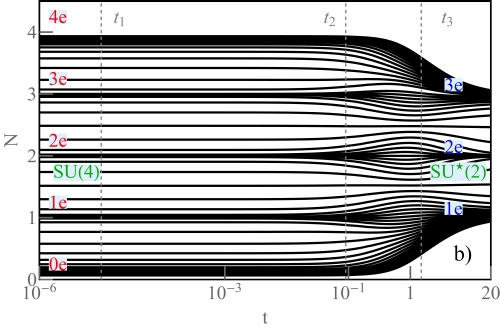}\\
\includegraphics[width=0.75\linewidth]{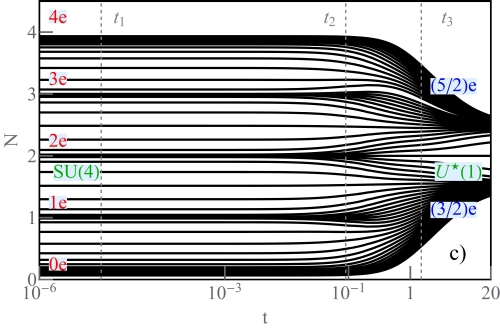}
\caption{\label{fig:epsart} (Color online) Total occupancy number $N$ for CNTQD coupled to 1MF (a), 2MFs (b) and 3MFs (c). The numbers indicate the integer and fractional electron charge in the system. $SU(4)$, $SU^{\star}(3)$, $SU^{\star}(2)$ and $U^{\star}(1)$ symbolize the spin-orbital, fractional, integer Kondo states and the charge degeneracy line.}
\end{figure}

In this section, we will now turn to discussing the charge fluctuations that occur in the CNTQD-TSC device. The total occupancy number of CNTQD is given by $N=\sum_{\nu}Q_{\nu}+\sum_{\nu'}Q_{\nu'}$. The quadratic charge fluctuation is evaluated by the difference of the expectation value of $\langle N^{2}\rangle$ and $\langle N\rangle^2$ in the following form: $\Delta N^{2}=\langle N^{2}\rangle-\langle N\rangle^2$. The magnitude $\Delta N^{2}$ can be expressed in terms of the boson fields: $\Delta N^{2}=\sum_{ls}p^{2}_{ls}+4\sum_{\nu}d^{2}_{\nu}+9\sum_{ls}\overline{t}^{2}_{ls}+16f^{2}
-(\sum_{ls}p^{2}_{ls}+2\sum_{\nu}d^{2}_{\nu}+3\sum_{ls}\overline{t}^{2}_{ls}+4f^{2})^{2}$. In Figures 10a and 10b we can see that  for the SU(4)  Kondo state, in the regime of weak and intermediate coupling strengths determined by $\ket{p_{ls}}_{4}$, $\ket{t_{ls}}_{4}$ and $\ket{d_{\nu}}_{6}$ ($\ket{q_{a(b)}}_{8}$ and $\ket{q_{w}}_{12}$) quantum states, the charge fluctuation $\Delta N^{2}$ is strongly suppressed, comparable to $T_{[\star]}$ (more precisely to $T_{K}$) \cite{Krychowski2020}. By increasing the coupling $t$ to the TSC for $Q=(3/2)(e^{2}/h)$, the charge fluctuation remains at the level $\Delta N^{2}=1/4$, which is the significant result for the fractional SU$^{\star}$(3) Kondo state (dark blue areas in Fig. 10a and blue curve in Fig. 10b). At the charge boundary for $Q=1$e and $Q=2$e, the charge fluctuation reaches 1/2. The main contribution to the large finite value of $\Delta N^{2}$ comes from the Majorana fermion-coupled channel. $N^2$ is determined by the expected values of the boson fields operators. For the SU$^{\star}$(3) Kondo state, $\Delta N^{2}$ reaches $1/4$ and the value is associated with $p^{2}_{+\downarrow(-s)}=d^{2}_{20,\uparrow\uparrow,\uparrow\downarrow}=1/6$ for $Q=(3/2)$e and $\overline{t}^{2}_{+\uparrow(-s)}=d^{2}_{02,\downarrow\downarrow,\downarrow\uparrow}=1/6$ for $Q=(5/2)$e. In contrast to the region where the MF-coupled channel dominates in the quantum transport, $\Delta N^{2}=1/4$ but is determined by the amplitudes $e^{2}=p^{2}_{+\uparrow}=1/2$
for the quantum states $\ket{q_{z}}_{2}$ and $f^{2}=p^{2}_{+\downarrow}=1/2$ for $\ket{q_{x}}_{2}$.

Fig. 11 shows the evolution of the occupancy number $N=\sum_{\nu}Q_{\nu}$ with the increase of the coupling strength to 1TSC, 2TSC and 3TSC. The black lines on the landscape plots represent the total occupancy number $N$ with an increment $\delta E_{d}=+0.15$ for $E_{d}$ in the range $-12$ to $+3$. We observed a significant change in the value of the total charge for $t>t_{2}$. The strong influence of the anomalous Green's function $\hat{G}^{R}_{12}=\langle\langle f^{\dagger}_{\nu'};f^{\dagger}_{\nu'}\rangle\rangle^{R}$ on the statistical value of the total charge $N$ is manifested by an increase in the value of the correlator $i\widetilde{t}_{\nu'}\langle\gamma_{\nu'}f^{\dagger}_{\nu'}\rangle^{<}$. In the total charge number Q we observe the leakage of the charge by finite value of the tunnel correlator and additional degrees of freedom of the Majorana fermion.  Formally, by decomposing the tunneling term into two parts (see Sec. IIIA), we can say that the local isospin $\langle f^{\dagger}_{\nu'}f^{\dagger}_{\nu'}\rangle$ in the system increases and at the same time the value of the charge is modified by tunneling processes. In the CNTQD-TSC device we observe characteristic fractional charges $Q=(1/2)$e and $Q=(7/2)$e when the system is determined by two doublets $\ket{q_{x}}_{2}$ and $\ket{q_{z}}_{2}$. In terms of the fractional SU$^{\star}$(3) Kondo  state, the charge values are quantized to $3/2$e and $5/2$e, where two sextuplets $\ket{q_{g}}_{6}$ and $\ket{q_{y}}_{6}$ are the ground states. In the CNTQD-2TSC system the SU$^{\star}$(2) effect is realized in the strong coupling regime. For the Kondo state, the quantum conductance reaches $3(e^{2}/h)$ and the total charge is $N=Q=2$e (Fig. 11b), where the octuplet quantum states $\ket{q_{y}}_{8}$ are the ground state.
In the channel coupled to the Majorana fermion, the occupancy number reaches $Q=1$e and $3$e for the quantum states $\ket{q_{z}}_{2}$ and $\ket{q_{x}}_{2}$. The coupling strength to the three Majoranas with chirality $ls=+s$ and $ls=-\uparrow$ leads with increasing $t$ to the degeneracy line of the two octuplets $\ket{q_{x}}_{8}$ and $\ket{q_{z}}_{8}$. The transport here occurs through three Majorana fermion-coupled channels and one normal channel between the charges $Q=5/2$e and $Q=3/2$e, i.e. exactly for $Q=2$e (Fig. 11c). The CNTQD-3TSC system is determined by the U$^{\star}$(1) charge symmetry state.

Figure 12 presents the absolute values of the X- and Z-components of the spin and isospin as a function of the atomic level of the quantum dot ($E_{d}$). The spin components can be written as $\hat{S}_{X}=(1/2)\sum_{l}(d^{\dagger}_{l\downarrow}d_{l\uparrow}+d^{\dagger}_{l\uparrow}d_{l\downarrow})$ and $\hat{S}_{Z}=(1/2)\sum_{l}(n_{l\uparrow}-n_{l\downarrow})$
and the isospin components can be expressed by $\hat{I}_{X}=(1/2)\sum_{l}(d_{l\downarrow}d_{l\uparrow}+d^{\dagger}_{l\uparrow}d^{\dagger}_{l\downarrow})$ and $\hat{I}_{Z}=(1/2)\sum_{ls}n_{ls}-1$.
\begin{figure}[b!]
\includegraphics[width=0.75\linewidth]{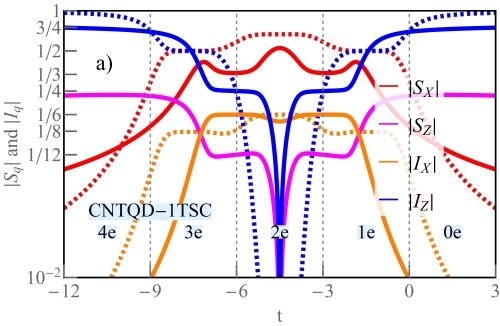}\\
\includegraphics[width=0.75\linewidth]{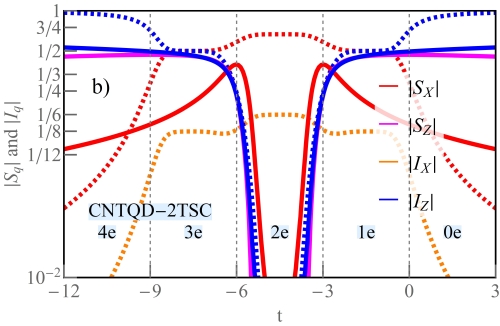}\\
\includegraphics[width=0.75\linewidth]{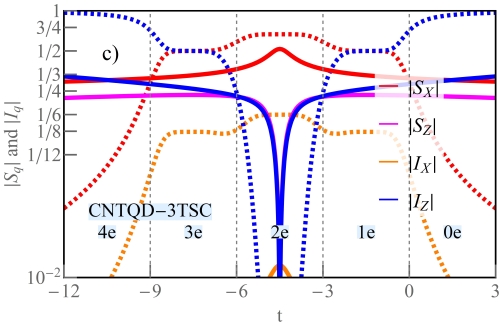}
\caption{\label{fig:epsart} (Color online) The absolute value of the spin $S_{q=X,Z}$ and the isospin $I_{q=X,Z}$ components plotted for $t=10^{-3}$ (dashed lines) and $t=20$ (solid lines) in: a) CNTQD-1TSC, b) CNTQD-2TSC and c) CNTQD-3TSC devices.}
\end{figure}
The operators of the local X- and Z-spin components are defined by the boson fields operators in the following terms:
\begin{eqnarray}
&&\nonumber \hat{S}_{X}=(1/2)(\sum_{ls}p^{\dagger}_{ls}p_{l\overline{s}}+\sum_{s}d^{\dagger}_{s\overline{s}}(d_{ss}+d_{\overline{s}\overline{s}})
\\&&\nonumber+\sum_{s}d^{\dagger}_{ss}(d_{s\overline{s}}+d_{\overline{s}s})+\sum_{ls}\overline{t}^{\dagger}_{ls}\overline{t}_{l\overline{s}})\\
&&\nonumber \hat{S}_{Z}=(1/2)(\sum_{l}(p^{\dagger}_{l\uparrow}p_{l\uparrow}-p^{\dagger}_{l\downarrow}p_{l\downarrow})
\\&&+2d^{\dagger}_{\uparrow\uparrow}d_{\uparrow\uparrow}-2d^{\dagger}_{\downarrow\downarrow}d_{\downarrow\downarrow}+\sum_{l}(\pm \overline{t}^{\dagger}_{l\uparrow}\overline{t}_{l\uparrow}\mp \overline{t}^{\dagger}_{l\downarrow}\overline{t}_{l\downarrow}))
\end{eqnarray}
The Majorana fermion leads to the local pairing-induced on CNTQD and modifies the X- and Z-isospin components in quantum dot. $\hat{I}_{X}$ and $\hat{I}_{Z}$ can be written in the bilinear form of the slave boson operators in the following way:
\begin{eqnarray}
&&\nonumber \hat{I}_{X}=(1/2)(-p^{\dagger}_{-\downarrow}p_{-\uparrow}+\sum_{s}(p^{\dagger}_{+s}\overline{t}_{+s}+p^{\dagger}_{-s}\overline{t}_{-\overline{s}})
\\&&\nonumber+d^{\dagger}_{20}(-e+f)+e^{\dagger}(d_{20}+d_{02})+d^{\dagger}_{02}f-f^{\dagger}d_{02}
\\&&\nonumber-d^{\dagger}_{\downarrow\downarrow}d_{\downarrow\uparrow}
-d^{\dagger}_{\uparrow\downarrow}d_{\uparrow\uparrow}-\overline{t}^{\dagger}_{-\uparrow}\overline{t}_{-\downarrow}-\sum_{s}\overline{t}^{\dagger}_{-s}p_{-\overline{s}})\\
&&\nonumber \hat{I}_{Z}=(1/2)(\sum_{ls}p^{\dagger}_{ls}p_{ls}+2\sum_{\nu}d^{\dagger}_{\nu}d_{\nu}
\\&&+3\sum_{ls}\overline{t}^{\dagger}_{ls}\overline{t}_{ls}+4f^{\dagger}f)-1
\end{eqnarray}
The dashed and solid lines in Fig. 12 represent the spin and isospin components for the intermediate coupling to the topological superconductor ($t=10^{-3}$) and for the strong coupling strength to a Majorana fermions (solid lines, $t=20$). The spin X-component in Fig. 12a leads to $1/2$ and $1/\sqrt{2}$ for the quantum states $\ket{q_{b(a)}}_{8}$ and $\ket{q_{w}}_{12}$, where in the SU$^{\star}$(3) Kondo state $|S_{X}|=1/3$ for $\ket{g_{y}}_{6}$ and at the e-h symmetry point the expectation value of the transverse spin leads to $|S_{X}|=1/2$. The spin Z-component reaches $\approx0$ for $t=10^{-3}$, in the strong coupling region $|S_{Z}|=1/12$ for $\ket{g_{y}}_{6}$ and $|S_{Z}|=1/4$ for $\ket{g_{x(z)}}_{2}$ (solid magenta line in Fig. 12a). The isospin $|I_{Z}|$ for $t=10^{-3}$ (dashed blue line) is quantized to $1/2$ for the octuplets $\ket{q_{b(a)}}_{8}$ and $|I_{Z}|=0$ for the duodecuplet state $\ket{q_{w}}_{12}$. The Z-component of the isospin for $t=20$ (solid blue line) reaches the value $|I_{Z}|=1/4$ for sextuplets $\ket{q_{g(y)}}_{6}$ and $|I_{Z}|=3/4$ for the doublet quantum states $\ket{q_{x(z)}}_{2}$. The transverse component of the isospin is also modified, under the weak coupling strength to TSC: $|I_{X}|$ approaches $1/8$ for octuplets and $|I_{X}|=1/6$ for duodecuplets. For the SU$^{\star}$(3) Kondo state, the transverse isospin component leads to a quantized value of $|I_{X}|=1/6$ for the fractional charges on the quantum dot $Q=(3/2)$e and $Q=(5/2)$e (orange solid line in Fig. 12a).

Figure 12b shows the values of $|S_{X(Z)}|$ and $|I_{X(Z)}|$ for the CNTQD-2TSCs system. We can see that, in terms of coupling strengths for the octuplet states $\ket{q_{y}}_{8}$ and the SU$^{\star}$(2) Kondo effect: $|S_{Z}|=|I_{Z}|=0$, and for the quartets $\ket{q_{x(z)}}_{4}$, $|S_{Z}|=|I_{Z}|=1/2$. The X-components for $\ket{q_{x(z)}}_{4}$ reach the values
$|S_{X}|=1/12$ and $|I_{X}|=0$. Figure 12c presents the expected values of the local spin and isospin for the CNTQD coupled to the three Majorana fermions $\{\gamma_{+s},\gamma_{-\uparrow}\}$.
The Z-components of the spin and isospin vanish $|S_{Z}|=|I_{Z}|=0$ in the e-h symmetry point at the boundary of the octuplets $\ket{q_{x(z)}}_{8}$. At this point, the charge state U$^{\star}$(1) is realized and the local transverse spin leads to $|S_{X}|=1/2$. For the quantum states $\ket{q_{x(z)}}_{8}$, when $Q=(5/2)$e and $Q=(3/2)$e,  the spin and isospin components reach $|S_{X}|=1/3$, $|S_{Z}|=1/4$ and $|I_{X}|=0$, $|I_{Z}|=1/3$.

Since the Majorana mode is assumed to be coupled to the spin-orbital energy level of the CNTQD with fourfold degeneracy, it breaks the spin and orbital symmetry of the system and is manifested in the temperature dependence of the entropy. Figure 13a shows the total entropy $S_{tot}$ for the CNTQD-1TSC system as the sum of the quantum dot entropy $S_{QD}$, the tunneling entropy $S_{T}$ and the entropy of the topological superconductor $S_{TS}$. We can express these quantities following the author \cite{Smirnov2015} in the form $S_{tot}=-\frac{\partial\widetilde{F}_{f}}{\partial T}$, using the thermodynamic potential $\widetilde{F}_{f}$ and Matsubara Green's functions  \cite{Coleman1987,Coleman2015}:
\begin{eqnarray}
&&\nonumber S_{tot}=S_{QD}+S_{T}+S_{TS}\\
&&\nonumber S_{QD}=-\sum_{\alpha\nu}Im\left\{i\log\Gamma\left[\frac{1}{2}+\frac{z\pm V_{\alpha}}{2\pi iT}\right]+ \right.\\
&&\left.\frac{(z\pm V_{\alpha})\Psi_{0}\left[\frac{1}{2}+\frac{z\pm V_{\alpha}}{2\pi iT}\right]}{2\pi T}\right\}|^{z=\Lambda_{\nu}}_{z=\Lambda_{\nu}-W}\\
&&\nonumber S_{T}=\sum_{\alpha\nu'}\int^{+\infty}_{-\infty}Re\left\{\frac{(E\pm V_{\alpha})\ln[\delta G_{\nu'}]}{16\pi iT^{2}\cosh^{2}\left[\frac{E\pm V_{\alpha}}{2T}\right]}\right\}dE\\
&&\nonumber S_{TS}=-\sum_{\alpha\nu'}Im\left\{i\log\Gamma\left[\frac{1}{2}+\frac{z\pm V_{\alpha}}{2\pi iT}\right]+ \right.\\
&&\left.\frac{(z\pm V_{\alpha})\Psi_{0}\left[\frac{1}{2}+\frac{z\pm V_{\alpha}}{2\pi iT}\right]}{2\pi T}\right\}|^{z=iT_{1}}_{z=iT_{1}-2T_{2}}
\end{eqnarray}
where $\log\Gamma[z]$ is the logarithm of the Euler gamma function, $T_{1}=\delta$ and $T_{2}$ are the characteristic temperatures in the entropy of the isolated TSC ($T_{2}$ is a characteristic temperature found in the tunneling entropy $S_{T}$ and is related to the vanishing of the first derivative of $\delta G_{\nu'}$).  $\delta G_{\nu'}$ in the tunneling entropy is given by \cite{Smirnov2015}:
\begin{eqnarray}
\delta G_{\nu'}=\frac{\sum_{k=1,2}\pm\hat{G}^{A}_{1k}(-E)\hat{G}^{R}_{k1}(E)}
{\sum_{k=1,2}\pm\hat{G}^{R}_{1k}(-E)\hat{G}^{A}_{k1}(E)}.
\end{eqnarray}
where $\hat{G}^{R(A)}_{ik}(-E)$ are the Green's function of the matrix in Eq. (6). In the evolution of $S_{T}$ as a function of T, we observed the third characteristic temperature $T_{3}=\widetilde{t}^{2}_{\nu'}/2|\widetilde{E}_{\nu'}|$, where $S_{T}$ goes to zero.

Figure 13a shows the saturation of the total entropy at the value $S_{tot}=\ln[4]$, which corresponds to the fraction of the four quantum states in the high temperature limit above $T_{K}$. The value of the coupling is $t=10^{-3}$, and the SU(4)-like Kondo state is realized by the octuplet quantum states $\ket{q_{a}}_{8}$, where $S_{tot}=0$. It is an SU(4)-like state, because ${\mathcal{G}}\approx2(e^{2}/h)$ and the octuplet states $\ket{q_{a(b)}}_{8}$ are different from the pure quantum states $\ket{p_{ls}}_{4}$ ($\ket{t_{ls}}_{4}$) (even if $a\gg a'$, see Eq.(7)). At intermediate temperatures between $T_{K}$ and $T_{2}$, the first entropy plateau $S_{tot}=\ln[4]/4$ is observed . This is strictly related to the Majorana fermion-coupled channel in the CNTQD-TSC device. Analogous results
are reported in the literature  for  the SU(2) Kondo dot coupled to a single Majorana fermion \cite{Zitko2011,Smirnov2015}, where $S_{tot}=\ln[2]/2$. Below the temperature $T_{2}$ the sign of $S_{T}$ changes to $S_{T}=-\ln[4]/4$. In this case the  contribution of the tunneling entropy is compensated by the entropy of  the topological superconductor $S_{TS}=\ln[4]/4$. The characteristic temperature $T_{1}=\delta$ is directly related to the lifetime of the Majorana fermion. In an analogous way the problem was defined in the paper \cite{Smirnov2015}, but there the author assumed the finite value of the overlap strength $\Delta_{(0)}$ between the two Majorana fermions in the TSC wire, and $S_{T}$ has an opposite sign. $S_{T}$ is positive in the range between $T_{1}$ and $T_{2}=\Delta_{(0)}$ and takes a negative sign below $T_{2}$. This is a consequence of the self-energy expression in \cite{Smirnov2015}. For the model with the overlapping between Majorana fermions, the self-energy can be expressed as $\widetilde{\Sigma}^{R}_{t}=(\widetilde{t}^{2}z)/(z^2-\Delta^{2}_{(0)})$. In our calculations we assumed $\Delta_{(0)}=0$ (highly coherent TSC wire, where $\lambda_{K}\approx0$) and therefore $\widetilde{\Sigma}^{R}_{t}=\widetilde{t}^{2}/z=\widetilde{t}^{2}/(E+i\delta)$, where $\delta$ is a finite lifetime of the Majorana fermion (see \cite{Liu2015}). The sign reversal of  $S_{T}$ is observed for the SU$^{\star}$(2) Kondo effect (Fig. 13d), where $T_{2}<T_{3}$ and $T_{K}<T_{2}$ in opposite to the SU(4)-like  Kondo state for the low coupling regime, where $T_{2}<T_{3}$ and $T_{K}>T_{2}$ (the ground state is determined by the octuplet $\ket{q_{a}}_{8}$).
\begin{figure}[t!]
\includegraphics[width=0.75\linewidth]{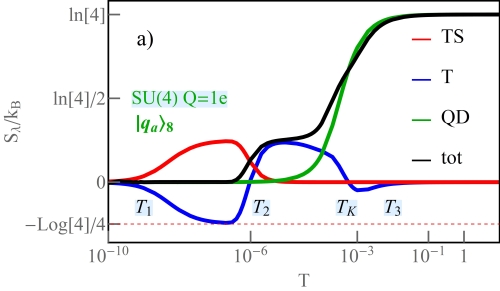}\\
\includegraphics[width=0.75\linewidth]{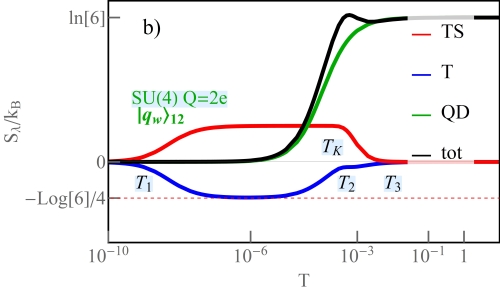}\\
\includegraphics[width=0.75\linewidth]{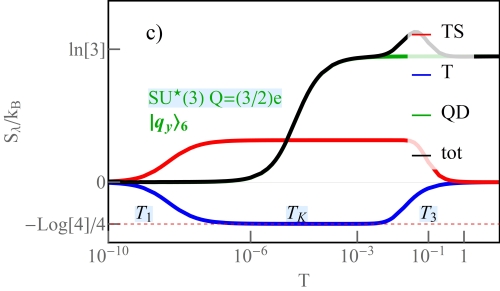}\\
\includegraphics[width=0.75\linewidth]{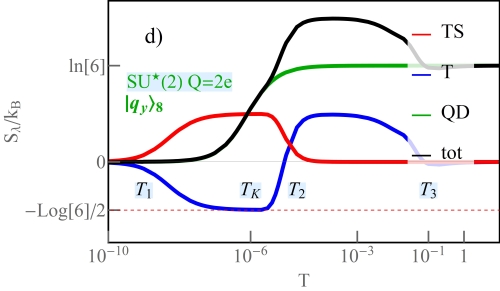}
\caption{\label{fig:epsart} (Color online) The total entropy $S_{tot}$, the tunneling entropy $S_{T}$, the entropy of the isolated topological superconductor $S_{TS}$ and the entropy of the quantum dot $S_{QD}$ as a function of temperature $T$ for a-c) CNTQD-1TSC device and d) CNTQD-2TSC device. Figures a, b) and Figs. c, d) are plotted for $t=10^{-2}$ and $t=20$.}
\end{figure}

Figure 13b shows the entropy for the weak coupling to a single Majorana fermion ($t=10^{-3}$) with two electrons on the quantum dots $Q=2$e. The ground state is determined by the duodecuplet $\ket{q_{w}}_{12}$ (Eq. (9)). The conductance reaches ${\mathcal{G}}=(7/2)(e^{2}/h)$, in contrast to the fully SU(4) Kondo state, where ${\mathcal{G}}$ is quantized to $4(e^{2}/h)$ (here the Kondo effect is determined by the six quantum states $\ket{d_{\nu}}_{6}$). For the duodecuplet state, the tunneling entropy is negative $S_{T}<0$ and is compensated by the $S_{TS}$ contribution. In $S_{tot}$ we observe a small boost above $\ln[6]$, around $T_{2}$.

For the strong coupling strength ($t=20$), the SU$^{\star}$(3) Kondo state is realized in the system for a fractional charge on the quantum dot $Q=(3/2)$e. The quantum conductance reaches ${\mathcal{G}}=(11/4)(e^{2}/h)=(1/2)(e^{2}/h)+(9/4)(e^{2}/h)$. In the high temperature limit $S_{tot}=\ln[3]$ and is dominated by the  $S_{QD}$ contribution. In the low temperature regime, the SU$^{\star}$(3) Kondo state is formed and the contribution of $S_{T}$ is completely compensated by $S_{TS}$ (blue and red lines in Figure 13c).
For the CNTQD-2TSC system with strong coupling $t$ (Fig. 13d)  we observe the sign reversal of the tunneling entropy $S_{T}= \mp\ln[6]/2$ (blue line). The negative entropy of $S_{TS}$ is the consequence of a high order topological state. The ordered part below the temperature $T_{2}$ is compensated by the entropy $S_{TS}$, and for $T>T_{2}$ we observe an increase of the entropy to a value of $S_{tot}=\ln[6]+\ln[6]/2$ (between $T_{2}$ and $T_{3}$) at the expense of the tunneling entropy $S_{T}$ (transient states effect). This is due to the mechanism of expanding the Hilbert space by a topological segment and the realization of the SU$^{\star}$(2) Kondo state by the octuplet state $\ket{q_{y}}_{8}$ (Eq. (12)). The Hilbert space is expanded and allows the higher entropy than $\ln[6]$ for 2e on the quantum dots.

Figure 14 shows the influence of the number of Majorana fermions on the SU(4)-like Kondo effect for 1e on QD, i.e. in terms of the weak coupling strength $t=10^{-3}$, where $T_{K}>T_{2}$. The Kondo state for the CNTQD-1TSC system is determined by the octuplet state. In Figure 14 the dark lines symbolize the tunneling entropies $S_{T}$ and the light curves are associated with the total entropy $S_{tot}$. For $\ket{q_{a}}_{8}$, in the temperature range between $T_{2}$ and $T_{K}$, the total entropy reaches a quantized value of $S_{tot}=\ln[4]/4$ (blue curves in Fig.14). For $T<T_{2}$, the tunneling entropy is compensated by the $S_{T}$ contribution. The SU(4)-like Kondo state is observed from $T_{1}=\delta$ to $T_{K}$. In the CNTQD-2TSC system the value of the total entropy reaches $S_{tot}=\ln[4]/2$, which is related to the quantum states defined by $\ket{q_{a}}_{16}$ and the expanding of the Hilbert space via the topological segments $\{\ket{\underline{0}},\ket{\underline{\Uparrow}}\}$ and $\{\ket{\overline{0}},\ket{\overline{\Uparrow}}\}$.
The inclusion of a third Majorana state ${\gamma_{+\downarrow}}$ in the CNTQD-3TSC device increases the value of the total entropy in the range between $T_{2}$ and $T_{K}$. The total entropy is quantized to $S_{tot}=(3\ln[4])/4$.
The tunneling entropy in this case changes the sign below $T_{2}$, and reaches a maximum positive value of $S_{T}=S_{tot}=(3\ln[4])/4$ above the characteristic temperature. In the high temperature limit $S_{tot}$ is saturated for all devices and the quantum limit $S_{tot}=\ln[4]$ is reached. The first plateau in $S_{tot}$ is associated with the Majorana-coupled channels. The normal channels are temperature resistant and participate in the Kondo effect ($T_{K}>T_{2}$). The number $N_{TS}$ of Majorana fermions \cite{Smirnov2015}, determines the value of the total entropy as follows $S_{tot}=N_{TS}\ln[4]/4$, and contains the information about the SU(4) symmetry of the Kondo state and about the spin-orbital degrees of freedom of the MF state.
\begin{figure}[t!]
\includegraphics[width=0.75\linewidth]{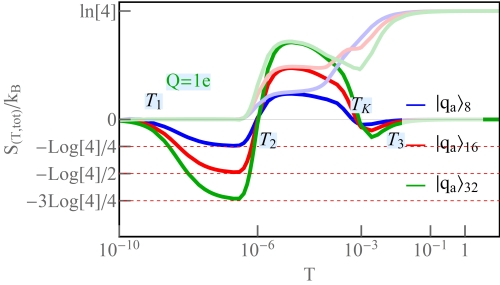}
\caption{\label{fig:epsart} (Color online) The total entropy $S_{tot}$, the tunneling entropy $S_{T}$, the entropy of the isolated topological superconductor $S_{TS}$ and the entropy of the quantum dot $S_{QD}$ as a function of temperature $T$ for Q=1e and $t=10^{-2}$. $S_{T}$ and $S_{tot}$ are represented by dark and light curves. Colors are assigned to the model: CNTQD-1TSC (blue lines), CNTQD-2TSC (green lines) and CNTQD-3TSC (red lines).}
\end{figure}

There are already several papers in the literature investigating the thermoelectric power in a QD system with a Majorana fermion using the equation of motion technique \cite{Lopez2014} and the renormalization group \cite{Majek2022,Wojcik2020,Majek2021}. In the first article, the authors analyze a single quantum dot device coupled to a Majorana fermion. The thermoelectric power in this system undergoes a transverse modification, and the authors observe a change in the sign of ${\mathcal{S}}$.
In the paper \cite{Lopez2014}, the authors analyzed the non-interacting QD-TSC system ($U=0$) and showed that the thermoelectric transport measurements can be used to detect the Majorana fermion. The authors pointed out that a finite phase shift for the Kondo SU(4) effect could significantly affect the value and sign of the thermal conductivity \cite{Lopez2014}.
In the second paper \cite{Majek2022}, the authors discussed the spin-resolved thermal signatures of the Majorana-Kondo effect in the DQD-T-shaped system. Using the numerical renormalization group method the authors focus on the two-stage Kondo effect and  the leakage of Majorana quasiparticles into the double dot system. Majorana-induced interference with strong electron correlations on the DQD system and is observed in the spin-Seebeck effect. For these problems, the authors have modified the linear response current $I_{s}=e^{2}L^{(0)}_{s}\delta V\pm e^{2}L^{(0)}_{s}\delta V_{S}-(e/T)L^{(1)}_{s}\delta T$ by adding the spin-dependent voltage in the form $\delta V_{S}=\delta V_{\uparrow}-\delta V_{\downarrow}$. This introduces the possibility to detect spin (orbital) contributions of the TEP, which is particularly interesting for spin (valley) dependent quantum calorimetry.

In terms of the linear response at small temperature $\delta T$ and voltage difference $\delta V$, the electric current $I$ and the thermal current $I_{Q}$ obey the linear equations presented in Sec. IIIC,  where $L^{(0)}_{\nu}$ and $L^{(1)}_{\nu}$ are the kinetic transport coefficients. In this paper we study the thermoelectric power (TEP), which can be expressed as follows ${\mathcal{S}}=(\delta V/\delta T)|_{I=0}=-(1/eT)(\sum_{\nu}L^{(1)}_{\nu}/\sum_{\nu}L^{(0)}_{\nu})$. Since we mainly want to relate the value of the thermoelectric power to a symmetry of the Kondo effect \cite{Krychowski2020}, we introduced the linear coefficient of the thermoelectric power $\gamma_{(S)}=({\mathcal{S}}T_{K})/(2\pi T)$. In general, we can express the TEP for the CNTQD-TSCs device by the following Mott's formula $\mathcal{S}=-(\pi^{2}/3|e|)[(\sum_{\nu}d\widetilde{\varrho}_{\nu}/dE|_{E=0}+\sum_{\nu'}d\widetilde{\varrho}_{\nu'}/dE|_{E=0})
/(\sum_{\nu}\widetilde{\varrho}_{\nu}+\sum_{\nu'}\widetilde{\varrho}_{\nu'})]$, where $\nu'$ is associated with the channels coupled to the Majorana fermions. Considering $T\ll T_{K}$ and $t\approx0$, the thermoelectric power satisfies the main prediction of FL theory, where the linear coefficient of the specific heat is independent of the quasiparticle interactions ($\gamma_{N}=\frac{\pi^{2}}{3}\sum_{\nu=ls}\widetilde{\varrho}_{\nu}$). Based on this assumption ${\mathcal{S}}$ is proportional to the two-body correlation function (Eq. (42)).
\begin{figure}[t!]
\includegraphics[width=0.75\linewidth]{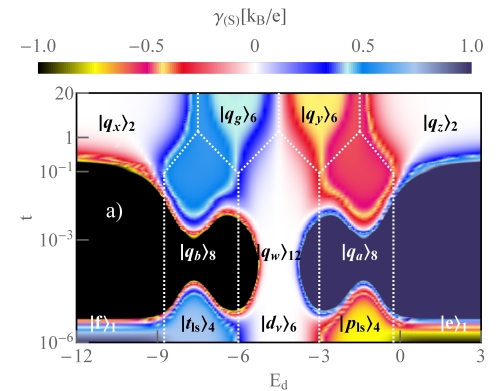}\\
\includegraphics[width=0.75\linewidth]{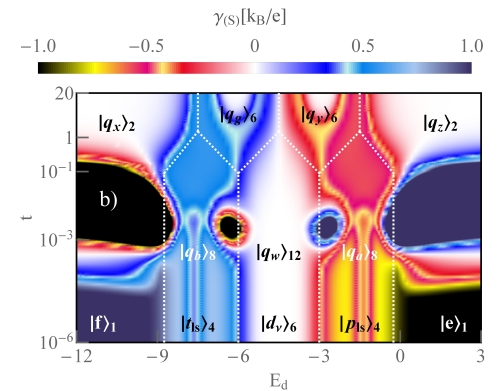}\\
\includegraphics[width=0.75\linewidth]{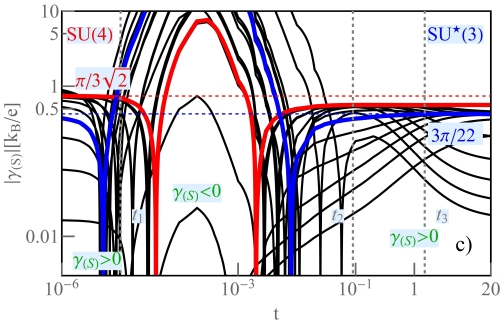}\\
\includegraphics[width=0.75\linewidth]{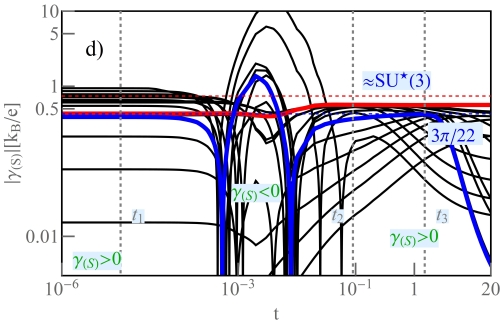}
\caption{\label{fig:epsart} (Color online) Thermoelectric power in the CNTQD-1TSC device: a, b) The density plot of the linear thermoelectric power coefficient $\gamma_{(S)}={\cal{(S)}}/2\pi T_{[\star]}$ as a function of $E_{d}$ and $t$ for $T=10^{-8}$ and $T=10^{-3}$. c, d) The landscape plots of $\gamma_{(S)}$ as a function of $t$ ($\delta E_{d}=0.15$).}
\end{figure}

For decoupled CNTQD to TSCs, the linear coefficient of thermoelectric power reaches a value of $\pm\pi/(3\sqrt{2})$ for the quantum states $\ket{p_{ls}}_{4}$ and $\ket{t_{ls}}_{4}$ (yellow and blue areas in Fig. 15a and red line in Fig. 15c), as we predicted earlier \cite{Krychowski2020}. ${\mathcal{S}}$ probes the Kondo resonance close to the Fermi level $E_{F}$, and due to the position of the quasiparticle resonance, the linear coefficient of the thermoelectric power in terms of the SU(4) Kondo effect approaches to the finite value $\gamma_{(S)}=-(k_{B}/|e|)(\pi/3)(\widetilde{E}_{\nu}/T_{K})=-(k_{B}/|e|)(\pi/3)\cos[\delta_{\nu}]$.
In the range of the fractional charges $Q=(5/2)e$ and $Q=(3/2e)$ for the strong Majorana-coupled channel, the thermal transport is determined by low and high energy sextuplets $\ket{q_{g}}_{6}$ and $\ket{q_{y}}_{6}$. For the SU$^{\star}$(3) Kondo effect, $\gamma_{(S)}$ has reached the quantized value $\pm3\pi/22$ and the system is still the FL state (light blue and light yellow regions in Fig. 15a, and blue line in Fig. 15c).

The density plot of $\gamma_{S}$ shows the transition around $t_{1}$ to the two doublet quantum states $\ket{q_{x}}_{2}$ and $\ket{q_{z}}_{2}$.
The transition line is gate independent and appears as the inverse sign of $\gamma_{S}$. The sign of $\gamma_{S}$ changes a second time around $t_{2}$, where the Majorana-coupled channels dominate over the normal thermoelectric transport.
In the weak coupling regime for $Q=1(3)$e, we observe the sharp sign reversal for the octuplets $\ket{q_{a(b)}}_{8}$.
The transition line is gate-dependent and the inhomogeneous sharp transition line is associated with the compensation of two processes in the Majorana-coupled channel $-\pi T_{K}(\cot[\delta_{\nu'}][\widetilde{\Gamma}_{\nu'}\delta^{2}+\widetilde{t}^{2}\delta])\approx-\pi T_{K}(-\cot[\delta_{\nu'}]\widetilde{\Gamma}_{\nu'}\widetilde{t}^{2})$.
For $Q=2$e, the linear thermoelectric coefficient approaches zero in a statement of charge neutrality. This is the consequence of the Friedel sum rule and the phase shift $\delta_{\nu}=\pi/2$ at the energy of the Fermi level. The gate-dependent behavior in the NFL phase is also indicated by the slight indentation for $Q=2$e. In the region of strong coupling strength, when the system goes to the SU$^{\star}$(3) Kondo state, for $E_{d}=-3$ and $E_{d}=-6$, we observe two flat areas: light yellow and light blue for $Q=3/2$ and $Q=5/2$. The $\gamma_{S}$ reaches a value of $\gamma_{(S)}=\mp\frac{\pi\sin[\delta_{\nu}]\sin[2\delta_{\nu}]}{-4+3\cos[2\delta_{\nu}]}=\mp3\pi/22$, which is significantly different from the value for SU(3) Kondo state, as shown in Figure 15a. This difference is shown by the dark violet line in Fig. 16a ($\gamma_{S}=\pm \pi/6$), where we have subtracted the contribution to ${\mathcal{S}}$ from the Majorana fermion channel $\nu'$. This result is confirmed in the literature \cite{Krychowski2020}.
$\gamma_{S}=\pm 3\pi/22$ is observed for the sextuplets, and due to the Onsager relations is associated with the finite topological value of the quantum conductance quantized at ${\mathcal{G}}=(11/4)(e^{2}/h)$. A finite temperature gradient $T=10^{-3}$ changes the picture in $\gamma_{S}$ (Fig.15b and Fig. 15d). The finite temperature violates the quantum conductance
in the crossover region, where the energy ground state is defined by high-degenerate quantum states $\ket{q_{a(b)}}_{8}$ (Fig. 9c). The effect of
reversal sign for $Q=1(3)$e in TEP disappears (red line on Fig. 15d) and reduced to the spots under these temperature conditions (Fig. 15b). For $\ket{p_{ls}}_{4}$ and $\ket{t_{ls}}_{4}$ we observe decrease of the TEP to $\gamma_{S}\approx\pm1/2$.
Increasing the temperature gradient (red line in Fig. 15d), the SU(4) Kondo effect is suppressed for $Q=1(3)$e, therefore $\pm1/2<\pm\pi/(3\sqrt{2})$.
The transition between $\ket{e}_{1}$ ($\ket{f}_{1}$) and the doublets $\ket{q_{x(z)}}_{2}$ is shifted (Fig. 15b).
The same effect was observed for $Q=(3/2)$e and $Q=(5/2)$e, where $t_{1}$ changes its position on the axis of the coupling strength $t$ (blue curve in Fig. 15d).
The second point is stable for $T=10^{-3}$ and is associated with the second compensation processes, where $-\pi T_{K}(3\dot{\widetilde{\varrho}}_{\nu}(0))/[(3\widetilde{\varrho}_{\nu}(0)+\widetilde{\varrho}_{\nu'}(0))]=-\pi T_{K}(\cot[\delta_{\nu'}][\widetilde{\Gamma}_{\nu'}\delta^{2}+\widetilde{t}^{2}(\delta-\widetilde{\Gamma}_{\nu'})])/[(3\widetilde{\varrho}_{\nu}(0)+\widetilde{\varrho}_{\nu'}(0))(\pi\widetilde{\Gamma}_{\nu'}(2\widetilde{t}^{2}+\widetilde{\Gamma}_{\nu'}\delta\csc^{2}[\delta_{\nu'}])^{2})]$.
In the intermediate coupling strength between $t=10^{-2}$ and $t_{3}$ the linear coefficient exceeds the value of $\gamma_{S}\approx\pm3\pi/22$ (blue line in Fig. 15d).
For $T^{SU^{\star}(3)}_{K}<T$ and $t>t_{3}$ we observe the suppression of $\gamma_{S}$ to zero in the sextuplet regions (blue line in Fig. 15d).
The envelope of the NFL-like region, where we observe the strong enhancement of $\gamma_{S}$ (black, and dark violet area in Fig. 15a,b), depends strongly on the compensation conditions and the energy level of the quantum dot. An increase in temperature contributes to a gradual narrowing of the NFL crossover region.
\begin{figure}[t!]
\includegraphics[width=0.75\linewidth]{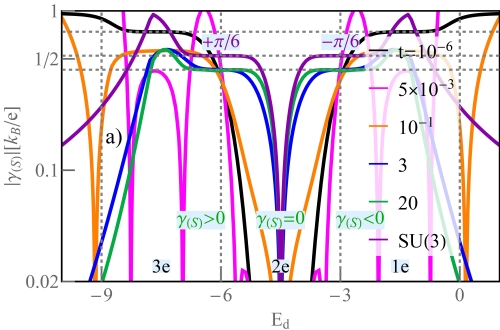}\\
\includegraphics[width=0.75\linewidth]{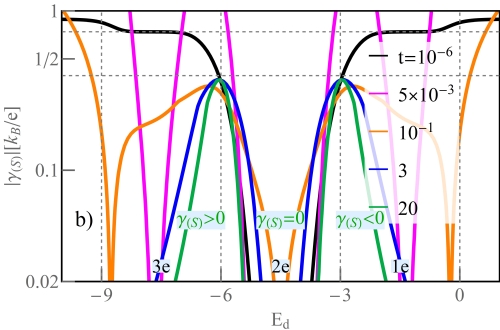}\\
\includegraphics[width=0.75\linewidth]{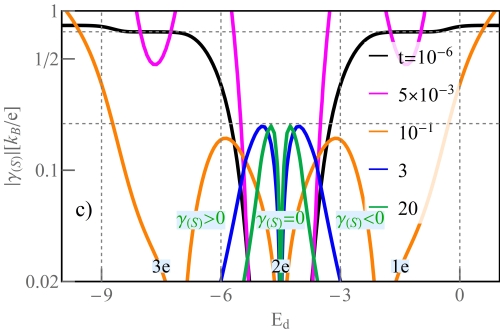}\\
\includegraphics[width=0.75\linewidth]{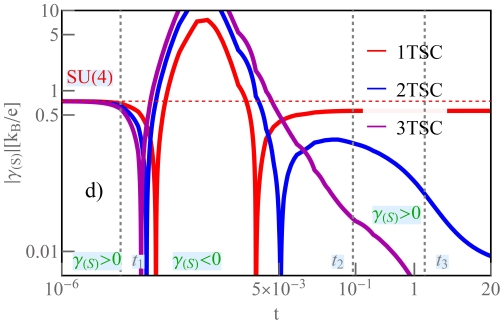}\
\caption{\label{fig:epsart} (Color online) a-c) $|\gamma_{S}|$ as a function of $E_{d}$ for CNTQD coupled to 1MF, 2MFs and 3MFs. The dark volet line in figure a shows $\gamma_{(S)}=\pm \pi/6$, the result for the  SU(3) Kondo effect (where the $+\uparrow$ channel in $\cal{S}$ is neglected) \cite{Krychowski2020}. $\gamma_{(S)}$ changes the sign at a half-filling point $Q=2$e d) $|\gamma_{S}|$ as a function $t$ for $Q=3$e.}
\end{figure}

Fig. 16 shows the cross sections of $|\gamma_{S}|$ on a logarithmic plot as a function of $E_{d}$ for different values of $t$.
In Figure 16a-c, the black lines present $|\gamma_{S}|$ for SU(4) Kondo state in CNTQD.
For $Q=1(3)$e we observed the quantized values $|\gamma_{S}|=\pi/(3\sqrt{2})$.
The ground state of the system $\ket{d_{\nu}}_{6}$  has on average two electrons ($Q=2$e) and the thermoelectric power in this region is strongly flattened and suppressed. $|\gamma_{S}|$ with $t=5\times10^{-3}$ leads to $3\pi/22$ for $1$e and $3$e charge sectors (magenta line in Fig.16a). Around the enhancement, we observed two compensation points, where $\gamma_{S}$ changed the sign.
For this cross section, we have observed the third point of compensation in $Q=2$e.
Orange line shows the saturation of $\gamma_{S}$ to $\pm\pi/6$ for $Q=1(3)$e.
For the strong coupling strength ($t=3$ and $t=20$), $|\gamma_{S}|$ quantized to $3\pi/22$, where formed SU$^{\star}$(3) Kondo state for the fractional charge on the quantum dot. The theory for this phase predicts the linear coefficient TEP in the form $\gamma_{(S)}=\mp\frac{\pi\sin[\delta_{\nu}]\sin[2\delta_{\nu}]}{-4+3\cos[2\delta_{\nu}]}$. At the e-h symmetry point we
observed a narrowing line of blocked thermoelectric transport $\gamma_{S}=0$ (blue and green lines in Fig. 16a).
Fig. 16b shows $|\gamma_{S}|$ as a function of $E_{d}$ for CNTQD-2TSC device.
For $t=3$ and $t=20$, we observed the area of the SU$^{\star}$(2) Kondo state, when the quantum dot is
in the octuplet state $\ket{q_{y}}_{8}$. The thermoelectric power remains strongly suppressed ($|\gamma_{S}|=0$), in contrast to the charge degeneracy point, for the Kondo state the region is strictly flattened, as indicated by the scaling energy $T_{[\star]}$.  Around the boundary between $\ket{q_{y}}_{8}$ and $\ket{q_{x(z)}}_{4}$, $|\gamma_{S}|$ reaches $\pi/8$ and this is associated with the U$^{\star}$(1) charge symmetry, where the quantum conductance leads to $2(e^{2}/h)$ (Fig. 6c). For the weak coupling strength to TSC ($t=5\times10^{-3}$), $\gamma_{S}$ vanishes and changes sign around $E_{d}=-(U/2)$ and $E_{d}=-(5U/2)$ (magenta curve in Fig. 16b). In Fig. 16d for the red and blue line we observe $\gamma_{S}=0$. At this point $\gamma_{S}$ shows a sharp sign reversal.
Red curved on Fig. 16d shows the saturation of $\gamma_{S}$ to $\pi/6$ for the strong coupling strength of CNTQD to TSC.
The value appears at the charge degeneracy point between the entangled quantum states $\ket{q_{x}}_{2}$ and $\ket{q_{g}}_{6}$.
A this point, the quantum conductance leads to ${\mathcal{G}}=(3/2)(e^{2}/h)$.
In the case of the CNTQD-3TSC system, TEP is suppressed  for $Q=2e$ and $\gamma_{S}=0$ for two degenerate octuplets $\ket{q_{x(z)}}_{8}$ at the e-h symmetry point. The switching around the charge degeneracy point reaches $|\gamma_{S}|=\pi/16$ (blue and green lines in Fig. 16c). Decreasing of the coupling strength $t$, shift the solution into the NFL-like behavior region, where $\gamma_{S}$ reverses the sign and is strongly increases (magenta curve in Fig.16c). The dark magenta line in Fig. 16d shows the evolution of $|\gamma_{S}|$ as the function of the coupling strength $t$ to the TSCs. Comparing the lines in Fig. 16d, we conclude that increasing the number of TSCs coupled to the quantum dot shifts the second compensation point into the region of  strong coupling strength. Even if, the evolution of $\gamma_{S}$ explains the Mott's formula, the behavior in the region for the weak Majorana coupling strength has not been scaled by the Kondo temperature, and in this sense we mean about NFL-like phase. The strong enhancement in the weak coupling regime is reflected in the temperature dependence of the entropies $S_{QD}$ and $S_{TS}$, where the channels $\nu$ associated with the Kondo state are  much more temperature resistant than the channels $\nu'$ interfering with the spin-orbital states in the quantum dots (Fig. 13a).

Before investigating the nonlinear current and shot noise, let's first discuss the influence of the coupling strength to Majorana fermions on the quantities of the two- and three-body correlation functions. Based on the thermodynamics of the Kondo state described in Sec. IIIB, we showed that the general static susceptibilities in Eqs. (21-24) can be expressed as the second and third derivatives of the thermodynamic potential and are related to the density of the quasiparticle states.
Using the results of \cite{Nishikawa2013,Oguri2022}, we have postulated the weak-coupling approach to find the off-diagonal two- and three-particle correlations, based on the Wilson ratio $W_{\nu\nu'}=1-\widetilde{\chi}_{\nu\nu'}/\sqrt{\widetilde{\chi}_{\nu\nu}\widetilde{\chi}_{\nu'\nu'}}=1
+\widetilde{U}_{\nu\nu'}\sqrt{\widetilde{\chi}_{\nu\nu}\widetilde{\chi}_{\nu'\nu'}}=1+\frac{\delta Q_{\nu\nu'}}{\Delta Q_{\nu\nu'}}$. At this point, we include the (residual) interaction between the quasiparticles $\widetilde{U}_{\nu\nu'}$, and we can express as an
invariant, the two-particle static susceptibility $\chi_{(z)}=(1/4)(\sum_{\nu}\widetilde{\chi}_{\nu\nu}-\sum_{\nu'\neq\nu}\widetilde{\chi}_{\nu\nu'})$, which is proportional to the Z-component
of the partial fluctuations in the pseudospin space. The same quantum metric can be defined for an odd three-body correlation function in the following way: $\chi^{[3]}_{(z)}=(1/4)(\sum_{\nu}\widetilde{\chi}_{\nu\nu\nu}-\sum_{\nu'\neq\nu}\widetilde{\chi}_{\nu\nu'\nu'})$.
$T\chi_{(z)}(T)$ is screened, when $T\chi_{(z)}(T)=0$ and the Kondo state is formed. At low temperatures in $\chi_{(z)}(T)$ we observe the saturated constant value, proportional to $1/T_{K}$ in the limit $T\mapsto0$. Therefore, the quantities $T_{[\star]}\chi_{(z)}(0)$ and $T^{2}_{[\star]}\chi^{[3]}_{(z)}(0)$ are the information about the frozen pseudospin and three-particle correlation. In the context of experimental results for single Kondo dot \cite{Piquard2023}, we suggest that more significant information about the symmetry of the Kondo state and all modifications of the SU(N)-Anderson model (i.e. by the coupling term with Majorana fermions) is hidden in the extremely low- temperature measurements of $\chi_{(z)}$ by the charge-sensing technique.

The magnitude of $\lim_{T\mapsto0}T\chi_{(z)}(T)$ is screened for the SU(4) Kondo state and is close to zero. For the high temperature limit $T\chi_{(z)}(T)$ reaches the value corresponding to the expectation value of the Z-component of the quadratic Casimir operator ($TC^{2}_{Z}$). Although $T\chi^{[3]}_{(z)}(T)$ in the high temperature limit is not equivalent to the Z-component of the cubic of the Casimir operator ($TC^{3}_{Z}$), for further analysis it is a good approach to investigate the contributions to the nonlinear current and the shot noise, where inelastic processes, beyond the e-h symmetry point play an important role. $\chi^{[3]}_{(z)}$ includes all three-body correlation functions, except $\widetilde{\chi}^{[3]}_{\sigma_{1}\sigma_{2}\sigma_{3}}$, which are non-zero only for the tunneling asymmetry between the left and right electrodes ($\widetilde{\Gamma}_{L}\neq\widetilde{\Gamma}_{R}$) \cite{Oguri2022} (last term in Eq.(18)). Gate-dependent three-particle correlators in quantum dot are the odd parity functions and change the sign when passing through the e-h symmetry point.
\begin{figure}[t!]
\includegraphics[width=0.75\linewidth]{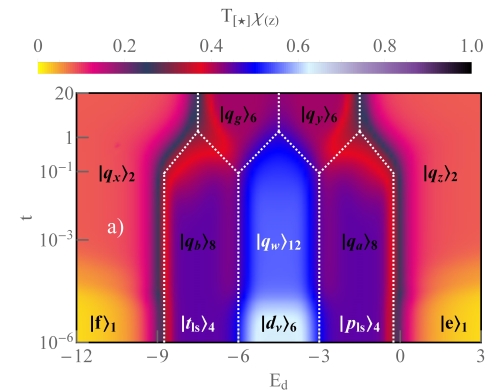}\\
\includegraphics[width=0.75\linewidth]{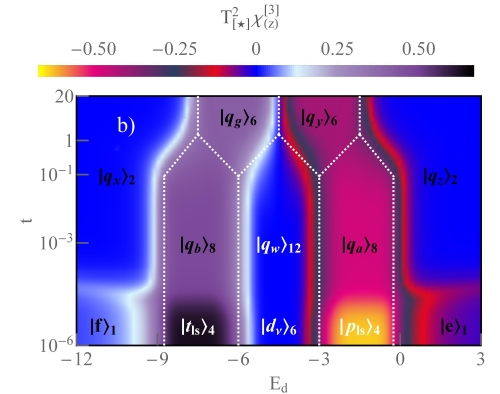}\\
\includegraphics[width=0.75\linewidth]{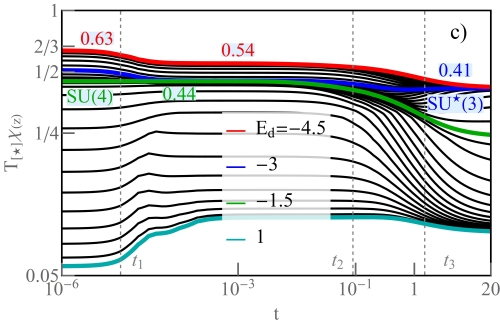}\\
\includegraphics[width=0.75\linewidth]{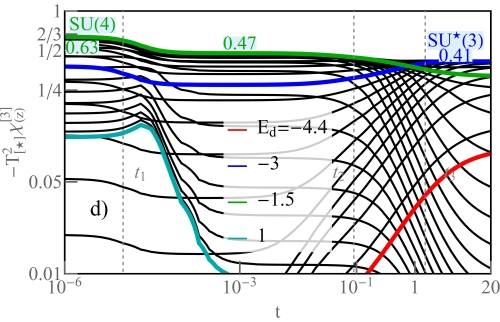}
\caption{\label{fig:epsart} (Color online) a, b) The density plots of the two-body ($T_{[\star]}\chi_{(z)}$) and three-body ($T^{2}_{[\star]}\chi^{[3]}_{(z)}$) frozen correlation functions versus $E_{d}$ and $t$ for the CNTQD-1TSC device. c, d) The landscape plots of $T_{[\star]}\chi_{(z)}$ and $-T^{2}_{[\star]}\chi^{[3]}_{(z)}$ as function of $t$. The black lines are plotted with an increment of $\delta E_{d}=0.15$ from $E_{d}=-4.5$ to $1$. The numbers indicate the values of the correlations in the Kondo phases.}
\end{figure}
\begin{figure}[t!]
\includegraphics[width=0.75\linewidth]{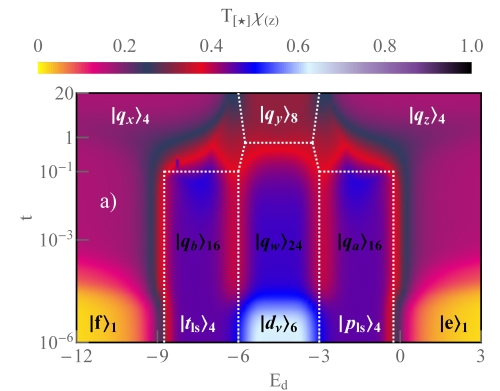}\\
\includegraphics[width=0.75\linewidth]{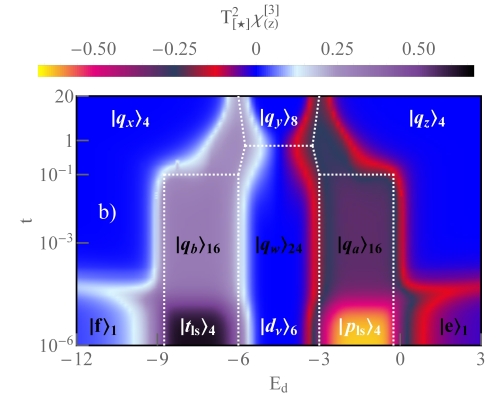}\\
\includegraphics[width=0.75\linewidth]{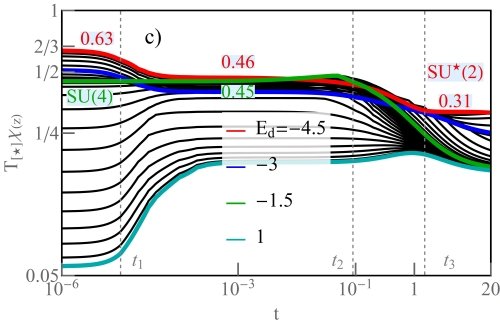}\\
\includegraphics[width=0.75\linewidth]{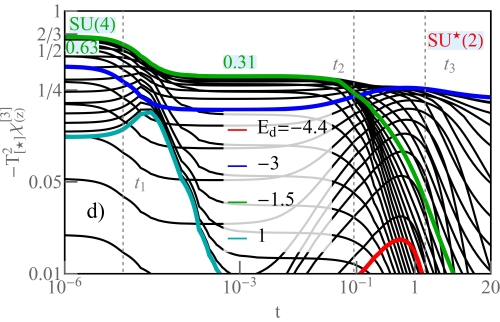}
\caption{\label{fig:epsart} (Color online) Two- and three-body correlations for the CNTQD-2TSC system:  a, b) $T_{[\star]}\chi_{(z)}$) and $T^{2}_{[\star]}\chi^{[3]}_{(z)}$ versus $E_{d}$ and $t$. c, d) the landscape plots of $T_{[\star]}\chi_{(z)}$ and $-T^{2}_{[\star]}\chi^{[3]}_{(z)}$ as function of $t$.}
\end{figure}

Fig. 17a, b shows the density plots of $T_{[\star]}\chi_{(z)}$ and $T^{2}_{[\star]}\chi^{[3]}_{(z)}$ as a function of $E_{d}$ and the coupling strength $t$. The magnitudes are multiplied by the characteristic temperature $T_{[\star]}$, to scale the two-body and three-body correlators proportional to $\sim1/T^{2}_{K,\nu}$ and $\sim1/T^{4}_{K,\nu}$ (Eqs. (21-24)). $T_{[\star]}\chi_{(z)}(0)$ and $-T^{2}_{[\star]}\chi^{[3]}_{(z)}(0)$ correspond to the fluctuations of the frozen second and third moments of the pseudospin in CNTQD.
In the strongly correlated phase, we observed the Kondo cloud and $T_{[\star]}$ is equal to the Kondo temperature $T_{K}$.
As we have shown in Fig. 9a, $T_{[\star]}$ varies with the increase of the coupling strength $t$ and reaches characteristic values for a given phase. Fig. 17a presents $T_{[\star]}\chi_{(z)}$ for the SU(4) Kondo state, via  the intermediate phase to the SU$^{\star}$(3) Kondo effect. In the weak coupling regime, $T_{[\star]}\chi_{(z)}$ leads to $0.44$ (green line in Fig. 17c), and remains constant for the quantum states $\ket{q_{a(b)}}_{8}$ (purple area in Fig. 17a). With increasing $t$, $T_{[\star]}\chi_{(z)}$
reaches to the quantized value $1/4$. $T_{[\star]}\chi_{(z)}=1/4$ is related to the charge degeneracy line between two quantum states $\ket{q_{y}}_{6}$ and $\ket{q_{z}}_{2}$.
The transition between $\ket{p_{ls}}_{4}$ and the octuplet $\ket{q_{a}}_{8}$ is visible in the three-body correlators (Fig. 17b, d). The value of $-T^{2}_{[\star]}\chi^{[3]}_{(z)}$ changes from $0.63$ via $0.47$ for the weak coupling strength to $1/3$ for the strong coupling limit. In Figure 17b, we observe the color change in this region from orange to purple for $Q=1$e, and from black to violet for $Q=3$e, respectively. The response in the three-particle correlator $-T^{2}_{[\star]}\chi^{[3]}_{(z)}$ is much more sensitive for odd charges than in the two-body susceptibilities when the ground state configuration changes, because the high-order correlator, is the derivative of the two-particle correlators (Eq. (22) and Eq. (24)).

For $Q=2$e ($E_{d}=-4.5$) we observe in the two-body correlation a transition between $\ket{d_{\nu}}_{6}$ and
$\ket{q_{w}}_{12}$ (light and dark blue regions in Fig. 17a). $T_{[\star]}\chi_{(z)}$ changes from $0.63$ to $0.54$ at the transition line around $t_{1}$ (red line in Figure 17c). The three-particle correlation function for the e-h symmetry point is equal to zero (red line in Figure 17d), due to the oddity and mirror symmetry of $-T^{2}_{[\star]}\chi^{[3]}_{(z)}$. The three-body correlator changes the sign at this point. Even if, the transition is observed in $T_{[\star]}\chi_{(z)}$ (red line in Fig. 17c), the argument of the enhancement in the derivative is weaker than vanishing by reciprocal zeroing of the partial three-body correlators in $-T^{2}_{[\star]}\chi^{[3]}_{(z)}$. The dependence of $\widetilde{E}_{\nu(\nu')}$ is responsible for reducing the three-body correlators in Eqs. (22) and (24).
The early experiment confirmed and showed, that the three-body correlators are capable of being measured using the nonlinear current and the shot noise detection \cite{Hata2021}.
The important issue about the zeroing of higher-order susceptibilities suggests that they will be helpful in the experimental measurements, where the e-h symmetry point is shifted e.g. by the Zeeman compensating field in CNTQD Kondo dot attached to ferromagnetic electrodes \cite{Hauptmann2008}. In this device, the exchange field disappears at the e-h symmetry point, the spin Kondo state is restored, and this point is shifted by applying the compensating magnetic field.
In terms of the transition between $\ket{f}_{1}$($\ket{e}_{1}$), and $\ket{q_{x}}_{2}$($\ket{q_{z}}_{2}$), we observe around $t_{1}$ a slight boost in two-body correlation function and suppression in $-T^{2}_{[\star]}\chi^{[3]}_{(z)}$ (dark cyan line in Figures 17c, d).
For the fractional charges $Q=(3/2)$e and $Q=(5/2)$e, the SU$^{\star}$(3) Kondo state is formed in the strong coupling region, and both correlators, which are significant, are quantized to $T_{[\star]}\chi_{(z)}=|T^{2}_{[\star]}\chi^{[3]}_{(z)}|=0.41$. The system is determined by the two sextuplets $\ket{q_{y}}_{6}$ and $\ket{q_{g}}_{6}$. The equality of these quantities is due to the expectation values of the boson fields operators, where $|p_{+\downarrow(-s)}|^{2}=|d_{20,\uparrow\uparrow,\uparrow\downarrow}|^{2}=1/6$ ($|t_{+\uparrow(-s)}|^{2}=|d_{02,\downarrow\downarrow,\downarrow\uparrow}|^{2}=1/6$).

The figures 18 show the higher-order correlation functions for the CNTQD-2TSC device. For $Q=1$e, we observe a transition from $\ket{p_{ls}}_{4}$ to $\ket{q_{a}}_{16}$, and in the strong coupling region
to the quantum quartet $\ket{q_{z}}_{4}$. In the same way as before, it is difficult to separate the transition in the weak coupling regime, because $a'\ll a$ in the quantum state, and finally for $\ket{q_{a}}_{16}$ the SU(4)-like Kondo state is realized (even if ${\mathcal{G}}_{\nu}={\mathcal{G}}_{\nu'}=(1/2)(e^{2}/h)$, the transmissions are different). Although, in the same weak coupling region,
$-T^{2}_{[\star]}\chi^{[3]}_{(z)}$ demonstrates two values $0.63$ and $0.31$ at the transition line between the two states $\ket{p_{ls}}_{4}$ and $\ket{q_{a}}_{16}$ (green line around $t_{1}$ in Fig. 18d). The frozen effective Kondo spin features a constant value $T_{[\star]}\chi_{(z)}=0.45$. In the strong coupling to two Majorana fermions $\gamma_{l\uparrow}$, $T_{[\star]}\chi_{(z)}$ sloped down to $1/8$. The similar result is observed for $E_{d}=1$, where in both cases the quantum quartets $\ket{q_{z}}_{4}$ dominate (purple region in Fig.18a and dark cyan line in Fig. 18c). For $Q=2$e in the two-particle correlation function, we observe three characteristic values: $T_{[\star]}\chi_{(z)}=0.63$ for the sextuplet $\ket{d_{\nu}}_{6}$, $0.46$ for $\ket{q_{w}}_{24}$ and finally constant quantized value $0.31$ for the octuplet $\ket{q_{y}}_{8}$ (red line in Fig. 18c). $T_{[\star]}\chi_{(z)}=0.31$ is associated with the restoration of the SU$^{\star}$(2) Kondo state in the system.
In the density plot of $T^{2}_{[\star]}\chi^{[3]}_{(z)}$ we observe an enhancement at the transitions between the states $\ket{q_{z}}_{4}$ and $\ket{q_{a}}_{16}$, similarly for $\ket{q_{x}}_{4}$ and $\ket{q_{b}}_{16}$.
In the weak coupling strength region, the three-body correlation is quantized to $-T^{2}_{[\star]}\chi^{[3]}_{(z)}=\pm0.31$.
$T^{2}_{[\star]}\chi^{[3]}_{(z)}$ approaches to $1/3$ around $\ket{f}_{1}$, where empty $\ket{e}_{1}$ and fully occupied state $\ket{f}_{1}$ evolve to quartets, and strongly decreases with increasing of the number of the topological segments (dark cyan line in Fig. 18d). The opposite tendency is observed for two-body correlators, where $T_{[\star]}\chi_{(z)}$ saturates to a constant value (dark cyan lines in Figs. 17-19c).

Fig. 19 shows the evolution of the frozen effective pseudospin moment $T_{[\star]}\chi_{(z)}$ and three-body correlation $T^{2}_{[\star]}\chi^{[3]}_{(z)}$ versus $E_{d}$ with increasing the coupling strength $t$ for the CNTQD-3TSC hybrid device.
One of the most significant results is the shogun helmet-like gate dependence of $T_{[\star]}\chi_{(z)}$ and $T^{2}_{[\star]}\chi^{[3]}_{(z)}$ in the strong coupling region. Formally, for CNTQD-3TSC, the quantum conductance is suppressed by increasing the tunneling term between the Kondo dot and three Majorana fermions (Fig. 7d). The ${\mathcal{G}}$ in all three Kondo-like channels, is determined by the interference effect with Majorana fermions. The single normal channel preserves the quantized value ${\mathcal{G}}=(e^{2}/h)$ for $Q=2$e at the e-h symmetry point, where
the ground state is defined by two octuplets $\ket{q_{x}}_{8}$ and $\ket{q_{z}}_{8}$. Beyond this line, the total quantum conductance reaches $(3/2)(e^{2}/h)$ and the transport through the normal channel is blocked. In contrast to the static high-order correlations, where we observed the enhancement of $T_{[\star]}\chi_{(z)}$ and the sign reversal in $T^{2}_{[\star]}\chi^{[3]}_{(z)}$ around $E_{d}=-U$ and $E_{d}=-2U$. There is an amplification associated with a leakage of the quantum states $\ket{q_{w}}_{48}$, into the
forbidden charge region $Q=(5/2)$e and $Q=(3/2)$e, where $\ket{q_{x}}_{8}$ and $\ket{q_{z}}_{8}$ are the new energy ground states.
The effect requires further analysis, but we can tentatively conclude that the mechanism is due to  the presence in two octuplets $\ket{q_{x}}_{8}$ and $\ket{q_{z}}_{8}$ the states (the basis vectors) from the forbidden charge region in the normal phase.
We have marked in red two significant quantum states $\ket{2\uparrow n_{1}n_{2}n_{3}}$ and $\ket{0\downarrow n_{1}n_{2}n_{3}}$ in Eqs. (13-14), which are responsible for the quantum leakage. The physics behind this effect can be explained by the entanglement mechanism with opposite charge-leaking states.
\begin{figure}[t!]
\includegraphics[width=0.75\linewidth]{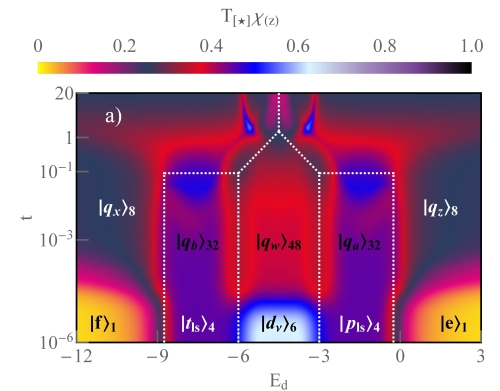}\\
\includegraphics[width=0.75\linewidth]{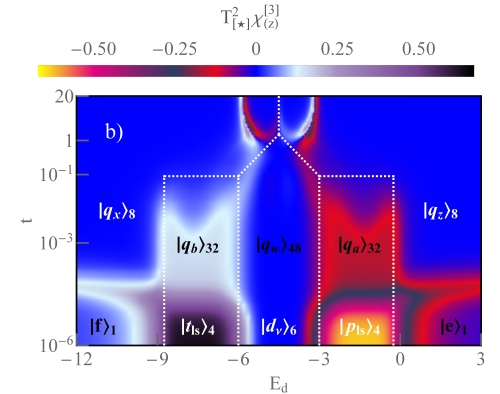}\\
\includegraphics[width=0.75\linewidth]{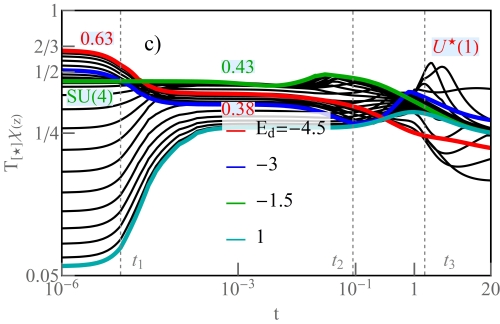}\\
\includegraphics[width=0.75\linewidth]{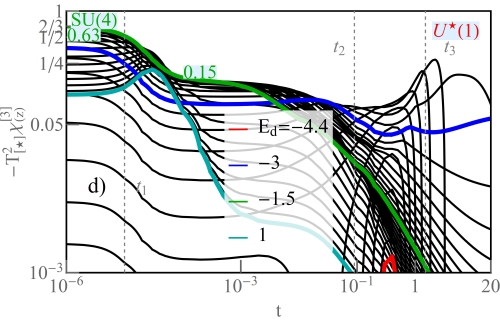}
\caption{\label{fig:epsart} (Color online) Two- and three-body correlations for the CNTQD-3TSC device: a, b) $T_{[\star]}\chi_{(z)}$ and $T^{2}_{[\star]}\chi^{[3]}_{(z)}$ as a function of $E_{d}$ and $t$. c, d) The landscape plots of $T_{[\star]}\chi_{(z)}$ and $-T^{2}_{[\star]}\chi^{[3]}_{(z)}$ as function of $t$.}
\end{figure}

The density plots in Fig. 19a, b show the transition with increasing of $t$ between the empty(fully) occupied states $\ket{e}_{1}$($\ket{f}_{1}$) and the low energy octuplets $\ket{q_{x(z)}}_{8}$. In the weak coupling regime for the high degenerate quantum states $\ket{q_{a(b)}}_{32}$, the frozen two-body susceptibilities lead to $0.43$ (green line in Fig. 19c). For the strong coupling strength, $T_{[\star]}\chi_{(z)}$ approaches to $1/4$ for both energies $E_{d}=-1.5$ and $E_{d}=1$, where one of the octuplet states dominates (green and dark cyan curves in Fig. 19c).
The red curve in Fig. 19c shows three characteristic values: $T_{[\star]}\chi_{(z)}=0.63$, $0.38$ and $0.15$.
We observed the enhancement of $T_{[\star]}\chi_{(z)}$ to $1/2$ above $t_{3}=U/2$. The charge-leaking states form the shogun helmet-like shape in the density plot of the susceptibilities.
Fig. 19d shows $-T^{2}_{[\star]}\chi^{[3]}_{(z)}$. For $E_{d}=-1.5$, the three-body correlation function changes from $0.63$ to $0.15$ at the transition line between $\ket{p_{ls}}_{4}$ and $\ket{q_{a}}_{32}$ (green curve in Fig. 19d). For the strong coupling region, where two octuplets $\ket{q_{z}}_{8}$ dominate, $-T^{2}_{[\star]}\chi^{[3]}_{(z)}$ is reduced, except for the charge-leaking lines (Fig. 19d).
In Fig. 19d, the sign of $\chi^{[3]}_{(z)}$ changes between $\pm1/4$ along the shogun helmet-like shape.
For $E_{d}=1$, with increasing the coupling strength between quantum dot and Majoranas, the device demonstrates a decrease in $-T^{2}_{[\star]}\chi^{[3]}_{(z)}$ to zero and a saturation in $T_{[\star]}\chi_{(z)}$ to $1/4$.

Figures 20 show the frozen higher-order quantities: the effective pseudospin $T_{[\star]}\chi_{(z)}$,
frozen charge susceptibility $T_{[\star]}\chi_{(c)}$ and its three-particle correlations $T^{[3]}_{[\star]}\chi_{(z)}$ and $T^{[3]}_{\star}\chi_{(c)}$ for CNTQD coupled to multi-Majorana device.
The charge susceptibility can be expressed as follows $\chi_{(c)}=\sum_{\nu\nu'}\widetilde{\chi}_{\nu\nu'}$, and in the analogous way the charge three-body correlator $T^{[3]}_{[\star]}\chi_{(z)}=\sum_{\nu\nu'}\widetilde{\chi}^{[3]}_{\nu\nu'\nu'}$.
Fig. 20a compares two-body correlators in three types of nanodevices as a function of $t$ for different gate voltages applied to CNTQD ($E_{d}$).
The dark, light and lighter lines are the results for CNTQD device coupled  to 1TSC, 2TSC and three Majoranas (3TSC).
For $E_{d}=-4.5$ and $E_{d}=-3$, the frozen effective pseudospin decreases with increase of the coupling strength to the topological segments. The quantum steps observed in $T_{\star}\chi_{(z)}$ correspond to ground states that determine the specific phase in the strongly correlated hybrid device. In the strong coupling limit, SU$^{\star}$(3) Kondo state is formed for the fractional charge $Q=(3/2)$e (dark blue line).
For the fractional Kondo phase, the two-body correlator leads to $T_{[\star]}\chi_{(z)}=0.41$ and converges to the same value at the e-h symmetry point, for $Q=2$e as mentioned before. At high degeneracy point of two sextuplets $\ket{q_{g}}_{6}$ and $\ket{q_{y}}_{6}$, the phase is determined by the $U^{\star}$(1) charge symmetry. The difference between two states, SU$^\star$(3) Kondo and charge $U^{\star}$(1) phase, is visible in the quantized value of $T_{[\star]}\chi_{(c)}=0$ and $T_{[\star]}\chi_{(c)}=1/3$. Formally, the frozen charge susceptibility is related to the sum of the charge fluctuations $\sum_{ls}\delta n_{ls}^{2}$, and for the screened Kondo cloud $T_{[\star]}\chi_{(c)}\approx0$ (dark blue and light red curves in Fig. 20b).
For $E_{d}=+1$ we observe a gradual increase of $T_{[\star]}\chi_{(z)}$ with the growth of the number of Majorana fermions $N_{TS}$ coupled to the CNTQD device (cyan lines in Fig. 20a). $T_{[\star]}\chi_{(z)}$ is the magnitude proportional to quadratic of the spin-orbital fluctuations (where $\chi_{(z)}=\sum_{\nu}\int^{1/k_{B}T}_{0}d\tau\langle\delta n_{\nu}(\tau)\delta n_{\nu}(0)\rangle^{<}-\sum_{\nu'\neq\nu}\int^{1/k_{B}T}_{0}d\tau\langle\delta n_{\nu}(\tau)\delta n_{\nu'}(0)\rangle^{<}$), which increases with $N_{TS}$ for $E_{d}=1$, in contrast to the local isospin Z-components shown in Fig. 12.
In the strong coupling region, for $E_{d}=-1.5$, $T_{\star}\chi_{(z)}$ converges to $1/4$ in CNTQD coupled to: 1TSC and 3TSC devices.
For the strongly coupled CNTQD-3TSC system, the frozen two-body correlator is reduced to $T_{[\star]}\chi_{(z)}=1/6$. $T_{[\star]}\chi_{(c)}$ vanishes for the SU(4), SU$^{\star}$(3) and SU$^{\star}$(2) Kondo states (Fig. 20b).
For $E_{d}=+1$, the two-body charge susceptibilities start from $1/8$ for uncoupled QD to TSCs. In the intermediate and strong coupling region, $T_{[\star]}\chi_{(c)}$ is enhanced by the number of topological wires $N_{TS}$ (cyan lines in Fig. 20b).
$T_{[\star]}\chi_{(c)}$ leads to $0.34$ for $\ket{q_{z}}_{2}$, $0.62$ for $\ket{q_{z}}_{4}$ and $T_{[\star]}\chi_{(c)}=0.87$ for $\ket{q_{z}}_{8}$.
In particular the partial susceptibilities indexed by $\nu'$ channels coupled to Majorana fermions for $E_{d}=+1$ raise the value of $T_{[\star]}{(c)}$. In the charge susceptibility we observe a dip around $t_{3}$ due to the opposite contributions of diagonal and off-diagonal parts in the sum of $\chi_{(c)}=\sum_{\nu\nu'}\widetilde{\chi}_{\nu\nu'}$.
Between $t_{1}$ and $t_{2}$ for $E_{d}=-4.5$ and $-1.5$, when CNTQD is determined by the high degenerate states, we observe low fluctuations of $T_{[\star]}\chi_{(c)}$, characterized for the Kondo states. Above $t_{2}$, for strong coupling strength, the two-body charge correlator increases and saturates at finite values, when the quantum states are in the U$^{\star}$(1) charge phases (all lines except the dark blue and light red curves in Fig. 20b).
The three-body correlation function $T^{[3]}_{[\star]}\chi_{(z)}$ versus $t$ is shown in Fig. 20c. $T^{[3]}_{[\star]}\chi_{(z)}$ shows three quantized values, related to the change of the quantum ground states (blue and green lines in Fig. 20c). In the strong coupling region, $T^{[3]}_{[\star]}\chi_{(z)}$ vanishes for the CNTQD-2TSC and CNTQD-3TSC devices, where the quartets $\ket{q_{z}}_{4}$ and the octuplets $\ket{q_{z}}_{8}$ determine the lowest energy solution in the system. The transition between the empty occupied state and the entangled doublet, quartet and octuplet for $E_{d}=1$ is manifested by raising the value around $t_{1}=\delta$. For the fractional SU$^{\star}$(3) Kondo state, $T^{[3]}_{[\star]}\chi_{(z)}$ is quantized to $0.41$ (dark blue line in Fig. 20c). We observed a similar evolution in the frozen three-body charge correlation $T^{[3]}_{\star}\chi_{(c)}$ (Fig. 20d). For CNTQD coupled to 2TSC and 3TSC, in particular for the weak coupling region, $T^{[3]}_{[\star]}\chi_{(c)}$ changes sign due to the dominant role of the off-diagonal three-body correlations (light and lighter green curves in Fig. 20d). At the e-h symmetry point, both correlations $T^{[3]}_{[\star]}\chi_{(z)}$ and $T^{[3]}_{[\star]}\chi_{(c)}$ are suppressed and obtains to zero.
\begin{figure}[t!]
\includegraphics[width=0.75\linewidth]{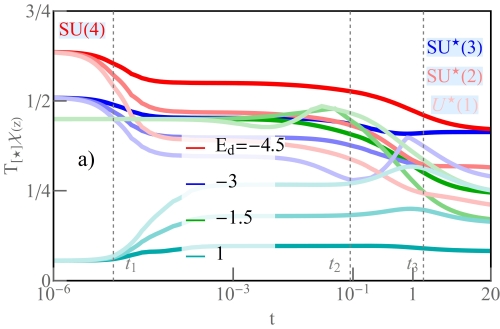}\\
\includegraphics[width=0.75\linewidth]{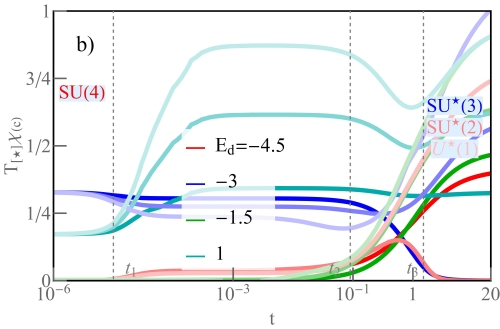}\\
\includegraphics[width=0.75\linewidth]{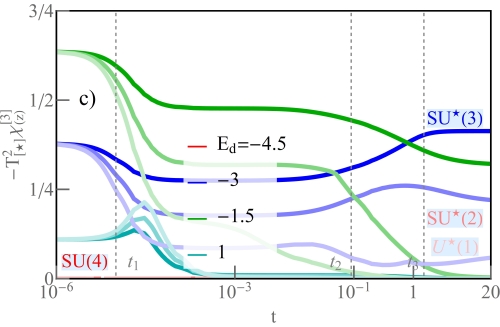}\\
\includegraphics[width=0.75\linewidth]{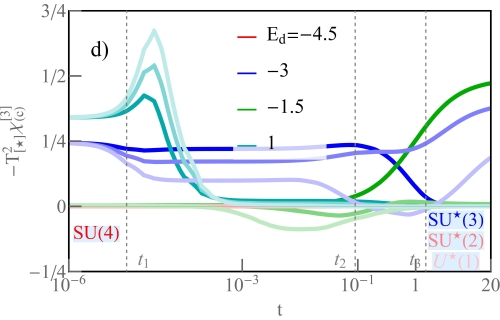}
\caption{\label{fig:epsart} (Color online) a, b) $T_{[\star]}\chi_{(z)}$ and the frozen charge susceptibilities $T_{[\star]}\chi_{(c)}$
against $t$. c, d) $-T^{2}_{[\star]}\chi^{[3]}_{(z)}$ and $-T^{2}_{[\star]}\chi^{[3]}_{(c)}$ as a function of the coupling strength $t$. Dark, light and brightest color of the lines show the results for CNTQD coupled to MF, 2MFs and 3MFs.}
\end{figure}
\begin{figure}[t!]
\includegraphics[width=0.75\linewidth]{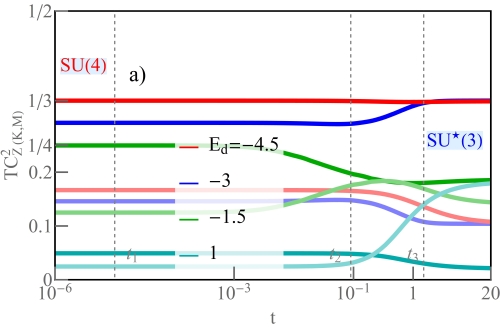}\\
\includegraphics[width=0.75\linewidth]{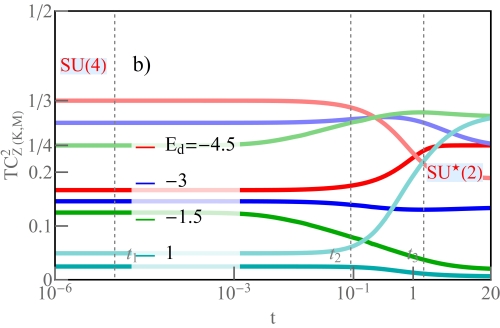}\\
\includegraphics[width=0.75\linewidth]{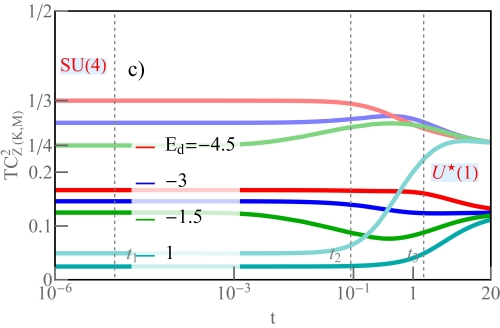}
\caption{\label{fig:epsart} (Color online) a-c) The expected value of the quadratic of the Casimir operator $C^{2}_{Z}$ as a local spin fluctuations versus $t$ for CNTQD-TSC, CNTQD-2TSC and CNTQD-3TSC devices. Dark and light lines present $TC^{2}_{Z(K)}$ and $TC^{2}_{Z(M)}$ in Kondo and Majorana-coupled channels.}
\end{figure}
\begin{figure}[t!]
\includegraphics[width=0.75\linewidth]{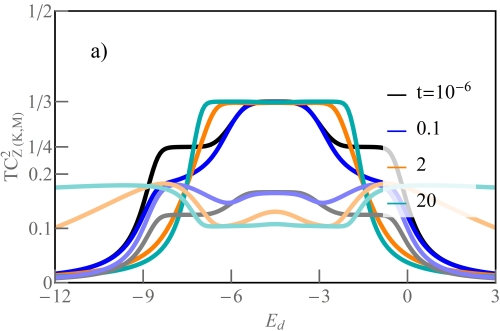}\\
\includegraphics[width=0.75\linewidth]{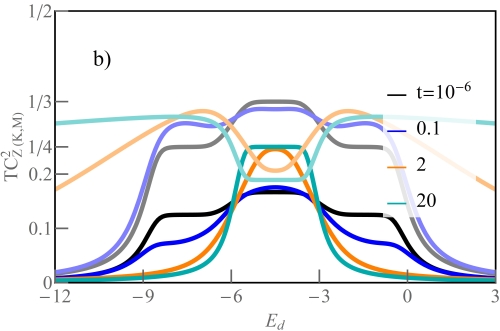}\\
\includegraphics[width=0.75\linewidth]{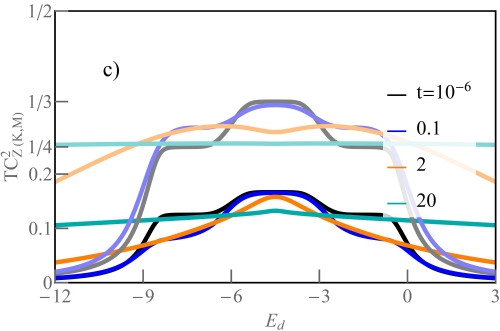}
\caption{\label{fig:epsart} (Color online) a-c) The expected value of the quadratic of the Casimir operator $C^{2}_{Z}$ as a function of $E_{d}$ with increasing of $t$ for the CNTQD-TSC, CNTQD-2TSC and CNTQD-3TSC devices. Dark and light lines present $TC^{2}_{Z(K)}$ and $TC^{2}_{Z(M)}$ in Kondo and Majorana-coupled channels.}
\end{figure}

Figures 21 and 22 show the fluctuations of the local effective pseudospin $TC^{2}_{Z}=TC^{2}_{Z(K)}+TC^{2}_{Z(M)}$, expressed by the
expected value of the Z-component of the quadratic Casimir operator \cite{Iachello2007}. The quadratic Casimir operator is written as a sum of two parts $C^{2}_{Z(K)}$ and $C^{2}_{Z(M)}$ (dark and light lines in Figs. 21-22). The first contribution describes the local fluctuations in the $\nu$ channel associated with the normal states and the second $C^{2}_{Z(M)}$, represents  the two-particle correlations in the $\nu'$ Majorana-coupled channel. We have expressed the Casimir operator by the boson fields operators in Eqs. (29-34). For the CNTQD-TSC device the fluctuations in the normal channels are higher than the contribution in the $\nu'$ channel: $TC^{2}_{K}>TC^{2}_{M}$. For $E_{d}=-4.5$, $TC^{2}_{Z(K)}=1/3$ and is constant with increasing the tunneling strength, in contrast to $TC^{2}_{Z(M)}$, which changes from $0.16$ to $0.1$ in the strong coupling limit (dark red and light red curves in Fig. 21a). For this behavior is responsible the transition from $\ket{q_{w}}_{12}$ in the Kondo state to the U$^{\star}$(1) charge symmetry (dark and light cyan lines in Fig. 22a). For both quantum states $\ket{q_{g}}_{6}$ for $Q=(5/2)$e and $\ket{q_{y}}_{6}$ for $Q=(3/2)$e, $TC^{2}_{Z(K)}$ reaches $1/3$, and $TC^{2}_{Z(M)}=0.1$, which is the fingerprint of the SU$^{\star}$(3) Kondo state \cite{Zitko2013}. In practice, the states in the $\nu$ channel are screened by the nonlocal fluctuation $TC^{2}_{s-d}=\lim_{T\mapsto0}T\chi_{(z)}(T)-TC^{2}_{Z(K)}(T)=-1/3$ consisting of the quantum dot and electrode states ($TC^{2}_{s-d}$ is called the nonlocal quadratic Casimir operator). In $T\chi_{(z)}(T)$, in the strong coupling limit, with increasing temperature we should observe two quantum steps, first for the Majorana-coupled channel $\nu$, where $T\chi_{(z)}(T)=0.1$ and second for the high temperature limit $T\chi_{(z)}(T)=0.1+1/3$, in contrast to the entropy, where the tunneling entropy is compensated by the topological part (Fig. 13c). For $E_{d}=+1$, the transport is determined by the Majorana-coupled channel, and $TC^{2}_{Z(M)}>TC^{2}_{Z(K)}$. In the strong coupling limit, the $TC^{2}_{Z(M)}$ saturates and reaches to $0.18$ (light cyan lines in Fig. 21a and 22a). In this region, $\ket{q_{z}}_{2}$ is the ground state and the spin and isospin components achieve $|S_{Z}|=1/4$ and $|I_{Z}|=3/4$ (Fig. 12a).

In CNTQD-2TSC, the tendencies of the fluctuations are opposite $TC^{2}_{Z(M)}>TC^{2}_{Z(K)}$, except for $E_{d}=-4.5$, in the strong coupling limit, where the SU$^{\star}$(2) Kondo state is realized. With increasing the coupling strength $t$, for $Q=2$e, we observed the transition between the Kondo state for $\ket{q_{w}}_{12}$ and SU$^{\star}$(2) Kondo effect. This occurs, when $TC^{2}_{Z(M)}<TC^{2}_{Z(K)}$ and the fluctuation in the Majorana-coupled channel achieves $TC^{2}_{Z(M)}=0.2$ and in the channel associated with the Kondo state $TC^{2}_{Z(K)}=1/4$ (dark and light cyan lines in Fig. 22b and both red curves in Fig. 21b). For these quantities are responsible the expected values of the boson fields operators Eqs. (31-32). As we see, in the topological qubit states $\ket{q_{y}}_{8}$ in Eq. (12), only two bosons of $d_{\nu}$, single $p_{ls}$ and one $t_{ls}$ contribute to the increase of the fluctuations. The spin and the isospin are zero for the SU$^{\star}$(2) Kondo effect, due to the symmetry of the strongly correlated state and are determined by the values of the boson fields operators in Eq. (44-45). Topological states in the octuplets $\ket{q_{y}}_{8}$ contribute to the increase of the total entropy for the intermediate temperature (black line in Fig. 13d). For $\ket{q_{z}}_{4}$ states, in the strong coupling region for $E_{d}=-1.5$ and $E_{d}=+1$, the transport is determined by the channels coupled to Majorana fermions (light green and cyan lines in Fig. 21b), and only $TC^{2}_{Z(M)}=0.29$ contributes to the pseudospin fluctuations (light cyan line in Fig. 22b). In this region we observe a sharp switch in the Z-component of the spin and isospin, $|S_{Z}|=|I_{Z}|=1/2$ (Fig. 12b). The topological qubits $\ket{q_{-z_{n(\overline{n})}}}$ are determined by the entanglement of the empty, two single occupied states and one double occupied state in the auxiliary slave boson representation (Eq. (11)).

For the CNTQD-3TSC device, the contributions from the $\nu'$ channels coupled to the three Majorana fermions exceed the fluctuation from the normal channel: $TC^{2}_{Z(M)}>TC^{2}_{Z(K)}$ (Fig. 21c). In the strong coupling limit, the octuplet states $\ket{q_{x}}_{8}$ and $\ket{q_{z}}_{8}$ dominate on the both sides of the e-h symmetry point. With increasing $t$ we observe the completely flat and constant gate-dependent behavior for both fluctuations  $TC^{2}_{Z(M)}=1/4$ and $TC^{2}_{Z(K)}=0.1$ (light and dark cyan curves in Fig. 22c). This is caused by the fact that topological superconductors use $N-1$ degrees of freedom of the QD.

Let's turn to the problem of the shot noise in the CNTQD-TSC devices.
According to the results in the paper \cite{Oguri2022}, the current and the shot noise (in terms of linear voltages) can be expressed by the transmission in the Landauer-B\"{u}tikker form Eqs. (37-38). The shot noise formula is derived using the Hartree-Fock approximation (HFA) for a two-particle Green's function \cite{Kadanoff1962,Aguado2004}.
Finally, in the context of the shot noise (the fluctuations of the current), we can introduce the linear Fano factor $F_{0}=\lim_{V\mapsto0}S_{0}/2|e|I_{0}$ in the low bias region. The quantum magnitude for certain quantum systems yields to the quantized values less than or greater than 1. For the non-interacting particles, $F_{0}$ is equal to 1.
$F_{0}$ manifests two types of statistical behavior: sub-Poissonian noise ($F_{0}<1$) and super-Poissonian noise ($F_{0}>1$).
The particles in the nanodevice (mostly electrons) are bunched or anti-bunched, by the repulsive and attractive Coulomb interaction.
The type of interaction between the quasiparticles determines the value of the Fano factor, e.g. Cooper pairs in BCS superconductors demonstrate $2$, the Fano factor of the Dirac fermions is equal to $1/3$, and for the Kondo singlet quasiparticle, the experiments show $5/3$. In the CNTQD-TSCs device, we couple the Kondo quasiparticle to Majorana fermions (real half-fermion state), and by increasing of the tunneling terms, the Kondo cloud is modified by the interference effects. We observe the coexistence of two states, the Majorana-Kondo state.
In the first part of the discussion, we analyze the ballistic transport (for low bias voltage), where the current fluctuation $S_{0}$ is described by the transmission. In this picture, the quasiparticle in the Kondo state, behaves like non-interacting particles (in particular for the SU(2) Kondo effect, where $\delta_{\nu}=\pi/2$). This quantity is observed in the linear coefficient of the specific heat, where in the frame of FL theory $\gamma_{N}=\frac{\pi^{2}}{3}\sum_{\nu=ls}\widetilde{\varrho}_{\nu}$ \cite{Nishikawa2013}. In general, the non-interacting particles are modified by the renormalization, in this sense the symmetry of the SU(4) Kondo effect and the interaction reveal in low-bias measurements.
Finally, the Fano factor consists of the linear and nonlinear parts in the following form:
\begin{eqnarray}
&&F=\frac{S}{2|e|I}=\frac{S_{0}V+S_{K}V^{3}+0[V^{5}]}{2|e|(I_{0}V+I_{K}V^{3}+0[V^{5}])}
\end{eqnarray}
where the nonlinear part is measured by subtracting the linear parts $S_{0}$ and $I_{0}$ from the noise and the currents:
\begin{eqnarray}
&& F_{K}=\frac{|S_{K}|}{2|e||I_{K}|}=\frac{\delta S_{K}}{\delta I_{K}}\approx\frac{d^{2}S/dV^{2}}{d^{2}I/dV^{2}}
\end{eqnarray}
$F_{K}$ is the nonlinear contribution to the shot noise and includes the elastic and inelastic scattering processes, which develops from the high-order correlations and the interaction between dressed Kondo quasiparticles ($eV,T<T_{K}$). In the further calculations, we have adopted the Eqs. (39-40) from \cite{Oguri2022}.

Fig. 23a shows the density plot of $F_{0}$ as a function of $E_{d}$ and the coupling strength $t$. In Fig. 23a, for the CNTQD-1TSC device, with respect to the transition from the empty (full) occupied states $\ket{e}_{1}$($\ket{f}_{1}$)
to the doublets $\ket{q_{z}}_{2}$ ($\ket{q_{x}}_{2}$), we observe a reduction of the shot noise from $F_{0}=1$ (red area in Fig.23a) to a value of $1/2$ (dark orange region in Fig.23a). Comparing the figures in section 23, the reduction is quantized to $F_{0}=1/2$, regardless of the number of TSC segments in the hybrid devices. This is a consequence of the geometry of the measurements (T-shaped like device). The $N_{TS}$ dependence in $F_{0}$ is detectable, only in the direct coupling geometry, where the TSC is one of the transport electrodes, where we measure the Andreev reflection contributions to the shot noise \cite{Lutchyn2014}. Formally, in the systems with direct coupling, where the quartet $\ket{q_{x(z)}}_{4}$ and the octuplet $\ket{q_{x(z)}}_{8}$ are the ground states, we should observe $F_{0}\gg1$ in the strong coupling limit.
\begin{figure}[t!]
\includegraphics[width=0.75\linewidth]{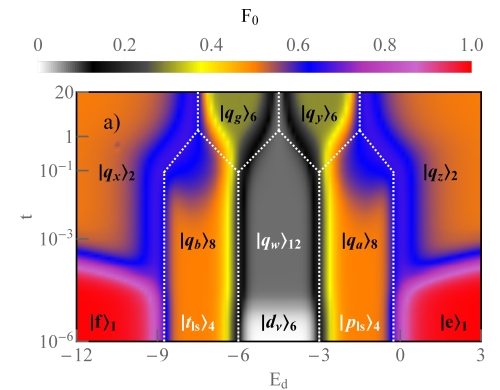}\\
\includegraphics[width=0.75\linewidth]{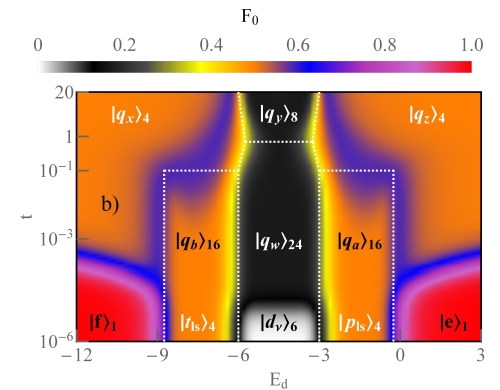}\\
\includegraphics[width=0.75\linewidth]{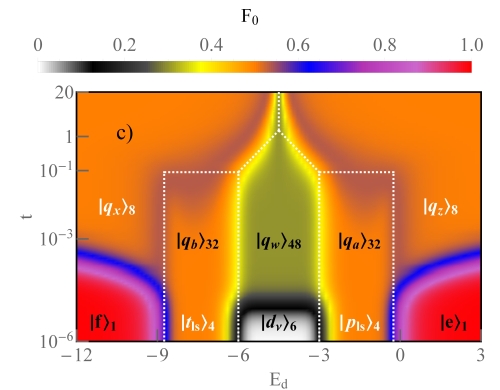}
\caption{\label{fig:epsart} (Color online) a-c) The density plot of the linear Fano factor $F_{0}=S_{0}/(2eI_{0})$ versus $E_{d}$ and $t$ for the CNTQD-MF, CNTQD-2MFs and CNTQD-3MFs systems.}
\end{figure}
\begin{figure}[t!]
\includegraphics[width=0.75\linewidth]{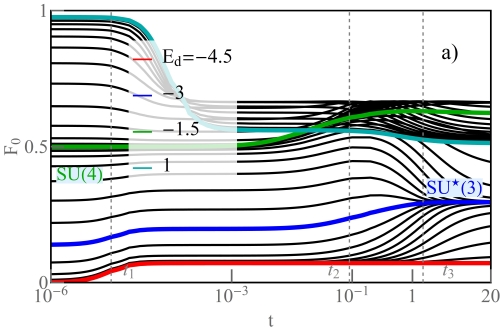}\\
\includegraphics[width=0.75\linewidth]{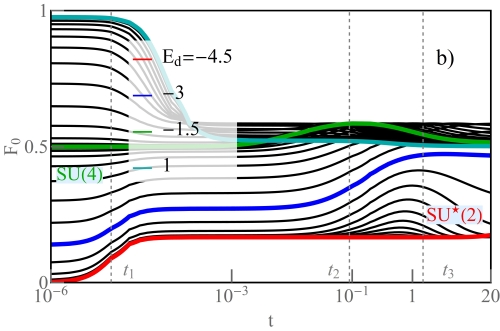}\\
\includegraphics[width=0.75\linewidth]{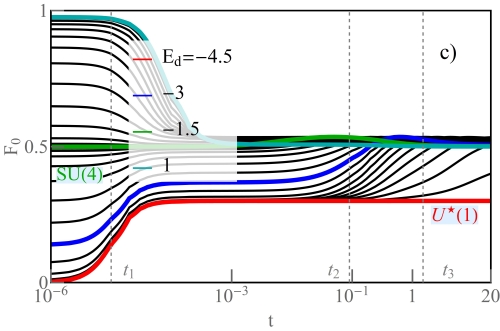}
\caption{\label{fig:epsart} (Color online) a-c) The landscape plots of $F_{0}$ as a function $t$ with increment $\delta E_{d}=0.15$ for CNTQD-TSC, CNTQD-2TSC and CNTQD-3TSC devices.}
\end{figure}

$F_{0}=1/2$ has been confirmed for the QD-TSC circuit \cite{Lutchyn2014,Smirnov2015}, and is related to the fact that the sum of transmissions in the Majorana-coupled channel reaches $(1/2)(e^{2}/h)$. One comment is necessary here, $F_{0}$ leads to $1/2$, only beyond the e-h symmetry point, at the charge degeneracy line the device is in the U$^{\star}$(1) charge symmetry phase, the quantum conductance is equal to ${\mathcal{G}}=(3/2)(e^{2}/h)$ and the linear Fano factor converges to $1/6$ (the increase of the quantum conductance is also confirmed by NRG calculations \cite{Weymann2017}).

The difference between $F_{0}$ in CNTQD-TSCs devices is significant in the measurements of $S_{0}$ and $I_{0}$. $F_{0}=1/2$ has a different source for  the CNTQD-1TSC circuit. For $\ket{q_{x(z)}}_{2}$ we observe the noise $S_{0}=(1/4)(e/h)$ and the current quantized to $2|e|I_{0}=(1/2)(e/h)$. For CNTQD-2TSC, $F_{0}=1/2$ is related to two quartets $\ket{q_{x(z)}}_{4}$, where $S_{0}=(1/2)(e/h)$ and $2|e|I_{0}=1(e/h)$. For a system coupled to three Majorana fermions $F_{0}=1/2$, but $S_{0}=(3/4)(e/h)$ and $2|e|I_{0}=(3/2)(e/h)$, and the transport is determined by two octuplets $\ket{q_{x(z)}}_{8}$. The value $F_{0}=1/2$ is observed for $t>t_{1}$ and is symbolized by the orange area in Fig. 23 and by the dark cyan lines in Figures 24a-c. The black lines in Fig. 24 are plotted in the range from $E_{d}=-4.5$ to $1$ with the increment $\delta E_{d}=+0.15$.

Let us discuss $F_{0}$ for the e-h symmetry point ($E_{d}=-4.5$). Increasing $t$ we observe the transition in $F_{0}$ between the quantum state $\ket{d_{\nu}}_{6}$, where $F_{0}=0$, to the finite quantized values, i.e. $F_{0}=\frac{S_{0}}{2|e|I_{0}}=\frac{1/4}{7/2}=1/14$ for $\ket{q_{w}}_{12}$, $F_{0}=\frac{1/2}{3}=1/6$ for $\ket{q_{w}}_{24}$ and $F_{0}=\frac{3/4}{5/2}=3/10$ for $\ket{q_{w}}_{48}$ (Fig. 23 and red curves in Fig. 24). $F_{0}=1/6$ is related to the SU$^{\star}$(2) Kondo state, $F_{0}=1/14$ and $F_{0}=3/10$ are determined by the U$^{\star}$(1) charge symmetries with twelvefold and sixteenfold degeneracy point. For the fractional SU$^{\star}$(3) Kondo state, $F_{0}=13/44$, where the noise is quantized to $S_{0}=(13/16)(e/h)$ and the current leads to $2|e|I_{0}=(11/4)(e/h)$ (dark yellow region in Fig. 23a and blue line in Fig. 24a).
For the quantum states with $Q=1(3)$e on the dot, in the range of weak coupling strength $F_{0}\approx1/2$ (orange region in Fig. 23 and green lines in Fig. 24). The transition between the states $\ket{p_{ls}}_{4}$($\ket{t_{ls}}_{4}$) and $\ket{q_{a}}_{8(16,32)}$($\ket{q_{b}}_{8(16,32)}$)  appears in the shot noise, exactly like in the quantum conductances (Figs. 5-7a). Finally, we can conclude that increasing the number of $N_{TS}$ topological superconducting wires coupled to the CNTQD  squeezes the Fano factor in the landscape plots to $1/2$ in the limit $N_{TS}=N$.

In the $V^{3}$ bias range, the nonlinear Fano factor dominates, which consider the information of the interaction between Kondo quasiparticles.
The effective sbMFA Hamiltonian, includes the Coulomb interaction, but formally to calculate the inelastic scattering processes, we take into account the fluctuations in the free energy $\Delta\widetilde{F}$. On this basis we can introduce into the system the quantity called the Wilson coefficient, which for the interacting quasiparticles leads to $W_{\nu\nu'}-1>0$, and exactly in the backscattering processes with the Kondo singlet, takes the value of $1/(N-1)$ for the N-orbital Anderson model \cite{Oguri2018,Oguri2022}.
CNTQD-TSC devices with N-1 channels coupled to the Majorana fermion behave like the Fermi liquid systems, where $n_{\nu}=\delta_{\nu}/\pi$.
Based on this assumption, and on the fact that the shot noise and the current have similar linear and the third power bias contributions, we can formally adopt in the first approach the result of the paper \cite{Oguri2022}. Previous papers, using the Fermi liquid approach, showed that the nonlinear current and shot noise can be expressed by the sum of the elastic and inelastic contributions \cite{LeHur2009,Mora2008,Mora2009}. In \cite{Oguri2022} the authors
showed for the first time that the FL coefficient can be expressed by the two- and three-body correlation functions ($\chi_{\nu\nu'}$ and $\chi^{[3]}_{\nu\nu'\nu'}$), and the factors $c_{V,\nu}$ and $c_{S,\nu}$ at $V^{3}$ in the current and shot noise series are given by the sum of the high-order correlation functions consisting of the backscattering processes found by using the Ward identities \cite{Zawadowski1978} and including the vertex functions \cite{Oguri2022}.

Figure 25 shows the shot noise and current in the nonlinear regime for $V<T_{K}$, which includes the information about the quasiparticle interaction in the Kondo state. The quantities are expressed in terms of two- and three-particle correlation functions. The $\delta I_{K}=T_{[\star]}^{2}|I_{K}|$ is the excess current multiplied by the square of the characteristic temperature $T_{[\star]}$. The two-body correlation functions determine the $\delta I_{K}$. The static susceptibilities are expressed by Eqs. (21-24), and are thus inversely proportional to the square of $T_{K}$, so that the characteristic temperature scales the quantities to quantum values in the range between $0$ and $1$ for $\delta I_{K}$, and from $0$ to $3/2$ for $\delta S_{K}=T_{[\star]}^{2}|S_{K}|$. The $\delta I_{K}$ is directly expressed by the sum of the factors $c_{V,\nu}$ and $c_{V,\nu'}$ for the uncoupled and coupled channels to the topological superconductor. The two-particle and three-particle scattering in $c_{V,\nu}$ have opposite signs, and in certain ranges of parameters $t$ and $E_{d}$ the processes can be equivalent. Figure 25a shows the density plot of $\delta I_{K}$ and Fig. 25b presents the landscape plot of the nonlinear current as a function of $t$. The $\delta I_{K}$ in Figure 25b is plotted with an increment of $\delta E_{d}=0.15$. The colored lines represent $\delta I_{K}$ in the different ground state regions. For $E_{d}=1$ (dark cyan line in Fig. 25b), with increasing $t$ we evolve from the empty state $\ket{e}_{1}$ to the doublet $\ket{q_{z}}_{2}$. The $\delta I_{K}$ in terms of the weak and strong coupling regime to the Majorana fermion takes the value $\delta I_{K}\approx0.08$ and for $t=0$ the value $\delta I_{K}=0.018$. Between intermediate and strong coupling strength we observe a clear point where the current is extinguished $\delta I_{K}=0$ (yellow line in Figure 25a and dark cyan curve in Figure 25b). Above this point, the contribution of the two-particle correlation dominates over the three-particle correlation, and $\delta I_{K}$ reverses the sign. For $E_{d}=-1.5$ (green line in Fig. 25b), we observe three characteristic values of $\delta I_{K}=0.33$ for the quartet $\ket{p_{ls}}_{4}$, $\delta I_{K}=0.25$ for the octuplet $\ket{q_{a}}_{8}$ and $\delta I_{K}\mapsto0$ at the boundary of the charge areas $Q=(3/2)$e and $Q=(1/2)$e. For $Q=2$e ($E_{d}=E_{e-h}$, red line in Fig. 25b), the current takes on the values: $\delta I_{K}=0.89$ for the $\ket{d_{\nu}}_{6}$ states, $\delta I_{K}=0.69$ for the duodecuplet states $\ket{q_{w}}_{12}$ and $\delta I_{K}=0.525$ in the strong coupling regime at the degeneracy point between the sextuplets $\ket{q_{g}}_{6}$ and $\ket{q_{y}}_{6}$. For SU$^{\star}$(3) (the blue line in Fig. 25b), the nonlinear current reaches a value of $\delta I_{K}=0.515$ for the ground state $\ket{q_{y}}_{6}$. Fig. 25a shows that the zeroing of the current occurs at the boundary between the doublets $\ket{q_{x}}_{2}$ and $\ket{q_{z}}_{2}$.

Figures 25c and 25d illustrate the nonlinear shot noise $\delta S_{K}=T_{[\star]}^{2}|S_{K}|$ as a function of $E_{d}$ and $t$. In contrast to the current factor Eqs. (39-40), the number of scattering processes in $\delta S_{K}$ leads to a double line of the reduction of the shot noise (light yellow lines in Fig. 25c). The $\delta S_{K}$ is a fluctuation of the current, called second cumulant \cite{Nazarov2009}, which zeroes predominantly at other points than the current. For $E_{d}=+1$, $\delta S_{K}$, has double zero points and the nonlinear shot noise between the compensation points is negative $\delta S_{K}<0$ (dark cyan line in Fig. 25d). This follows from a simple fact, the coefficient $c_{S,\nu}$ is expressed by doubling the phase shift $2\delta_{\nu}$ and quadrupling $4\delta_{\nu}$ in the trigonometric functions. The zeroing effect, as before, occurs on the boundary with the doublet states $\ket{q_{x(z)}}_{2}$.
The $\delta S_{K}$ for $E_{d}=+1$, with increasing the coupling strength takes values of $\delta S_{K}=0.016$ for the empty state $\ket{e}_{1}$, in the weak and strong coupling range it is $\delta S_{K}=0.07$. A pronounced suppression of the noise $\delta S_{K}=0$ occurs for the two coupling values $t=0.25$ and $t=1.5$ (yellow line in Figure 25c, and dark cyan curve in Figure 25d). At these points, the transport is noiseless. The $\delta S_{K}<0$ occurs in the interval between these specific lines. The magnitude of the negative noise, is affected by the static three-particle correlators, which increases the negative value of the shot noise.
This is a very interesting result considering that the noise is a variance of the nonlinear current \cite{Nazarov2009}. For $E_{d}=-1.5$ (the green line in Fig. 25d), we observe a constant value of $\delta S_{K}=0.11$ for the quartet $\ket{p_{ls}}_{4}$ and the octuplet $\ket{q_{a}}_{8}$. The $\delta S_{k}=0$, at $t=4.5$, and the noise takes on a negative value of $\delta S_{K}=-0.013$ in the strong coupling regime, i.e. for the quantum state $\ket{q_{z}}_{2}$, with dominant role of the three-body correlators.
For $Q=2$e ($E_{d}=E_{e-h}$, red line in Fig. 25d), the shot noise is quantized to $\delta S_{K}=1.33$ for $\ket{d_{\nu}}_{6}$, $\delta S_{K}=1.02$ for the duodecuplet quantum state $\ket{q_{w}}_{12}$ and reaches $0.74$ for the strong coupling regime at the degeneracy point of two sextuplets $\ket{q_{g}}_{6}$ and $\ket{q_{y}}_{6}$.
The $\delta S_{K}$ for $E_{d}=-3$ (blue line in Fig. 25d) at the boundary of the quantum states of $\ket{d_{\nu}}_{6}$, $\ket{p_{ls}}_{4}$ and $\ket{q_{w}}_{12}$, $\ket{q_{a}}_{8}$ takes on values of $0.44$ and $0.35$. Finally, for the SU$^{\star}$(3) Kondo state, the shot noise reaches $\delta S_{K}=0.35$. The blue line shows a characteristic minimum around $t\approx0.25$, most likely due to a quantum state transition from U$^{\star}$(1) charge symmetry to threefold SU$^{\star}$(3) symmetry with a reduction of $c_{S}$ by contributions from the three-body correlation functions. This is the main argument why the three-particle interactions are significant in the current and shot noise for the states with broken symmetry.
\begin{figure}[t!]
\includegraphics[width=0.75\linewidth]{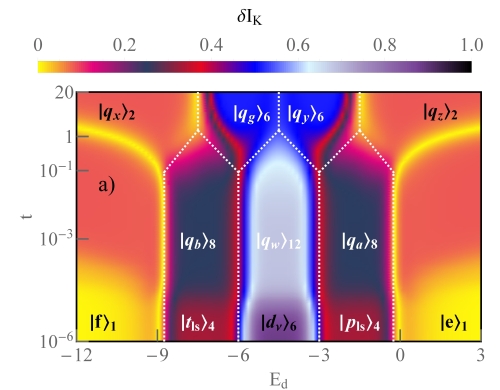}\\
\includegraphics[width=0.75\linewidth]{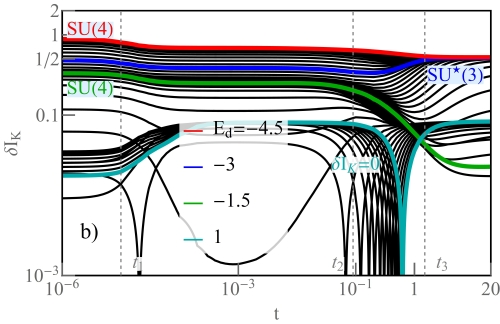}\\
\includegraphics[width=0.75\linewidth]{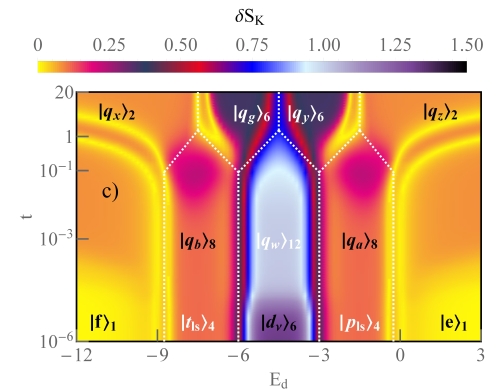}\\
\includegraphics[width=0.75\linewidth]{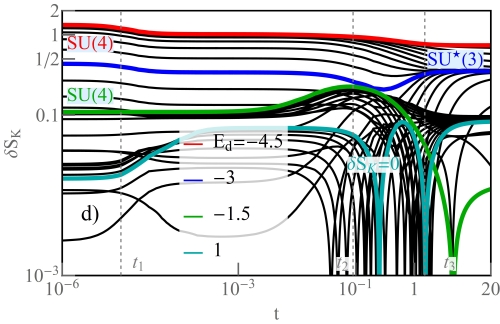}
\caption{\label{fig:epsart} (Color online) CNTQD-TSC device: a, c) The density plot of the rescaled nonlinear current $\delta I_{K}=T^{2}_{[\star]}I_{K}$ and shot noise $\delta S_{K}=T^{2}_{[\star]}S_{K}$ versus $E_{d}$ and $t$. b, d) The landscape log-log plot of $\delta I_{K}$ and $\delta S_{K}$ as a function of $t$ with increment $\delta E_{d}=0.15$. Yellow lines indicate the vanishing of $\delta I_{K}$ and $\delta S_{K}$, respectively.}
\end{figure}
\begin{figure}[t!]
\includegraphics[width=0.75\linewidth]{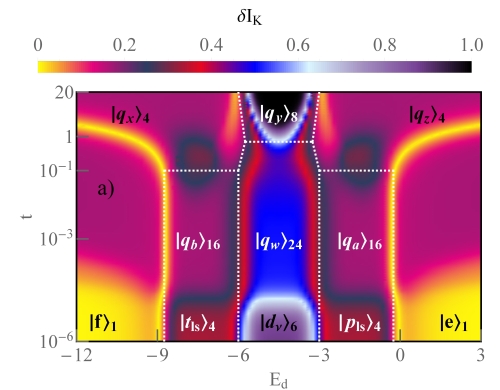}\\
\includegraphics[width=0.75\linewidth]{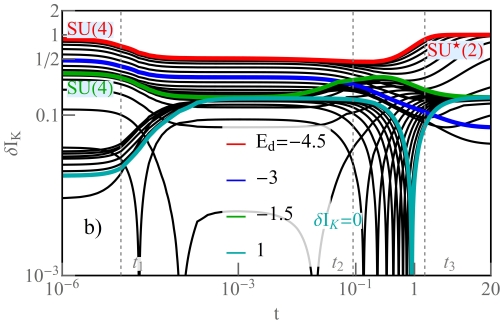}\\
\includegraphics[width=0.75\linewidth]{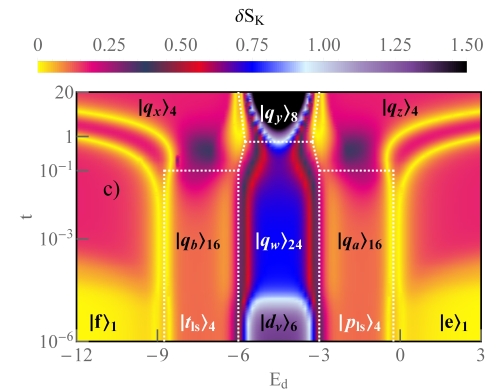}\\
\includegraphics[width=0.75\linewidth]{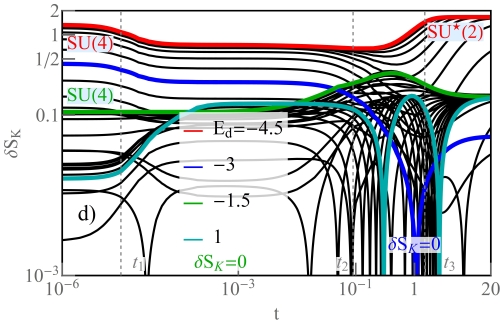}
\caption{\label{fig:epsart} (Color online) CNTQD-2TSC device: a, c) The density plot of the rescaled nonlinear current $\delta I_{K}$ and shot noise $\delta S_{K}$ versus $E_{d}$ and $t$. b, d) The landscape log-log plot of $\delta I_{K}$ and $\delta S_{K}$ as a function of $t$ with an increment of $\delta E_{d}=0.15$.}
\end{figure}
\begin{figure}[t!]
\includegraphics[width=0.75\linewidth]{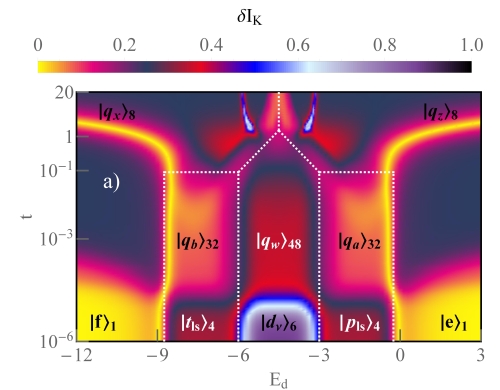}\\
\includegraphics[width=0.75\linewidth]{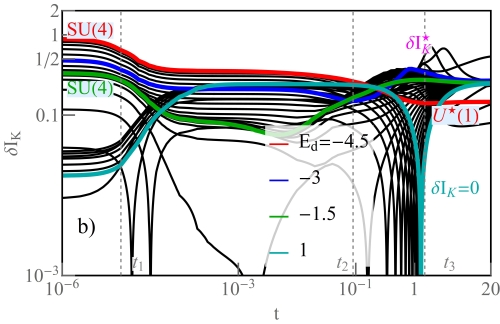}\\
\includegraphics[width=0.75\linewidth]{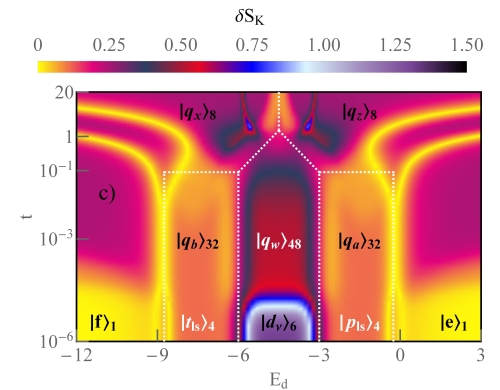}\\
\includegraphics[width=0.75\linewidth]{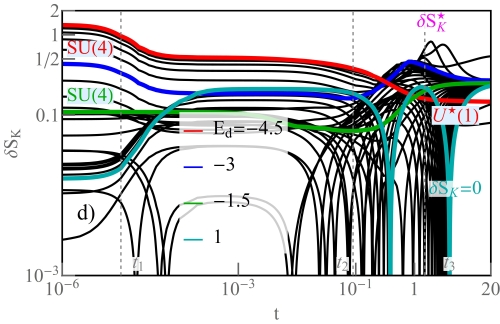}
\caption{\label{fig:epsart} (Color online) CNTQD-3TSC system: a, c) The density plot of the rescaled nonlinear current $\delta I_{K}$ and shot noise $\delta S_{K}$ versus $E_{d}$ and $t$. b, d) The landscape log-log plot of $\delta I_{K}$ and $\delta S_{K}$ as a function of $t$ with increment $\delta E_{d}=0.15$.}
\end{figure}

Fig. 26 shows the shot noise and current in the nonlinear regime for the CNTQD-2TSC device. For $E_{d}=+1$ (dark cyan line in Fig. 26b), $\delta I_{K}$  reaches $0.16$ with increasing $t$, where the quantum state changes from the $\ket{e}_{1}$ to the quartet $\ket{q_{z}}_{4}$. The $\delta I_{K}$ between the weak and strong coupling range to two Majorana fermions shows a pronounced extinction for $t=1$ ($\delta I_{K}=0$ presents yellow line in Figure 26a, and dark cyan curve in 26b). The mechanism for changing the sign is identical to the CNTQD-1TSC system. For $E_{d}=-1.5$ (green line in Fig. 26b), we observe three characteristic values of $\delta I_{K}$: $0.33$ for the quartets $\ket{p_{ls}}_{4}$ (this corresponds to the SU(4) Kondo symmetry), $\delta I_{K}=0.16$ for $\ket{q_{a}}_{16}$ and $\ket{q_{z}}_{4}$. It is interesting to note that for $\delta I_{K}=1/3$ and $\delta S_{K}=1/10$ the Fano factor is equal to $1/3$, which is similar to the result for Dirac fermions. The analogy of these systems derives from the bispinor of the Dirac fermion state and the spin-orbital SU(4) Kondo state with one electron in the system.
In $\delta I_{k}$ we observe the boost of the current between the configuration change from $\ket{q_{a}}_{16}$ to the quartet states $\ket{q_{z}}_{4}$ (green line in Fig. 26b). For $E_{d}=-3$, the current with increasing the coupling strength $t$ is quantized to: $\delta I_{K}=0.5$, $0.3$ and $0.07$ at the boundaries of areas with different charge numbers on the dot. For $Q=2$e (i.e. $E_{d}=E_{e-h}$, red line in Fig. 26b), the current takes on values: $\delta I_{K}=0.89$ for the sextuplets $\ket{d_{\nu}}_{6}$ (SU(4) Kondo symmetry), $\delta I_{K}=0.51$ for $\ket{q_{w}}_{24}$ and $\delta I_{K}=1$ in the region of strong coupling strength, where the octuplet $\ket{q_{y}}_{8}$ determines the SU$^{\star}$(2) Kondo phase (red line in Fig. 26b, and the black area in Fig. 26a).

Figures 26c, d present the nonlinear shot noise $\delta S_{K}$ versus $E_{d}$ and $t$. $\delta S_{K}$  has double zero points for $E_{d}=+1$, where we observe the sign reversal of the shot noise ($\delta S_{K}<0$). The $\delta S_{K}$ for $E_{d}=+1$ in the weak and strong coupling range reaches a value of $\delta S_{K}=0.14$ for the quartet states $\ket{q_{z}}_{4}$ (dark cyan curve in Fig. 26d). A pronounced suppression of the shot noise ($\delta S_{K}=0$) occurs for the two coupling strength values $t=0.3$ and $t=2.7$ (yellow line in Fig. 26c, and dark cyan curve in Fig. 26d). At these points, the transport is noiseless, which can be very attractive for quantum measurements. For $E_{d}=-1.5$ (green line in Fig. 26d), we observe a constant value of $\delta S_{K}=0.11$ for $\ket{p_{ls}}_{4}$ and $\ket{q_{a}}_{16}$, in contrast to $\delta I_{K}$ where the transition is detectable. With increasing the coupling $t$, the shot noise saturates to $0.14$, which corresponds to the quartet state $\ket{q_{z}}_{4}$ in the quantum dot.
For $E_{d}=-3$ (blue line in Fig. 26d), the noise leads to three quantized values $0.44$, $0.24$ and $0.06$, which occur at the boundary of the different charge sectors. We observe the noiseless transport at the point $t=1$ for $E_{d}=-3$, and for $t>1$ the shot noise takes on a negative value $\delta S_{K}=-0.06$. For $Q=2$e ($E_{d}=E_{e-h}$, red line in Fig. 26d), the nonlinear noise is quantized to the values: $\delta S_{K}=1.33$ for $\ket{d_{w}}_{6}$ (the SU(4) Kondo state), $\delta S_{K}=0.75$ for $\ket{q_{w}}_{24}$ and $\delta S_{K}=1.65$ in terms of strong coupling strength, where the Kondo SU$^{\star}$(2) effect is realized by the octuplet $\ket{q_{y}}_{8}$ - as the ground state in the system.

Figure 27 shows the $\delta S_{K}$ and $\delta I_{K}$ for the CNTQD-3TSC system. For $t=0$, the Kondo effect with SU(4) symmetry is realized in the system and $\delta S_{K}$ and $\delta I_{K}$ are identical to the previously presented numbers.
In the weak and strong coupling regime to the TSC for $E_{d}=+1$ $\delta I_{K}$ reaches $1/4$ for the octuplet state $\ket{q_{z}}_{8}$ (dark cyan line in Fig. 27b).
The sign reversal appears above $t=1.3$ for $E_{d}=+1$, but as we can see from Figure 27a, the suppression effect depends on the atomic level of the quantum dot $E_{d}$ and the coupling strength to the TSC (yellow line shows $\delta I_{K}=0$). For $E_{d}=-1.5$ (green line in Fig. 27b) in the weak coupling regime, we observe a value of $\delta I_{K}=0.08$ for $\ket{q_{a}}_{32}$ with saturation in the strong coupling strength range up to a value of $\delta I_{K}=0.25$ for $\ket{q_{z}}_{8}$. For $E_{d}=-3$, we observe three values of the current with increasing the coupling $t$: $\delta I_{K}=0.5$, $0.21$ and $0.27$. The $\delta I_{K}=0.21$ appears at the boundary of the regions with different charge numbers on the quantum dot (between $\ket{q_{w}}_{48}$ and $\ket{q_{a}}_{32}$). In the strong coupling region there is an increase in the nonlinear current, labeled $\delta I^{\star}_{K}$ in Figure 27b, where the quantum state $\ket{q_{z}}_{8}$ is close enough to the e-h symmetry point to enhance the current by the entanglement mechanism with opposite charge-leaking states. This is closely related to the quantum states marked in red in Eqs. (13-14). The state $\ket{2\uparrow n_{1}n_{2}n_{3}}$ is entangled with states below the e-h symmetry point (from a different charge region). The identical situation occurs on the other side of the e-h symmetry point, for the octuplet $\ket{q_{x}}_{8}$. In this case, the state $\ket{0\downarrow n_{1}n_{2}n_{3}}$ in Eq. (14) is responsible for the charge leakage effect. This seems to be the first report in the literature that indicates such a mechanism, and at the same time suggests the possibility of verifying with the lock-in technique in the noise measurement. To emphasize this result, let us call this shogun helmet-like state. For $Q=2e$ (i.e. $E_{d}=E_{e-h}$, red line in Fig. 27b), the current takes on the values: $\delta I_{K}=0.89$ for the states $\ket{d_{w}}_{6}$ (the Kondo state with SU(4) symmetry), $\delta I_{K}=0.35$ for $\ket{q_{w}}_{48}$ and $\delta I_{K}=0.5$ in the range of strong coupling $t$ at the e-h symmetry point where the octuplet states degenerate: $\ket{q_{x}}_{8}$ and $\ket{q_{z}}_{8}$ (red line in Fig. 27b, and the shogun helmet-like state in Fig. 27a). Figures 27c, d  show the nonlinear shot noise $\delta S_{K}$, which is rescaled by the characteristic energy $T_{[\star]}$ for a CNTQD system coupled to three Majorana fermions.
For $E_{d}=+1$, $\delta S_{K}$ (dark cyan curve at 27d) has double zeros.  The $\delta S_{K}=0$ appears for two coupling values $t=0.4$ and $t=4$ (yellow line in Figure 27c, and dark cyan curve in Figure 27d). Between these points $\delta S_{K}$ is negative. The $\delta S_{K}$  for $E_{d}=+1$, in the range of the weak and strong coupling strength takes the value around $0.25$, when the ground state is determined by the octuplet $\ket{q_{z}}_{8}$. For $E_{d}=-1.5$ (the green line in Fig. 27d) we observe a constant value of $\delta S_{K}=0.11$ for $\ket{p_{ls}}_{4}$ and $\ket{q_{a}}_{32}$. As the coupling strength increases, the shot noise saturates to $\delta S_{K}=0.25$, which corresponds to the octuplet $\ket{q_{z}}_{8}$ as the ground state energy. For $E_{d}=-3$ the noise reaches three quantized values of $0.44$, $0.19$ and $0.25$ (blue line in Fig. 27d), which occur at the boundary of the different charge sectors. At the the shogun helmet-like point there is an increase in shot noise to a maximum value denoted by $\delta S^{\star}_{K}$ in Fig. 27d. This is also related to the mechanism of entanglement with opposite charge-leaking states. For $Q=2$e ($E_{d}=E_{e-h}$, red line in Fig. 27d), the noise takes on the values: $\delta S_{K}=1.33$ for the $\ket{d_{\nu}}_{6}$, $\delta S_{K}=0.52$ for the quantum state $\ket{q_{w}}_{48}$ and $\delta S_{K}=0.15$ in the strong coupling regime for the U$^{\star}$(1) charge symmetry phase at the e-h symmetry point.

The analyzed quantity in the quantum transport measurements is the nonlinear Fano factor $F_{K}=\delta S_{K}/2e\delta I_{K}=e^{\star}/e$, whose value different from $1$, indicates for the influence of the residual interaction $\widetilde{U}_{\nu\nu'}$ between the quasiparticles. The main factors affecting these values are the second and third order fluctuations observed in the current $\delta I_{K}$ and in the shot noise $\delta S_{K}$. Figures 28, show the density plots of the nonlinear Fano factor as a function of the quantum dot level energy and the coupling strength to the topological superconductor. Figure 28a, shows $F_{K}$ for the CNTQD-1TSC model, in the density plot we observe bright areas resulting from the blocked current $\delta I_{K}=0$, with a non-zero negative shot noise value $\delta S_{K}<0$. On these lines, $F_{K}$ has an asymptotic behavior. In the density plot, we also see regions where the transport is noiseless $F_{K}=0$ (black colored lines). This occurs at the boundary between $\ket{q_{a}}_{8}$ and $\ket{q_{z}}_{2}$ and between $\ket{q_{y}}_{6}$ and $\ket{q_{z}}_{2}$, where $\delta S_{K}=0$ and $\delta I_{K}>0$. Similar behavior was observed in the results of the NRG method \cite{Oguri2020}. It is definitely not the effect of the absence of the higher order corrections in the coefficients at $V^{5}$, since these contributions are insignificant below the Kondo energy scale, where current and shot noise are described by the Fermi liquid theory assumptions and the vertex corrections for the current-current correlation. Could quaternary and higher-order fluctuations affect the values of the coefficients $c_{V,\nu}$ and $c_{S,\nu}$?  This is difficult to answer unambiguously, the authors in \cite{Oguri2022} have limited themselves to two- and three-particle correlations, pointing to low-energy excitation as the mechanism of the Kondo phase. The zeroing of $F_{K}$ also occurs  for a single quantum dot in the Kondo state  (we have only four states there $\{0,\uparrow,\downarrow,2\}$), where is difficult to imagine fourth-order dot correlators, which says something about the physical implications of this behavior.
\begin{figure}[t!]
\includegraphics[width=0.75\linewidth]{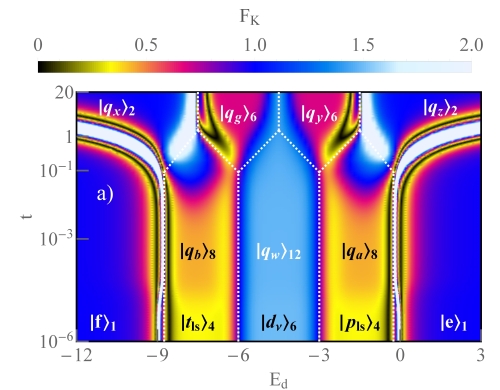}\\
\includegraphics[width=0.75\linewidth]{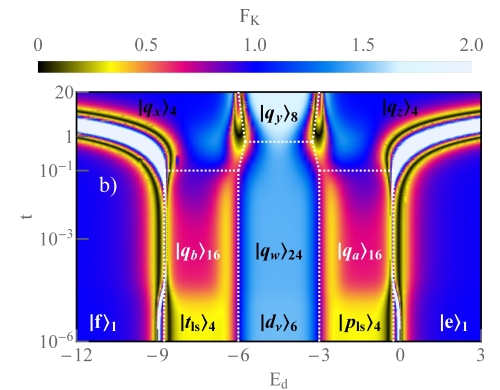}\\
\includegraphics[width=0.75\linewidth]{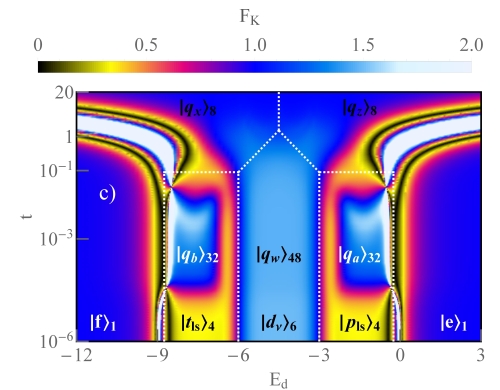}
\caption{\label{fig:epsart} (Color online) a-c) The density plot of the fractional nonlinear Fano factor $F_{K}=|c_{S}|/|c_{V}|$ versus $E_{d}$ and $t$ for the CNTQD-TSC, CNTQD-2TSC and CNTQD-3TSC devices. Black and white lines correspond to $S_{K}=0$ and $I_{K}=0$ respectively.}
\end{figure}
\begin{figure}[t!]
\includegraphics[width=0.75\linewidth]{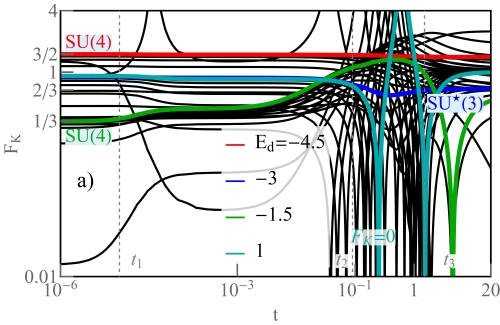}\\
\includegraphics[width=0.75\linewidth]{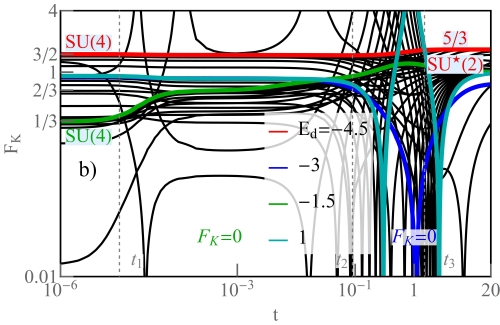}\\
\includegraphics[width=0.75\linewidth]{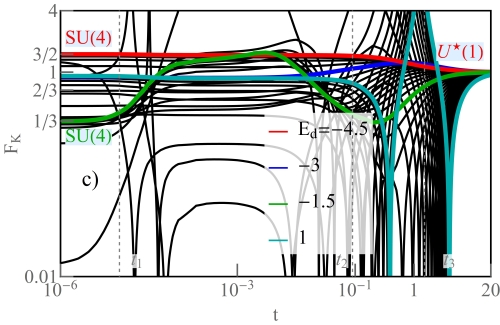}
\caption{\label{fig:epsart} (Color online) a-c) The landscape log-log plots of $F_{K}$ as a function of $t$ with increment $\delta E_{d}=0.15$ for the CNTQD coupled with MF, 2MFs, and 3MFs.}
\end{figure}
\begin{figure}[t!]
\includegraphics[width=0.75\linewidth]{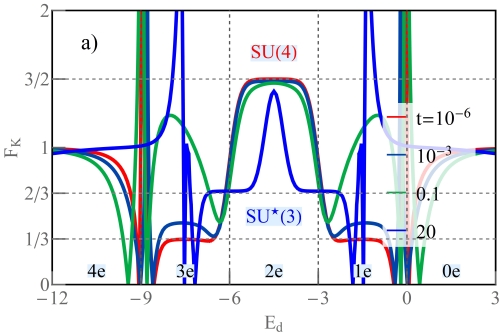}\\
\includegraphics[width=0.75\linewidth]{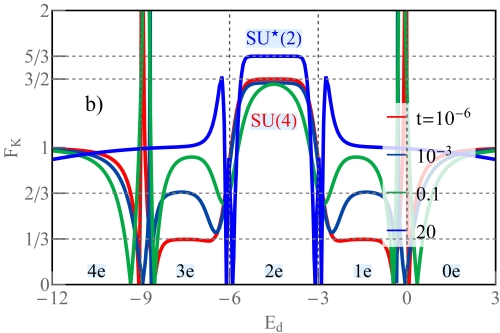}\\
\includegraphics[width=0.75\linewidth]{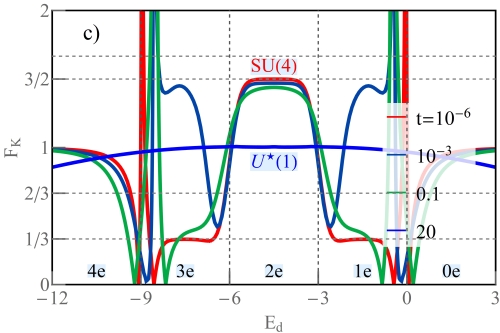}
\caption{\label{fig:epsart} (Color online) a-c) $E_{d}$ dependence of $F_{K}$ with increasing $t$ for CNTQD-TSC, CNTQD-2TSC and CNTQD-3TSC devices. Dashed vertical and horizontal lines divide the charge regions $Q=0e,1e,2e,3e,4e$ and show characteristic limits of $F_{K}$.}
\end{figure}

Figure 28a illustrates the characteristic values of $F_{K}$ in the density plot for the CNTQD-1TSC device. For SU(4) Kondo state, (a result known from the literature \cite{LeHur2009,Oguri2020}), as $t$ approaches $0$ we observe $F_{K}=3/2$ for $Q=2$e and ($\ket{d_{\nu}}_{6}$) $F_{K}=0. 33$ for $Q=1$e ($\ket{p_{ls}}_{4}$). For the empty and fully occupied states the Fano factor reaches $F_{K}=1$ and is negative $F_{K}=\frac{S_{K}}{(2eI_{K}}=-1$ for the doublet state $\ket{q_{z}}_{2}$.
In the weak coupling regime the device demonstrates  $F_{K}=3/2$ for $\ket{q_{w}}_{12}$) and $F_{K}=0.11/0.25\approx0.44$ for the octuplet quantum states $\ket{q_{a(b)}}_{8}$. For the SU$^{\star}$(3) Kondo state, the Fano factor reaches $0.62$. The influence of the Majorana fermion-coupled channel $\nu'=+\uparrow$ is negligible, and the result is consistent with the SU(3) Kondo state. In a sense this is understandable due to the fact that the residual interaction occurs only between the Kondo quasiparticles. Majorana fermions do not interact with each other (they are only represented by two states $\{\underline{0},\underline{\Uparrow}\}$). The interaction between  the Majorana fermions analyzes the authors in this papers  \cite{Chiu2015,Rahmani2019} and based on the concept of Wilson ratio, the type of interaction can be relevant for $F_{K}$.
At the level of the linear regime of the current and the shot noise $F_{0}$ is equal to $1/2$ in the channel coupled to the Majorana fermion state, in the nonlinear regime of bias voltage, it is probably insignificant \cite{Liu2015}. However, there is a paper \cite{Smirnov2015} where the author suggests for QD-TSC device with the infinite U on quantum dot, $F_{K}=3/2$ in the high voltage regime for $t>V>E_{d}\ll0$ \cite{Smirnov2015}, beyond the Kondo regime ($V\gg T_{K}$).

Figure 28b shows the density plot of $F_{K}$ as a function of $E_{d}$ and the coupling strength $t$ for the CNTQD-2TSC system. The SU$^{\star}$(2) Kondo effect arises in the strong coupling region for $Q=2$e. In this region $F_{K}=5/3$ and the channels associated with SU$^{\star}$(2) Kondo effect play a significant role in the nonlinear shot noise and current (Fig. 26). The fractional super-Poissonian Fano factor is identical to the prediction for the SU(2) Kondo state\cite{Sela2006} and the experimental results \cite{Yamauchi2011}.
Majorana channels in the strong coupling limit remain inactive in the nonlinear voltage regime. In the strong coupling limit, the channels coupled to the TSC contribute only to the linear Fano factor (see Fig. 23b). In the weak coupling limit, the ground state is represented by $\ket{q_{w}}_{24}$ and $\ket{q_{a(b)}}_{16}$. For this quantum state, Majorana fermions modify the value of the Fano coefficient for $Q=1(3)$e from $1/3$ to $F_{K}\approx2/3$ (green line in Fig. 29b). Super-Poissonian values of $F_{K}$ occur at the boundary of the states $\ket{q_{x(z)}}_{4}$ and $\ket{q_{a(b)}}_{16}$. Noiseless transport ($\delta S_{K}=0$) is represented by the black lines in the density plot of the nonlinear Fano factor. $F_{K}=0$ is realized at the boundary of the octuplet $\ket{q_{y}}_{8}$ and the two quartets $\ket{q_{x(z)}}_{4}$, and around the transition between $\ket{q_{x(z)}}_{4}$ and $\ket{q_{a(b)}}_{16}$.

Fig. 28c illustrates $F_{K}$ as a function of $E_{d}$ and the coupling strength $t$ for the CNTQD-3TSC hybrid device. In terms of the weak coupling to three Majorana fermion states, we observe an SU(4) Kondo-like effect, with a ground state $\ket{q_{a(b)}}_{32}$ in the charge regions $Q=1(3)$e. The $F_{K}$ in this area reaches $F_{K}\approx1.33$ (the light blue region in Fig. 28c). The $F_{K}$ shows the difference between the fully SU(4) Kondo phase and the SU(4) Kondo-like state.
For $Q=2$e, the area is determined by the quantum state $\ket{q_{w}}_{48}$, and $F_{K}$ approaches the quantized value $F_{K}=3/2$, which corresponds to the result for the SU(4) Kondo effect. The variability of $\delta S_{K}$ and $\delta I_{K}$ is  intersected as a red curves in Fig. 26b,d. The role of the Majorana fermion is significant in the current and the shot-noise, even if $F_{K}=|S_{K}|/2e|I_{K}|=3/2$ with increasing $t$, is constant and does not depend on the coupling strength $t$ in terms of weak coupling to the TSC.
In both magnitudes $S_{K}$ and $I_{K}$, the mechanism of entanglement with opposite charge-leaking states is evident, but in $F_{K}$ the mechanism is practically invisible, due to the comparable values of $I_{K}$ with $S_{K}$. In Fig. 28c, black lines present noiseless transport, where $\delta S_{K}=0$ and the white areas represent the blocked transport for $\delta I_{K}=0$.

Figure 29 shows the cross sections of $F_{K}$ as a function of $t$ with the increment $\delta E_{d}=0.15$ in the range $E_{d}$ from $-12$ to $3$. Figure 29 a-c presents $F_{K}=1$ for $E_{d}=+1$ in the weak coupling regime to TSC (dark cyan lines). In the intermediate region, between $t_{2}$ and $t_{3}$, we observe a sign reversal in $\delta S_{K}$, leading to an extinction of $F_{K}$ at two characteristic points. Significantly, for $t>t_{3}$, $S_{K}$ becomes positive again, however $I_{K}$ has a negative value in the strong coupling regime. We can say that the quasiparticles reverse the flow of the current. This is due to the two-body correlations in $\delta I_{K}$.  The red line shows $F_{K}=3/2$ for $Q=2$e and the green line reaches a value of $F_{K}=1/3$ in the weak coupling regime \cite{Sela2006,Mora2008}. For the octuplet $\ket{q_{a}}_{8}$, we observe $F_{K}\approx0.44$ with a small increase of the Fano factor around $t_{2}$ to the quantized value $F_{K}=3/2$. Above $t_{3}$, there is point of the noiseless transport and $F_{K}$ saturates with increasing $t$ to the value $F_{K}=-1/2$ (green line Fig. 29a). For the sextuplet $\ket{q_{y}}_{6}$, a fractional effect is realized in the strong coupling regime to the TSC. The SU$^{\star}$(3) Kondo state is formed and $F_{K}$ reaches $0.62$ (blue line Fig. 29a). In Fig. 29b and Fig. 29c for $E_{d}=-4.5$ we observe the Kondo effect with the SU$^{\star}$(2) symmetry and a charge-degenerate state with U$^{\star}$(1) symmetry  (between two octuplets $\ket{q_{x}}_{8}$ and $\ket{q_{z}}_{8}$). In terms of strong coupling strength, $F_{K}$ leads to values of $5/3$ and $1$, respectively. Of note is the fact that, for the CNTQD-3TSC system, for the states of $\ket{q_{a}}_{32}$, we observe a boost of the Fano factor to the value of $F_{K}\approx3/2$, which formally occurred for the case of full SU(4) symmetry at the half-filling (green line in Fig. 29c).

Figures 30 include the cross sections of $F_{K}$ as a function of $E_{d}$ for the uncoupled, intermediate and strong coupling regimes to the TSC. Fig. 30a shows the gate-dependent $F_{K}$ for a CNTQD system coupled to a single Majorana fermion $\gamma_{+\uparrow}$. The plot shows the absolute value of the Fano factor, there are regions where $F_{K}<0$, preceded by the noiseless points ($F_{K}=0$), as we have written, this is due to the dominance of two-body processes over three-particle correlators. For $t=10^{-6}$, the system is in the SU(4) Kondo state, and $F_{K}$ assumes two characteristic numbers: $F_{K}=3/2$ for $Q=2$e and $F_{K}=1/3$ for $Q=1(3)$e. At the transition between the charges $Q=0(4)$e and $Q=1(3)$e, we observe the suppression of the shot noise in two points with super-Poissonian behavior in the middle, when $\delta I_{K}=0$. In terms of weak coupling, there is a modification of $F_{K}$ (green line for $t=0.1$ in Fig. 30a), $F_{K}$
is close to $3/2$ for $Q=2e$ ($\ket{q_{w}}_{12}$) and increases to $\approx 5/4$ for $Q=1(3)$e. In $Q=1(3)$e, the value varies from sub- to super-Poissonian noise with increasing the coupling strength ($\delta I_{K}>\delta S_{K}\mapsto\delta I_{K}\ll\delta S_{K}$). For the strong coupling limit, we observe two contrasting behaviors around the U$^{\star}$(1) charge symmetry line, where  $F_{K}=0$ and $F_{K}\mapsto +\infty$ (black and white lines in Fig. 28a).

Figure 30b shows the cross sections of $F_{K}$ corresponding to the density plot of Fig. 28b for $t=10^{-6}$, $0.1$, $10^{-3}$ and $20$. In the figure, we observe a gradual change of $F_{K}$ from the full SU(4) Kondo effect through the intermediate crossover region ($\ket{q_{a}}_{16}$) to the strong coupling solution for the quartet $\ket{q_{z}}_{4}$. For $t=0.1$, the value of the $\delta I_{K}$ and $\delta S_{K}$ at the transition between the states $\ket{q_{a(b)}}_{16}$ and $\ket{q_{z(x)}}_{4}$ shows an asymptotic maximum and between $\ket{q_{a(b)}}_{16}$  and $\ket{q_{w}}_{24}$ leads to minimum $F_{K}=1/3$ (green line Fig. 29 b). In the strong coupling regime $F_{K}$ changes its sign to $-1$. For the quantum state $\ket{q_{a(b)}}_{16}$, the value of $F_{K}$ approaches to $2/3$ (dark blue line Fig. 29b).
For two electrons on the quantum dot $F_{K}$ evolves from $2/3$ to a quantized value of $5/3$ for the SU$^{\star}$(2) Kondo state. Although $F_{K}=5/3$, as for the full SU(2) Kondo effect \cite{Sela2006}, we denote the state by $\star$ because the strongly correlated state appears for an even number of electrons $Q=2$e. The full SU(2) Kondo state is observed for $Q=1$e.
Below $E_{d}=-6$ and above $E_{d}=-3$, where $Q=3$e and $Q=1$e, the ground state is defined by two quadruplets $\ket{q_{x(z)}}_{4}$, $F_{K}=-1$ and the reverse current $\delta I_{K}<0$ dominates in the nonlinear transport (blue line in Fig. 29b).

Fig. 30c shows the cross sections of $F_{K}$ from Fig. 28c for $t=10^{-6}$, $0.1$, $10^{-3}$ and $20$. The $F_{K}$ evolve from the full SU(4) Kondo state and reconstructs via the crossover region ($\ket{q_{a}}_{32}$) to the stable U$^{\star}$(1) charge phase in the range of the strong coupling strength ($\ket{q_{z}}_{8}$). Two transition points are observed for $t=10^{-3}$: the first one at the boundary between $\ket{q_{a(b)}}_{32}$ and $\ket{q_{z(x)}}_{8}$ shows a maximum and the second one between $\ket{q_{a(b)}}_{32}$ and $\ket{q_{w}}_{48}$ reaches a minimum close to $1/3$ (dark blue line in Fig. 29c). For half filling and $Q=1(3)$e $F_{K}$ approaches to $3/2$ for $\ket{q_{w}}_{48}$ and $\ket{q_{a(b)}}_{32}$ (dark blue line Fig. 29c). With increasing the coupling strength, $F_{K}$ reaches $1/3$ at $t_{2}$ and approaches to $-1$ in the broad region of $E_{d}$, in particular for U$^{\star}$(1) charge symmetry point. A line of degeneracy appears for $Q=2$e between the fractional charges $Q=5/2$e and $Q=3/2$e, therefore we denote this state by $\star$ (blue line in Fig. 29c). The quantum conductance at this point leads to ${\mathcal{G}}=(5/2)(e^{2}/h)$ and is negatively spin (orbital) polarized $\Delta {\mathcal{G}}_{s(o)}=-1/5$ (Fig. 7a and Fig. 8c). This corresponds to the result for the QD-TSC device \cite{Weymann2017}, where the quantum conductance in the strong coupling regime is narrowed to the e-h symmetry line and reaches the value ${\mathcal{G}}=(3/2)(e^{2}/h)$.

\subsection{SOI and transport properties in CNTQD-1TSC device}
\begin{figure}[t!]
\includegraphics[width=0.75\linewidth]{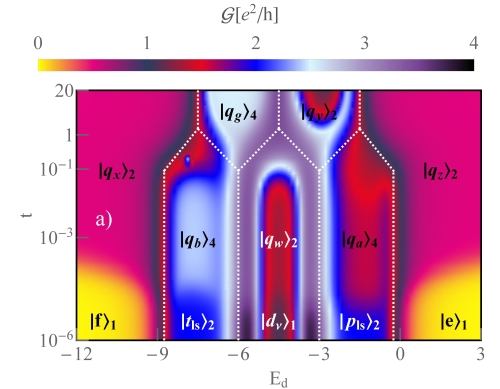}\\
\includegraphics[width=0.75\linewidth]{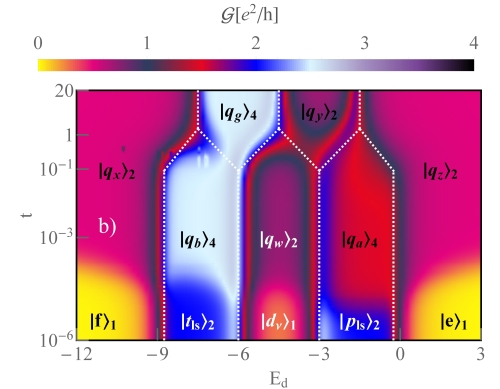}\\
\includegraphics[width=0.75\linewidth]{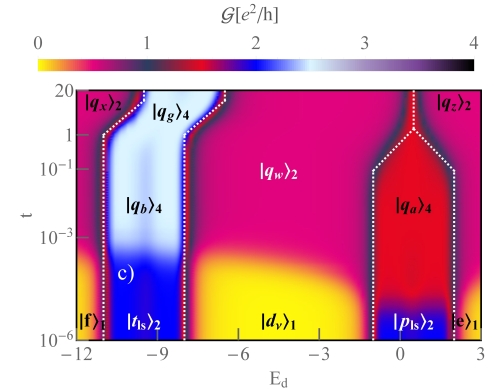}\\
\includegraphics[width=0.75\linewidth]{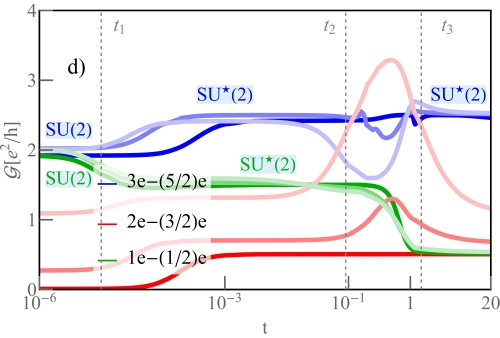}
\caption{\label{fig:epsart} (Color online) CNTQD-MF device with spin-orbit interaction $\Delta$: a-c) The density plot of ${\mathcal{G}}$ versus $E_{d}$ and $t$ for $\Delta=10^{-2},10^{-1}$ and $2$. d) ${\mathcal{G}}$ as a function of $t$. Brightest, light and dark color of the lines show the results for weak, intermediate and strong SOI.}
\end{figure}
\begin{figure}[t!]
\includegraphics[width=0.75\linewidth]{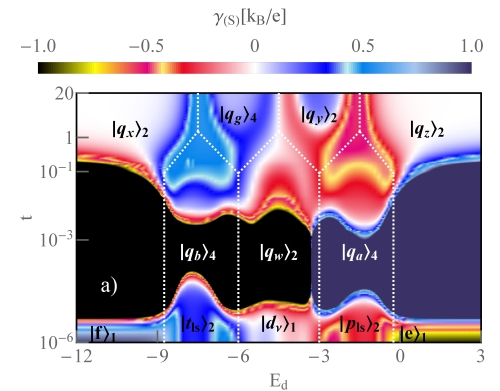}\\
\includegraphics[width=0.75\linewidth]{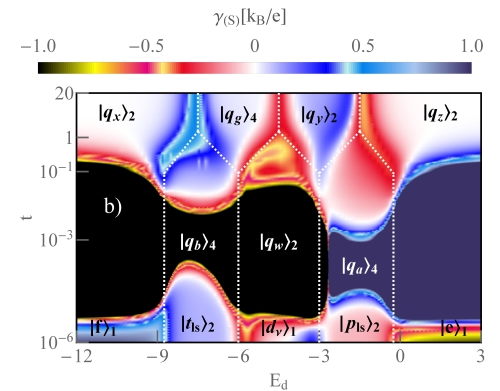}\\
\includegraphics[width=0.75\linewidth]{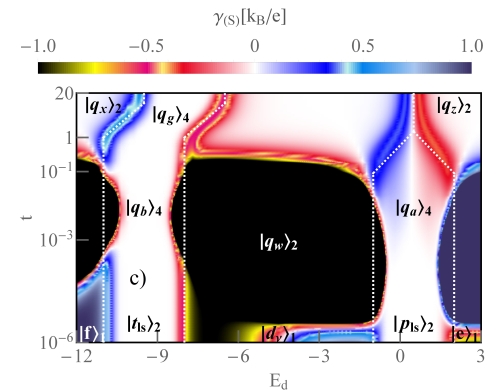}\\
\includegraphics[width=0.75\linewidth]{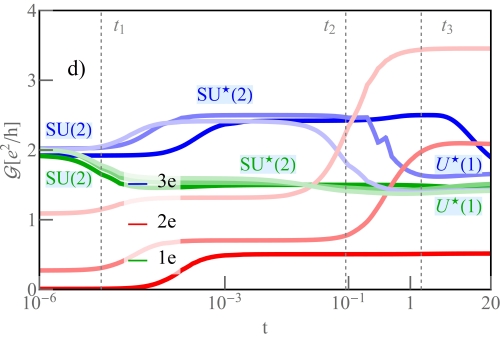}
\caption{\label{fig:epsart} (Color online) a-c) The density plot of $\gamma_{(S)}$ versus $E_{d}$ and $t$ for $\Delta=10^{-2},10^{-1}$ and $2$. d) ${\mathcal{G}}$ as a function of $t$. Brightest, light and dark color of the lines show the results for weak, intermediate and strong SOI.}
\end{figure}
\begin{figure}[t!]
\includegraphics[width=0.75\linewidth]{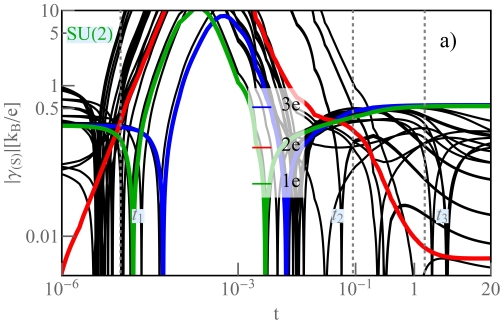}\\
\includegraphics[width=0.75\linewidth]{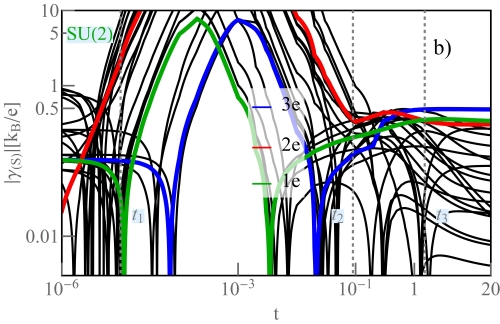}\\
\includegraphics[width=0.75\linewidth]{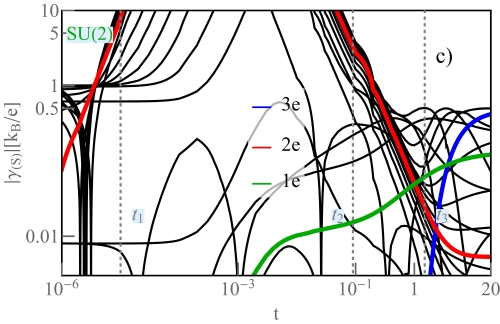}
\caption{\label{fig:epsart} (Color online) a-c) The landscape log-log plots of $|\gamma_{(S)}|$ as a function of $t$ for weak, intermediate, and strong SOI. The lines are plotted with an increment of $\delta E_{d}=0.5$ from $-10$ to $0.5$.}
\end{figure}

In the last subsection we will discuss the effect of the spin-orbit interaction ($\Delta$) on the transport quantities in the CNTQD-TSC hybrid system. The SOI in the CNTQD originates from the curvature of the nanotube and reaches the values ranging from thousands to a few meV \cite{Kuemmeth2008,Jespersen2011,Laird2015,Steele2013}.
The $\Delta$ removes the fourfold degeneracy of the states on the quantum dot, and forms two Kramers doublets: a low energy doublet $E_{+\downarrow}=E_{-\uparrow}$ and a high energy doublet $E_{+\uparrow}=E_{-\downarrow}$. The SU(4) Kondo state is broken by SOI and we observe the Kondo effect with SU(2) symmetry \cite{Galpin2010,Mantelli2016,Krychowski2018}. When we connect the quantum dot to a Majorana fermion of type $\gamma_{+\uparrow}$, one of the channels from the high-energy doublet, is operated by a topological superconductor. The other low-energy doublet is not directly connected to the topological state, but is indirectly capacitively coupled to the $\nu'=+\uparrow$ state through the Coulomb interactions. Figure 31 shows the quantum conductance maps of ${\mathcal{G}}$ for the weak SOI $\Delta=10^{-2}$ (a), intermediate $\Delta=10^{-1}$ (b) and for the strong spin-orbital interaction $\Delta=2$ ($\Delta\approx U$) (c) \cite{Laird2015,Steele2013}. Coupling with 1TSC changes the ground state configuration in the system. For $Q=1$e, with increasing $\Delta$ there is a transition from the quartet $\ket{p_{ls}}_{4}$ to the doublet $\ket{p_{ls}}_{2}$, in the weak coupling regime from the octuplet $\ket{q_{a}}_{8}$ to the quadruplet $\ket{q_{a}}_{4}$ and in the strong coupling range the  U$^{\star}$(1) charge symmetry phase is realized for two doublet states $\ket{q_{y}}_{2}$ and $\ket{q_{z}}_{2}$. It is noteworthy that for large SOI the $\ket{q_{y}}_{2}$ state is replaced by a lower energy state $\ket{q_{w}}_{2}$ for the $Q=2$e sector (Fig. 31c) . The degenerate line between the two doublets disappears under the influence of $\Delta$ (Fig. 31a).
This is particularly evident in Fig. 31d, where the sharp maximum between $\ket{q_{w}}_{2}$ and $\ket{q_{y}}_{2}$ is gradually suppressed (red curves for $Q=2$e). For $Q=1$e we observe the transitions between three quantum states. The first transition is observed between the doublet $\ket{p_{ls}}_{2}$ and the quadruplet state $\ket{q_{a}}_{4}$ and the conductance changes from ${\mathcal{G}}=2(e^{2}/h)$ to ${\mathcal{G}}=(3/2)(e^{2}/h)$. Finally, for the strong coupling strength, the quantum conductance ${\mathcal{G}}=(1/2)(e^{2}/h)$ determines the doublet state $\ket{q_{z}}_{2}$. However, in $\ket{q_{z}}_{2}$, the charge is reduced to $Q=(1/2)$e.
At the charge degeneracy between the two doublets: $\ket{q_{y(w)}}_{2}$ and $\ket{q_{z}}_{2}$, the system is in the U$^{\star}$(1) charge state and the conductance reaches the value ${\mathcal{G}}=(3/2)(e^{2}/h)$ (green lines in Fig. 32d, and   red narrow line in Fig. 31c). This is important, because for U(1) charge symmetry, the quantum conductance always approaches the value of the single quanta $(e^{2}/h)$.

The most significant result is the asymmetry relation in the quantum conductance, with respect to the e-h symmetry point, between $Q=1$e and $Q=3$e. The reason for the asymmetry is that the high energy is tunnel-coupled to the Majorana fermion state $\gamma_{+\uparrow}$, in contrast to the excited doublet, which is only capacitively coupled to $n_{+\uparrow}$ via the Coulomb interaction. In the weak and strong coupling regime for $Q=3$e-$(5/2)$e, the SU$^{\star}$(2) Kondo states are realized in the presence of Majorana state. The Kondo phases are determined by two quadruplets $\ket{q_{b}}_{4}$ and $\ket{q_{g}}_{4}$  (blue lines in Fig. 31d). In Fig. 31d, the tunneling term $t$ depends on $E_{d}$, using this dependence, we show the quantum conductance for the integer and fractional charges and we pass through the most important ground states of the system (in the contrast to Fig. 32d).

Figures 32-33 show the linear thermoelectric coefficient $\gamma_{(S)}$, defined
by Eq. (43) for the CNTQD-1TSC device. In the density plots, we see that for the weak coupling regime, depending on the value of $\Delta$, the thermoelectric power gradually reduces to the full SU(2) Kondo state, where the quasiparticle resonance is centered at the Fermi level,
hence $\gamma_{(S)}=0$ (bright white areas in Fig. 32c for $\ket{t_{ls}}_{2}$ and $\ket{p_{ls}}_{2}$ and green and blue lines in Fig. 33c). For $\Delta=10^{-2}$ and $\Delta=10^{-1}$, the ground state energy is determined by two doublets $\ket{t_{ls}}_{2}$ and $\ket{p_{ls}}_{2}$, however, we observe finite values of $|\gamma_{S}|\ll \pi/(3\sqrt{2})$, indicating that we are not exactly in the full SU(2) or SU(4) Kondo state. The value of $\gamma_{(S)}$ determines the symmetry and the quality of the Kondo effect.

In the range of intermediate coupling $t$ and $\Delta=10^{-2}, 10^{-1}$, we observe an enhancement of the thermoelectric power for $\ket{q_{a(b)}}_{4}$. The system behaves like a non-Fermi liquid in this region, because $T_{[\star]}$, does not scale the $\gamma_{(S)}$ to a constant FL number \cite{Krychowski2020}.
In Figures 33a, b, we observe the first and second compensation points in $\gamma_{S}$  (for $Q=1$e and $Q=3$e). The first zero in Eq. (43) occurs when $-\pi T_{K}(\cot[\delta_{\nu'}][\widetilde{\Gamma}_{\nu'}\delta^{2}+\widetilde{t}^{2}\delta])/[(3\widetilde{\varrho}_{\nu}(0)+\widetilde{\varrho}_{\nu'}(0))(\pi\widetilde{\Gamma}_{\nu'}(2\widetilde{t}^{2}+\widetilde{\Gamma}_{\nu'}\delta\csc^{2}[\delta_{\nu'}])^{2})]=-\pi T_{K}(-\cot[\delta_{\nu'}]\widetilde{\Gamma}_{\nu'}\widetilde{t}^{2})/[(3\widetilde{\varrho}_{\nu}(0)+\widetilde{\varrho}_{\nu'}(0))(\pi\widetilde{\Gamma}_{\nu'}(2\widetilde{t}^{2}+\widetilde{\Gamma}_{\nu'}\delta\csc^{2}[\delta_{\nu'}])^{2})]$. The first compensation point appears near the value of $t_{1}=\delta$ (where $\delta$ is the lifetime of the Majorana fermion) and is related to the contribution of the $\nu'$ channel. With increasing the coupling strength we then observe a maximum in the channel coupled to the Majorana fermion. The sign of $\gamma_{(S)}$ changes in the intermediate coupling regime. The similar effect of the sign reversal is presented in the paper \cite{Lopez2014}. The second compensation point is already a result of balancing of the normal contribution, coming from the $\nu$ channel, with the contribution from the $\nu'$ quantum channel for the condition $-\pi T_{K}(3\dot{\widetilde{\varrho}}_{\nu}(0))/[(3\widetilde{\varrho}_{\nu}(0)+\widetilde{\varrho}_{\nu'}(0))]=-\pi T_{K}(\cot[\delta_{\nu'}][\widetilde{\Gamma}_{\nu'}\delta^{2}+\widetilde{t}^{2}(\delta-\widetilde{\Gamma}_{\nu'})])
/[(3\widetilde{\varrho}_{\nu}(0)+\widetilde{\varrho}_{\nu'}(0))(\pi\widetilde{\Gamma}_{\nu'}(2\widetilde{t}^{2}
+\widetilde{\Gamma}_{\nu'}\delta\csc^{2}[\delta_{\nu'}])^{2})]$. In the strong coupling regime, we observe in the linear thermoelectric power coefficient a constant value of $\gamma_{s}\approx \pm(1/2)$ at the boundary between  $\ket{q_{x}}_{2}$ and $\ket{q_{g}}_{4}$ and between $\ket{q_{y}}_{2}$ and $\ket{q_{z}}_{2}$ (blue, green lines in Fig. 32a-b).
The results correspond to the quantum conductances ${\mathcal{G}}=(3/2)(e^{2}/h)$ for $Q=1$e and $Q=3$e (brightest and light green curves in Fig. 32d). In contrast to Fig. 15a, where we have shown the symmetric evolution of the density plot of $\gamma_{(S)}$ in the NFL phase for $\Delta=10^{-3}$, we observe a significant asymmetry in the NFL state. In terms of weak and intermediate SOI, $\gamma_{(S)}$ is negative for the doublet  and quartet states: $\ket{q_{w}}_{2}$ and $\ket{q_{a}}_{4}$. The sharp switch is observed at the charge degeneracy line between  $\ket{q_{w}}_{2}$  and $\ket{q_{a}}_{4}$. The position of the e-h symmetry line in the NFL phase is perturbed due to the dominant role of the hole states in Eqs. (8-9). For $\Delta=2$ in the intermediate coupling regime, the regions for $\ket{q_{a(b)}}_{4}$ are well defined, and we already observe the full SU$^{\star}$(2) Kondo states, where
$\gamma_{S}=0$ (flat white areas in Fig. 32c). Interesting was the intersection by the NFL state of the region for the doublet $\ket{q_{w}}_{2}$ (the enhancement for $t\approx10^{-6}$ in Fig. 32c). This suggests that we need a much weaker coupling to remove the NFL phase from the dependence of $\gamma_{(S)}$.  The quantum measurements of the thermoelectric power ${\mathcal{S}}$ give us the information about the quality of the Kondo effect, much more precisely than the linear quantum conductance ${\mathcal{G}}$.
If we look on the lines in Fig. 31d and Fig. 32d for $Q=1(3)$e, we observe a small fluctuation of the conductance around ${\mathcal{G}}=(5/2)(e^{2}/h)$ and ${\mathcal{G}}=(3/2)(e^{2}/h)$, but in $\gamma_{S}$ this is a drastic change in character from NFL (green and blue lines in Fig. 33a, b) to FL behavior (green and blue lines in Fig. 33c).
\begin{figure}[t!]
\includegraphics[width=0.75\linewidth]{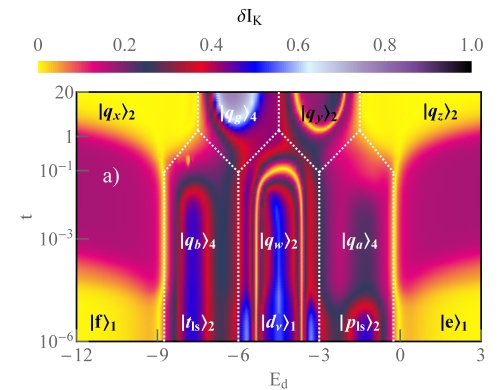}\\
\includegraphics[width=0.75\linewidth]{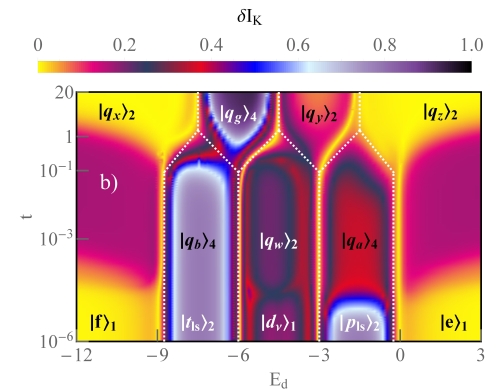}\\
\includegraphics[width=0.75\linewidth]{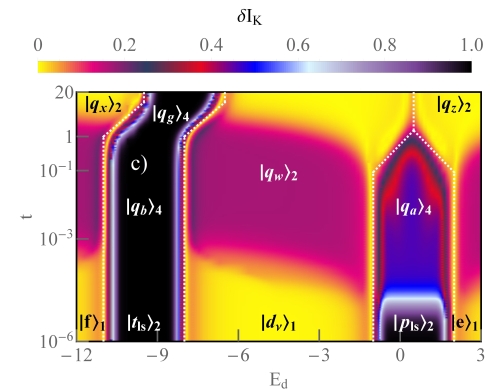}
\caption{\label{fig:epsart} (Color online) CNTQD-MF device with SOI: a-c) The density plot of the rescaled nonlinear current $\delta I_{K}$ versus $E_{d}$ and $t$ for $\Delta=10^{-2},10^{-1}$ and $2$.}
\end{figure}
\begin{figure}[t!]
\includegraphics[width=0.75\linewidth]{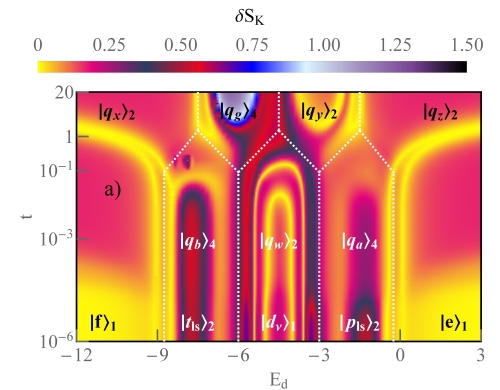}\\
\includegraphics[width=0.75\linewidth]{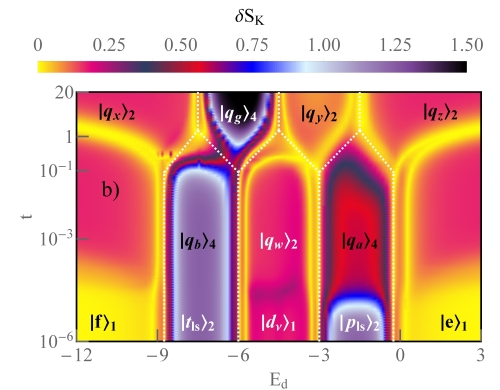}\\
\includegraphics[width=0.75\linewidth]{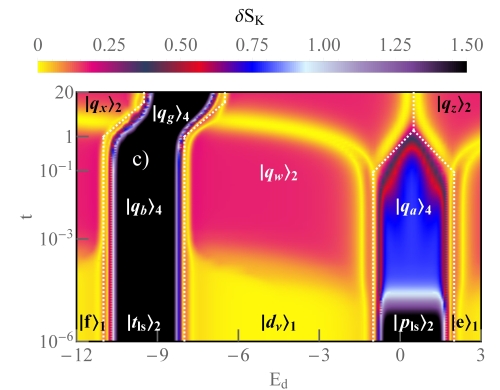}
\caption{\label{fig:epsart} (Color online) a-c) The density plot of the rescaled nonlinear shot noise $\delta S_{K}$ versus $E_{d}$ and $t$ for weak, intermediate and strong SOI. Yellow lines denote noiseless transport $\delta S_{K}\approx0$.}
\end{figure}
\begin{figure}[t!]
\includegraphics[width=0.75\linewidth]{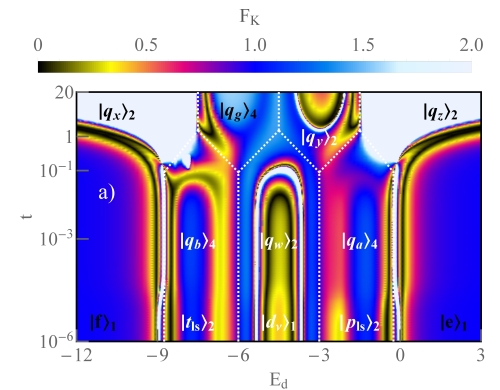}\\
\includegraphics[width=0.75\linewidth]{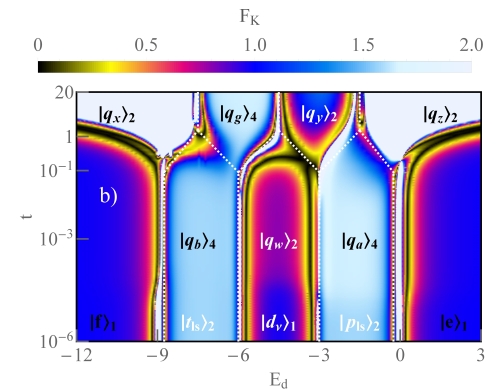}\\
\includegraphics[width=0.75\linewidth]{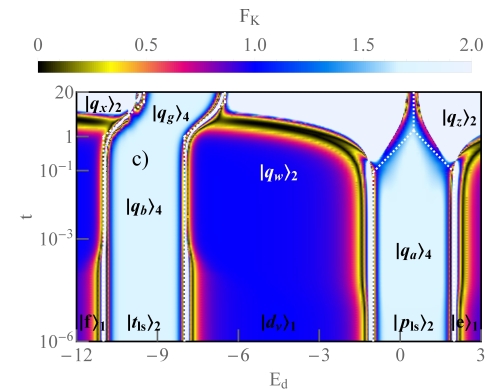}
\caption{\label{fig:epsart} (Color online) a-c) The density plot of $F_{K}$ as a function of $E_{d}$ and $t$ for $\Delta=10^{-2},10^{-1}$ and $2$. Black and white lines correspond to $S_{K}=0$ and $I_{K}=0$.}
\end{figure}

Fig. 34 shows the $\delta I_{K}$ as a function of $E_{d}$ and $t$ for weak (a), intermediate (b) and strong SOI on the quantum dot (c). The yellow lines and the yellow areas in Fig. 34a show the blocked nonlinear transport. The $\delta I_{K}\approx0$ occurs for $\Delta=10^{-2}$ in the region of the charge stability for two doublet quantum states $\ket{q_{w}}_{2}$ and $\ket{q_{y}}_{2}$. For $\Delta=0.1$, the zero lines merge into two lines at the charge degeneracy point between even and odd charges (Fig. 34b). For $t=10^{-6}$, when the system determines two doublets $\ket{t_{ls}}_{2}$ and $\ket{p_{ls}}_{2}$, $\delta I_{K}$ varies from $1/2$ for $\Delta=10^{-2}$, via $\delta I_{K}\approx0.75$ for $\Delta=0.1$, to $\delta I_{K}\approx1$ for $\Delta=2$.
For the strong SOI, the SU(2) Kondo states on the two quantum doublets are well defined and well separated (black regions in Fig. 34c). In the region of $Q=1(3)$e and $\Delta=0.1, 2$, for the weak tunneling rate, where $t=10^{-3}$ we observe an asymmetry between the SU$^{\star}$(2) Kondo effects in two quartet states $\ket{q_{b}}_{4}$ and $\ket{q_{a}}_{4}$ (Fig. 34b and Fig. 34c). In these domains, the $\delta I_{K}=T^{2}_{[\star]}|I_{K}|$ is quantized to $\delta I_{K}\approx1$ for $\ket{q_{b}}_{4}$ and $\delta I_{K}\approx1/2$ for $\ket{q_{a}}_{4}$. For $\Delta=0.1$, the states are also separated, but the nonlinear currents approach $\delta I_{K}\approx0.75$ and $0.35$. The transition between weak and strong coupling strength $t$ is clearly visible in $\delta I_{K}$ for the two doublets $\ket{q_{x(z)}}_{2}$. In these regions, with increasing tunneling strength we observe a change in the value from $\delta I_{K}=0.15$ (purple color in Fig. 34a-c) to $\delta I_{K}=10^{-3}\approx0$ (yellow area). In Fig. 25a, for CNTQD without spin-orbit interaction, the quantum states $\ket{q_{x(z)}}_{2}$ are shared by the yellow line representing $\delta I_{K}=0$, where two- and three-body correlators are compared. For the hybrid device with finite SOI, the line extends into the region with blocked nonlinear transport $\delta I_{K}\approx0$. The consequence of this effect is the super-Poissonian value of $F_{K}$. The transition between the Kondo effect with SU(2)symmetry and the SU$^{\star}$(2) Kondo state
is visible in the current, especially for $Q=1$e, where $\delta I_{K}\approx1$ for $\ket{p_{ls}}_{2}$ changes to $\delta I_{K}\approx(1/2)$ for $\ket{q_{a}}_{4}$. This is particularly evident, in the high energy doublet, due to the fact that the CNTQD is coupled to $\gamma_{+\uparrow}$. The interference effects between the topological and quantum dot states, significantly modify the value of the $\delta I_{K}$. The low energy doublet is  insensitive to increasing the coupling strength $t$, in contrast to the quantum conductance (black region in Fig. 34c), which changes from $2(e^{2}/h)$ to $(5/2)(e^{2}/h)$ (Fig. 31b, c).

Figures 35 show the density plots of the nonlinear shot noise $\delta S_{K}=T^{2}_{[\star]}|S_{K}|$ as a function of $E_{d}$ and $t$. The yellow lines represent the noiseless transport $\delta S_{K}\approx0$. The lines are doubled and between them the noise due to three-particle processes changes the sign $\delta S_{K}<0$ (yellow lines in Fig. 35a-c). In Fig. 35b we observe two noiseless lines separating the quantum states $\ket{q_{w}}_{2}$ from $\ket{q_{y}}_{2}$ and $\ket{q_{a(b)}}_{4}$ from $\ket{q_{z(x)}}_{2}$. In the intermediate and strong coupling regime for $\ket{q_{x(z)}}_{2}$, $\delta S_{K}$ reaches a finite value $\delta S_{K}\approx0.15$, leading to super-Poissonian behavior of $F_{K}$ (the bright white area in Fig. 36a-c). The double black lines where $F_{K}\approx0$ result from two types of phase shifts in Eq. (39) : $2\delta_{\nu}$ and $4\delta_{\nu}$. For $\Delta=10^{-2}$ the noiseless transport occurs in the charge stability regions for two quantum states $\ket{q_{w}}_{2}$ and $\ket{q_{y}}_{2}$ (Fig. 35a).  With increasing SOI there is a significant asymmetry in $\delta S_{K}$ between $Q=1$e and $Q=3$e.  In the high energy doublet, due to the coupling strength of the CNTQD to $\gamma_{+\uparrow}$, interference effects of the topological state significantly modify the value of the nonlinear shot noise. For $Q=1$e, we observe a change of the $\delta S_{K}$ for $\ket{p_{ls}}_{2}$ and $\ket{q_{a}}_{2}$ from $1.25$ to $0.55$ for $\Delta=10^{-1}$ and from $1.5$ to $0.75$ for $\Delta=2$.
\begin{figure}[t!]
\includegraphics[width=0.75\linewidth]{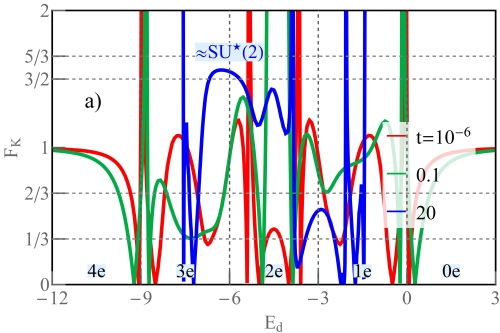}\\
\includegraphics[width=0.75\linewidth]{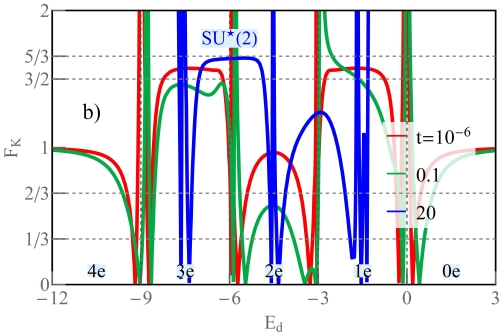}\\
\includegraphics[width=0.75\linewidth]{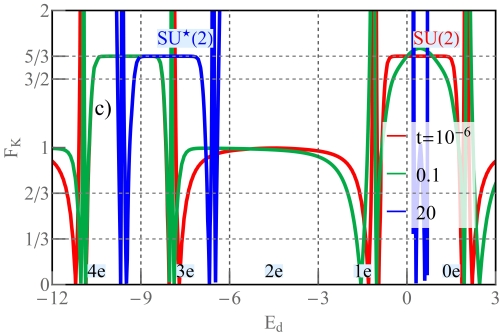}
\caption{\label{fig:epsart} (Color online) a-c) $E_{d}$ dependence of $F_{K}$ with increasing $t$ for weak, intermediate and strong SOI.}
\end{figure}

Figure 36 shows the density plots of the evolution of $F_{K}=\delta {S}_{K}/\delta {I}_{K}$ in functions $E_{d}$ and $t$ with increasing SOI in the quantum dot. Fig. 36a shows $F_{K}$ for $\Delta=0.01$. In the sector of the doublet states $\ket{q_{w}}_{2}$ and $\ket{q_{y}}_{2}$ and between $\ket{q_{a(b)}}_{4}$ and $\ket{q_{z(x)}}_{2}$, we observe the black lines symbolizing the noiseless transport $\delta S_{K}\approx0$, with strong enhancement when $\delta I_{K}\approx0$ and the nonlinear transport is blocked (bright white line and regions Fig. 36). For $Q=1$e and $Q=3$e, we observe the partially filled blue region where $F_{K}\approx1$ ($\Delta=0.01$), which is a precursor of the SU(2) and SU$^{\star}$(2)  Kondo states.  Figure 36a shows a narrow region for $Q=1$e and $Q=3$e with $F_{K}\approx1/3$, which reveals as a residual of the Kondo state with SU(4) symmetry (Fig. 28a). The asymmetry of $F_{K}$ manifests itself for the broken SU$^{\star}$(3) Kondo state, where finite $\Delta=0.01$ for $Q\approx(5/2)$e leads to an increase of the shot noise $F_{K}\approx(5/3)$ ($\ket{q_{g}}_{4}$) and for the
$Q\approx(3/2)$e we observe a decrease of the noise $F_{K}\approx0.5$ ($\ket{q_{y}}_{2}$). The transition from $\ket{q_{y}}_{2}$ to $\ket{q_{w}}_{2}$ is seen by changing the sign and reducing the nonlinear current $\delta I_{K}$, then $F_{K}$ reaches $1/2$, via $-1$ and to the super-Poissonian value $F_{K}\gg1$ (Fig. 36).

The  cross sections of $F_{K}$ are shown in Fig. 37a. The blue line in Fig. 37a, for $Q=(5/3)$e approaches to the value of $F_{K}=e^{\star}/e\approx(5/3)$ for the SU$^{\star}$(2) Kondo effect. For $Q=(3/2)$e and $t=20$, we observe the reduction of the Fano factor to $F_{K}=e^{\star}/e\approx1/2$ (Fig. 37a). In the region, where two doublets $\ket{q_{x(z)}}_{2}$ dominate, the Fano factor $F_{K}$ reaches the super-Poissonian values $F_{K}\gg1$, which is due to the finite value of the nonlinear shot noise $\delta S_{K}\approx0.15$, relative to the value of the current $\delta I_{K}\approx10^{-3}$. This behavior persists for the three values of the SOI $\Delta=0.01,0.1$ and $2$ (the white areas in Fig. 36, and the blue lines in Fig. 37). For $\Delta=0.1$ and $t=10^{-6}$, $F_{K}$ in the region of one and three electrons on the quantum dot is quantized close to $5/3$ for $\ket{p_{ls}}_{2}$ and $\ket{t_{ls}}_{2}$ states, which is the characteristic fingerprint of the SU(2) Kondo state (red line in Fig. 37b). With increasing the coupling strength $t$, the e-h symmetry is broken and  for $\Delta=0.1,2$, the SU$^{\star}$(2) Kondo state is realized  for the quadruplet $\ket{q_{g}}_{4}$ (blue curves in Fig. 37c). The high energy doublet evolves to the U$^{\star}$(1) charge symmetry, where between $\ket{q_{w}}_{2}$ and $\ket{q_{z}}_{2}$ we observe $F_{K}=1$ splitting two super-Poissonian regions.
\begin{figure}[t!]
\includegraphics[width=0.75\linewidth]{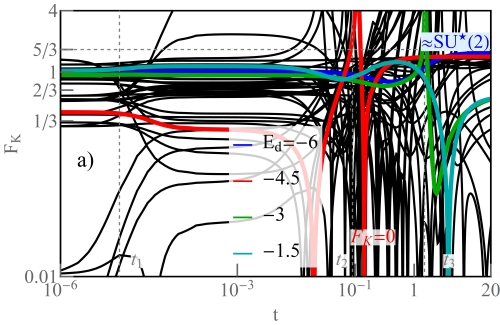}\\
\includegraphics[width=0.75\linewidth]{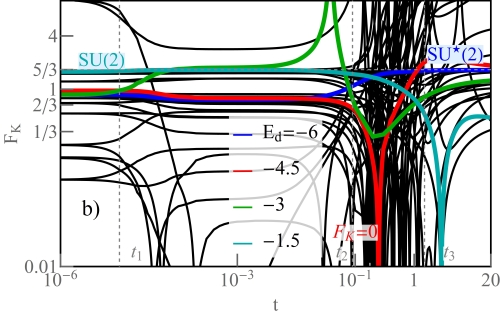}\\
\includegraphics[width=0.75\linewidth]{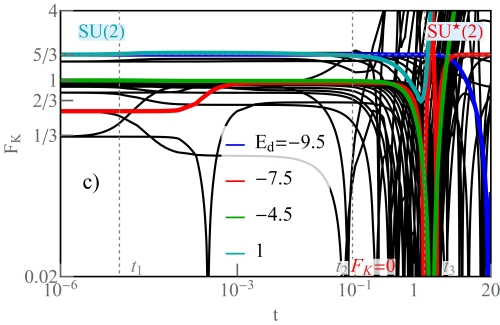}
\caption{\label{fig:epsart} (Color online) a-c) The landscape log-log plot of $F_{K}$ as a function of $t$ for $\Delta=10^{-2},10^{-1}$ and $2$.
$F_{K}=0$ corresponds the noiseless nonlinear transport ($S_{K}=0$).}
\end{figure}

Figure 38 shows the landscape plots of $F_{K}$ as a function of $t$. The black curves are plotted from $E_{d}=-10$ to $E_{d}=1$ with increment $\delta E_{d}=0.25$.  In Fig. 38a, b for $E_{d}=-6$ (blue line) and in Fig. 38c for $E_{d}=-7.5$ (red line), in the the strong coupling regime, $F_{K}$ approaches to $5/3$ for $Q=(5/2)$e, in contrast to $Q=(3/2)$e ($E_{d}=-3$, Fig. 38a, c), where $F_{K}\approx(1/2)$ for $\Delta=0.01$ and $F_{K}\approx1$  for $\Delta=0.1$. The changes in nonlinear current and shot noise are determined by the doublet quantum state $\ket{q_{y}}_{2}$. The coupling term to $\gamma_{+\uparrow}$ and $\Delta$ contributes to the asymmetry in $F_{K}$ between low and high energy doublets. The curve for $E_{d}=-4.5$ ($Q=2$e) shows two points of compensation in the Fano factor $F_{K}=0$. We observe the noiseless transport for $\ket{q_{w}}_{2}$. Between the point of the blocked nonlinear transport, we observe a negative shot noise $\delta S_{K}<0$ (red line in Fig. 38a). For $\Delta=0.1$, two compensation points are reduced to one noiseless point at the boundary of the doublet state (red curve in Fig. 38b).
For the quantum state $\ket{q_{y}}_{2}$, we observe the asymptotic enhancement of the Fano factor $F_{K}\gg1$ (green lines in Fig. 38a).
With increasing the SOI, for the intermediate spin-orbit coupling $\Delta=0.1$, the enhancement of $F_{K}$ shifts from $t_{3}$ to $t_{1}$. At this point we observe the super-Poissonian value of $F_{K}$, and the nonlinear current is blocked. For $\Delta=0.1$, $\delta I_{K}\approx0$ appears on the charge degeneracy line between two states $\ket{q_{w}}_{2}$ and the quadruplet $\ket{q_{a}}_{4}$. With increasing SOI, the super-Poissonian $F_{K}$ evolves to noiseless transport for $E_{d}=-4.5$ (green line in Fig. 38c). For $t=10^{-6}$ and intermediate and strong SOI, the SU(2) Kondo effect is realized by two doublets $\ket{p_{ls}}_{2}$ and the Fano factor reaches to the quantized value $F_{K}=(5/3)$ (dark cyan lines in Fig. 38b, c).
With increasing the coupling strength $F_{K}$ is constant and reaches $5/3$, but the effective charge e$^{\star}$  determines the quartet state $\ket{q_{a}}_{4}$. The dynamical changes to super-Poissonian values are observed in the strong coupling regime for $\ket{q_{x(z)}}_{2}$ and $Q=2$e, when $\ket{q_{w}}_{2}$ is the ground state (green line in Fig. 38c). The enhancement of $F_{K}$ is preceded by noiseless point around $t_{3}$ (black lines in Fig. 36 and green line in Fig. 38c).
\begin{figure}[h!]
\includegraphics[width=0.75\linewidth]{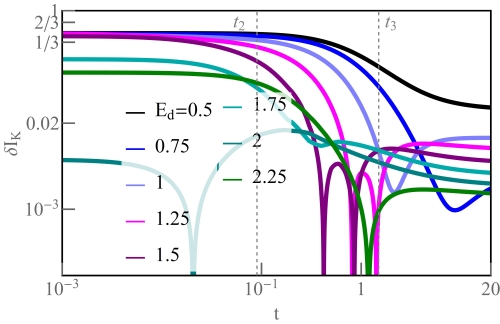}\\
\includegraphics[width=0.75\linewidth]{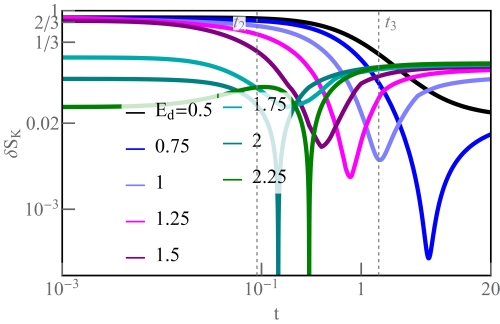}\\
\includegraphics[width=0.75\linewidth]{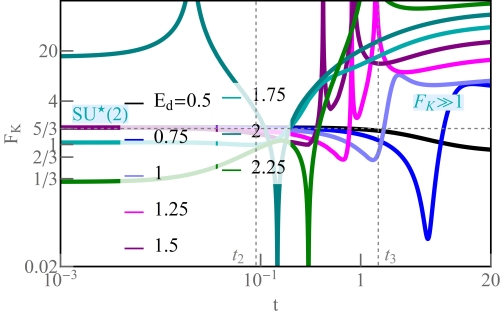}
\caption{\label{fig:epsart} (Color online) Super-Poissonian $F_{K}$: a-c) Gate-dependent $\delta I_{K}$, $\delta S_{K}$ and $F_{K}$ as a function of $t$ ($\Delta=2$). $F_{K}\gg1$ appears for $t$, where the nonlinear transport is determined by Majorana fermion-coupled channel.}
\end{figure}

Figure 39 shows the nonlinear current, shot noise and Fano factor as a function of the coupling strength $t$ for $\Delta=2$. The cross sections are drawn from $E_{d}=1/2$ to $1.75$ and intersect the quartet $\ket{q_{a}}_{4}$ for the weak coupling strength to TSC and the doublet $\ket{q_{z}}_{2}$ in the strong coupling regime. The curve for $E_{d}=2$ is plotted at the boundary of the $\ket{q_{a}}_{4}$ and $\ket{q_{z}}_{2}$ states.  At the charge boundary, $\delta I_{K}=0$ and $\delta S_{K}=0$ appear for two different values of $t$, hence we observe the logarithmic divergence of $F_{K}$ for $t=10^{-2}$, and noiseless transport near $t_{2}$ (darker cyan curve Fig. 39c). For $E_{d}=2.25$ the transport is determined by double $\ket{q_{z}}_{2}$ and in $F_{K}$ we observe the opposite tendency, the transport is blocked around $t_{2}$ and $F_{K}$ is  logarithmically divergent at $t_{3}$. For weak coupling and in the range between $E_{d}=0.5$ and $1.5$, the SU$^{\star}$(2) Kondo effect is realized (Fig. 38c) and $F_{K}$ reaches $5/3$. The super-Poissonian $F_{K}$ results from the blocking effect of the nonlinear current, and the presence of the spin-orbit interaction in the system. The $F_{K}$ for the doublet states show two types of the behavior:  $F_{K}\approx1$ for weak coupling strength to the TSC and $F_{K}\gg1$ in the strong coupling regime (Fig. 36c). The current and shot noise are modified by the SOI, in particular the contributions of the two- and three-body correlators in the current are compared to each other (Fig. 39a),  hence $\delta I_{K}\approx0$  we observe the logarithmic divergence of $F_{K}$ (Fig. 39c).
For $E_{d}=1.25$ and $E_{d}=1.5$ we observe two compensation points, where $\delta I_{K}=0$.
Between these points the current is negative and in the nonlinear regime, the current flows in the opposite direction to the applied voltage (the backward current of the quasiparticles), whereas before this behavior was reserved only for nonlinear noise. The super-Poissonian value of the nonlinear Fano is always accompanied by a sign reversal of the nonlinear current.
The calculations require a more detailed analysis in the future. However, it is worth mentioning that the result of the experiment \cite{Yamauchi2011}, where the authors obtained $F_{K}=3$ for the quantum dot system, is also surprising, perhaps the system is not in the full SU(2) Kondo state, and the symmetry is broken by an additional disorder. At the moment it is difficult to explain this with the existing theory.

\section{Conclusions}
In summary, we have studied  the transport properties of the novel type of the fractional Kondo effect with SU$^{\star}$(3) symmetry in the strong coupling regime with 1TSC. For the CNTQD system coupled with two Majorana fermions, we discovered the SU$^{\star}$(2) Kondo state with an even number of electrons on the quantum dot. The CNTQD-3TSC device showed that a state with charge symmetry U$^{\star}$(1) determines the quantum conductance ${\mathcal{G}}=(5/2)(e^{2}/h)$ in the strong coupling regime. For the octuplets $\ket{q_{x(z)}}_{8}$ we observed the charge leakage quantum effect. The effect was seen in the high order susceptibilities, nonlinear current and shot noise measurements. For SU$^{\star}$(3) Kondo phase, the charge fluctuations were finite and led to $\Delta N^{2}=1/4$, which is due to the coexistence of a coupled channel with a Majorana fermion.

In the linear thermoelectric power coefficient, in the range of weak coupling to TSCs, we observed NFL behavior with strong enhancement of the TEP with two compensation points, where $\gamma_{(S)}=0$.
In the range of weak and strong coupling, $\gamma_{(S)}$ led to FL-type behavior and the numbers  characterized the Kondo state with full symmetry. The extended KR sbMFA method showed the complementary results to the NRG calculations \cite{Teratani2020}, in particular with respect to the SU(4) Kondo effect. The new type of strongly correlated phases SU$^{\star}$(3) and SU$^{\star}$(2) showed $F_{K}=2/3$ and $F_{K}=5/3$. In this paper, a so-called the weak coupling ansatz is proposed to calculate the Wilson coefficients, and consequently the off-diagonal two- and three-body correlators. Measurements of the linear coefficient of the TEP and the nonlinear Fano factor, in the case of the broken SU(2) Kondo state, by SOI and most likely in the case of arbitrary other perturbations contain the information about the quality of the Kondo effect and its symmetry. The Kondo temperature with increasing coupling to TSCs, at the transition limit between the SU(2) and SU$^{\star}$(2) showed an enhancement controlled by the Coulomb interaction in CNTQD and by the tunneling rate to the normal electrodes. The total entropy as a function of temperature is quenched for the Kondo state. For a single electron in the CNTQD-1TSC device, the entropy has reached $\ln[4]/4$, which is closely related to the symmetry of the Kondo effect. By increasing the number of Majorana fermions coupled to the CNTQD, the entropy is raised in the intermediate temperature range to the following numbers $(N_{TS}\ln[4])/4$.

In the tunneling entropy we observed a sign reversal and a negative value, which is characteristic for TSC-coupled systems - indicating strong order. For the SU$^{\star}$(3) Kondo state, the fluctuations of the pseudospin moment increased with the temperature, and the high temperature limit of the entropy reached to the value of $\ln[3]$, suggesting the threefold degeneracy of the quantum states. The linear Fano effect led to the quantized fractional values. By measuring the linear coefficient of the TEP, we presented the detection of the lifetime of the Majorana fermion state and the moment of the transition from the NFL-like behavior to the FL phase. Hybrid devices with TSC showed the negative spin (orbital) polarization of the quantum conductance.
For 1TSC and 3TSC, the spin and orbital polarization are identical in sign and value. Shot noise and current measurements allowed indirect the determination of the pseudospin and charge susceptibilities. The SOI in the CNTQD-1TSC system, due to the formation of two quadruplets, led to an asymmetric behavior in the nonlinear current, shot noise and effective charge. The number of $N_{TS}$ topological sectors was introduced for the entanglement of the quantum states from even and odd charge regions, and the Hilbert space was extended to $2^{n+N_{TS}}$. In the strong coupling regime, for the doublet states as the ground state, we observed super-Poissonian values of the Fano factor, which is related to the damping of the nonlinear current by the SOI in CNTQD. In this paper, we have investigated the transport and correlation properties of the CNTQD system in the Kondo state coupled to a TSC. We have studied the quantum transport quantities such as quantum conductance, thermoelectric power, linear and nonlinear current, and shot noise in a wide range of the coupling term to the topological superconductor.
We have shown that CNTQD in the SU(4) Kondo state, can provide a very precise detector of the Majorana bound states, and the residual interaction between the quasiparticles.

\textbf{Acknowledgements}. This work received support from National Science Center in Poland through the research Project No. 2018/31/D/ST3/03965.
\newline
\bibliography{krych2024}{}
\end{document}